\documentclass{article}

\usepackage{graphicx}
\usepackage{amsmath}
\usepackage{amsfonts}
\usepackage{amssymb}
\usepackage{verbatim}

\renewcommand{\arraystretch}{1.3}

\catcode`\@=11
\def\marginnote#1{}

\newcount\hour
\newcount\minute
\newtoks\amorpm
\hour=\time\divide\hour by60
\minute=\time{\multiply\hour by60 \global\advance\minute by-\hour}
\edef\standardtime{{\ifnum\hour<12 \global\amorpm={am}%
        \else\global\amorpm={pm}\advance\hour by-12 \fi
        \ifnum\hour=0 \hour=12 \fi
        \number\hour:\ifnum\minute<10 0\fi\number\minute\the\amorpm}}
\edef\militarytime{\number\hour:\ifnum\minute<10 0\fi\number\minute}

\def\draftlabel#1{{\@bsphack\if@filesw {\let\thepage\relax
      \xdef\@gtempa{\write\@auxout{\string
          \newlabel{#1}{{\@currentlabel}{\thepage}}}}}\@gtempa \if@nobreak
    \ifvmode\nobreak\fi\fi\fi\@esphack} \gdef\@eqnlabel{#1}}
    \def\@eqnlabel{}
\def\@vacuum{}
\def\draftmarginnote#1{\marginpar{\raggedright\scriptsize\tt#1}}

\def\draft{
  \oddsidemargin -.5truein
  \def\@oddfoot{\footnotesize \sl preliminary draft \hfil
    \rm\thepage\hfil\sl\today\quad\militarytime}
  \let\@evenfoot\@oddfoot \overfullrule 3pt
    \let\label=\draftlabel
    \let\marginnote=\draftmarginnote
  \def\@eqnnum{(\theequation)\rlap{\kern\marginparsep\tt\@eqnlabel}%
    \global\let\@eqnlabel\@vacuum}

  }

\makeatletter
\newdimen\normalarrayskip              
\newdimen\minarrayskip                 
\normalarrayskip\baselineskip
\minarrayskip\jot
\newif\ifold             \oldtrue            \def\new{\oldfalse}
\def\arraymode{\ifold\relax\else\displaystyle\fi} 
\def\eqnumphantom{\phantom{(\theequation)}}     
\def\@arrayskip{\ifold\baselineskip\z@\lineskip\z@
     \else
     \baselineskip\minarrayskip\lineskip2\minarrayskip\fi}
\def\@arrayclassz{\ifcase \@lastchclass \@acolampacol \or
\@ampacol \or \or \or \@addamp \or
   \@acolampacol \or \@firstampfalse \@acol \fi
\edef\@preamble{\@preamble
  \ifcase \@chnum
     \hfil$\relax\arraymode\@sharp$\hfil
     \or $\relax\arraymode\@sharp$\hfil
     \or \hfil$\relax\arraymode\@sharp$\fi}}
\def\@array[#1]#2{\setbox\@arstrutbox=\hbox{\vrule
     height\arraystretch \ht\strutbox
     depth\arraystretch \dp\strutbox
     width\z@}\@mkpream{#2}\edef\@preamble{\halign
\noexpand\@halignto
\bgroup \tabskip\z@ \@arstrut \@preamble \tabskip\z@ \cr}%
\let\@startpbox\@@startpbox \let\@endpbox\@@endpbox
  \if #1t\vtop \else \if#1b\vbox \else \vcenter \fi\fi
  \bgroup \let\par\relax
  \let\@sharp##\let\protect\relax
  \@arrayskip\@preamble}
\def\eqnarray{\stepcounter{equation}%
              \let\@currentlabel=\theequation
              \global\@eqnswtrue
              \global\@eqcnt\z@
              \tabskip\@centering
              \let\\=\@eqncr

 \halign to \displaywidth\bgroup
    \eqnumphantom\@eqnsel\hskip\@centering
    $\displaystyle \tabskip\z@ {##}$%
    \global\@eqcnt\@ne \hskip 2\arraycolsep
         $\displaystyle\arraymode{##}$\hfil
    \global\@eqcnt\tw@ \hskip 2\arraycolsep
         $\displaystyle\tabskip\z@{##}$\hfil
         \tabskip\@centering
    &{##}\tabskip\z@\cr}
\newfont{\hr}{msbm10}
\newfont{\ams}{msam10}

\hoffset= - 0.9in        

\def\beq{\begin{equation}}
\def\eeq{\end{equation}}
\def\ba{\beq\new\begin{array}{c}}
\def\ea{\end{array}\eeq}
\def\be{\ba}
\def\ee{\ea}

\def\N2{${\cal N}=2$}

\def\1N{${\cal N}=1$}
\def\4N{${\cal N}=4$}
\def\nn{\nonumber}

\newdimen\linethick  \linethick=0.4pt
\newdimen\hboxitspace    \hboxitspace=5pt
\newdimen\vboxitspace    \vboxitspace=5pt

\def\fr#1{%
\beq\new
\vcenter{
\hrule height\linethick
          \hbox{\vrule width\linethick
                \kern\hboxitspace
                \vbox{\kern\vboxitspace
                      \hbox{$\begin{array}{c}\displaystyle#1
         \end{array}$}%
                      \kern\vboxitspace}%
                \kern\hboxitspace
                \vrule width\linethick}%
          \hrule height\linethick}%
\eeq}

\renewcommand{\tt}[1][mer]{\hbox{\tiny{#1}}}

\newcommand{\Tr}{\mathop{\rm Tr}\nolimits}

\def\tr{{\rm tr}\,}
\def\Tr{{\rm Tr}\,}

\def\SS{{^*\!S}}
\def\l[{\phantom.[}

\title{{\bf Character expansion
for HOMFLY polynomials. \\III. All 3-Strand braids in the first
symmetric representation} \vspace{.5cm}}
\author{{\bf H.Itoyama}\footnote{ {\small {\it
Department of Mathematics and Physics,
Osaka City University} and {\it Osaka City University Advanced Mathematical Institute (OCAMI), Osaka, Japan}};
itoyama@sci.osaka-cu.ac.jp}, \ {\bf A.Mironov}\footnote{ {\small {\it
Lebedev Physics Institute} and {\it ITEP, Moscow, Russia}};
mironov@itep.ru; mironov@lpi.ru}, \ {\bf A.Morozov}\thanks{{\small
{\it ITEP, Moscow, Russia}}; morozov@itep.ru}, \ {\bf
And.Morozov}\thanks{{\small {\it Moscow State University} and {\it ITEP, Moscow, Russia}};
Andrey.Morozov@itep.ru}\date{ }}

\begin{document}

\setcounter{footnote}{3}

\setcounter{tocdepth}{3}

\maketitle

\vspace{-6.5cm}

\begin{center}
\hfill FIAN/TD-05/12\\
\hfill ITEP/TH-15/12\\
\hfill OCU-PHYS-366
\end{center}

\vspace{3.5cm}

\begin{abstract}
We continue the program of systematic study of extended HOMFLY
polynomials, suggested in \cite{I,II}. Extended polynomials depend
on infinitely many time-variables, are close relatives of integrable
$\tau$-functions, and depend on the choice of the braid
representation of the knot. They possess natural character
decompositions, with coefficients which can be defined by
exhaustively general formula for any particular number $m$ of
strands in the braid and any particular representation $R$ of the
Lie algebra $GL(\infty)$. Being restricted to "the topological
locus" in the space of time-variables, the extended HOMFLY
polynomials reproduce the ordinary knot invariants. We derive such a
general formula, for $m=3$, when the braid is parameterized by a
sequence of integers $(a_1,b_1,a_2,b_2,\ldots)$, and for the first
non-fundamental representation $R = [2]$. Instead of calculating the
mixing matrices directly, as suggested in \cite{II}, we deduce them
from comparison with the known answers for torus and composite
knots. A simple reflection symmetry converts the answer for the
symmetric representation $[2]$ into that for the antisymmetric one
$[1,1]$. The result applies, in particular, to the figure eight knot
$4_1$, and was further extended to superpolynomials in arbitrary
symmetric and antisymmetric representations in \cite{IMMMfe}.
\end{abstract}

\bigskip

\bigskip

\section{Introduction}

In \cite{I,II} we started a program to promote the HOMFLY
polynomials \cite{HOMFLY} to character expansions, representing them
as linear combinations of the Schur functions $S_Q\{p_k\}$ (i.e. the
characters of linear groups) \cite{Fulton}. Such an expansion
depends on the choice of a braid realization of the
knot, thus,
its coefficients by themselves are not knot invariants, instead they
are pure group theory quantities and possess many nice properties.
For an $m$-strand braid ${\cal B}$ the HOMFLY polynomial in
representation $R$ is expanded as\footnote{Our calculus is based on the approach by \cite{RT},
though that of \cite{inds,Rama00,ZR} is, by essence, also very close.
The first part of this formula is related to Chern-Simons theory in the
temporal gauge \cite{MorSm}. The new element is a special emphasis on the
character expansion, which allows one to extend the knot polynomials
to arbitrary time variables \cite{I} and provides
very simple {\it general} formulas for entire classes of knots.
However,
following this line, we
omit the additional {\it factor}
$(A^\alpha q^\gamma)^{w^{\cal B}}$
in front on the trace in (\ref{1}), where $w^{\cal B}$ is the writhe
number of the braid, while $\alpha$ and $\gamma$ depend
on normalization of the ${\cal R}$-matrix.
Throughout this paper we use a special normalization of
${\cal R}$-matrices,
though in our normalization $\alpha$ and $\gamma$ are
actually non-vanishing: for most purposes (in the standard framing)
$\alpha = -|R|$ and $\gamma = -4\varkappa_R$.
We discuss this issue in a separate
subsection \ref{ff}, and actually restore the factors
in the formulas of the Appendix.
To simplify the notations we do not put star on $H(A)$
in this paper,
as we do in \cite{I,II,IMMMfe},
because the extended polynomials are almost not mentioned here.
They can be, however, obtained from the formulas of the Appendix
simply by removing the stars from $\SS_Q$, thus, promoting
them from the quantum dimensions to the Schur functions.  }
\be\label{1}
H_R^{\cal B} = \Tr_{R^{\otimes
m}}\Big((q^\rho)^{\otimes m}{\cal B}\Big) = \sum_{Q\vdash m|R|}
h_{RQ}^{\cal B}\,S_Q^*(A)
\ee
where
\be\label{tolo}
S_Q^*(A) = \Tr_{Q\in
R^{\otimes m}} (q^\rho)^{\otimes m} = S_Q\{p_k^*\}, \ \ \ \ \ \ \ \
\ \
p_k^* = {[kN]\over [k]} = \frac{A^k-A^{-k}}{q^k-q^{-k}}
\ee
are quantum dimensions of representations $Q$ of $SU(N)$,
where $A=q^N$ and $[x] = \frac{q^x-q^{-x}}{q-q^{-1}}$.
The coefficients $h_{RQ}^{\cal B}$ do not depend on $A$,
i.e. on $N$,
thus, they can be evaluated from analysis of an arbitrary group $SU_q(N)$.
Instead they can be represented as traces in auxiliary
spaces of intertwining operators ${\cal M}_{R^mQ}$, whose dimension
is the number ${\rm dim}\, {\cal M}_{R^mQ} = N_{R^mQ}$ of times
the irreducible representation $Q$ appears in the $m$-th tensor power
of representation $R$,
\be
R^{\otimes m} = \sum_{Q\vdash m|R|} {\cal M}_{R^mQ} \otimes Q
\label{decoRmQ}
\ee
The trace is taken of a product of the diagonal quantum ${\cal R}$-matrices
$\widehat{\cal R}$ acting in ${\cal M}_{R^mQ}$,
and the "mixing matrices" intertwining ${\cal R}$-matrices
that act on different pairs of adjacent strands in the braid.
These mixing matrices, in their turn, can be represented as products
of universal constituents, associated with a switch between two
"adjacent" trees describing various decompositions (\ref{decoRmQ}).

In \cite{II} we exhaustively described
such representations for the coefficients $h_{RQ}^{\cal B}$
for {\it arbitrary} $m=2,3,4$-strand braids
and for the simplest fundamental representation $R=[1]$:
\be\label{4}
\begin{array}{ccc}
m=2, & {\cal B} = {\cal R}^a: &
H_{[1]}^{(a)} = q^a S_2^*(A) + \left(-\frac{1}{q}\right)^a S_{11}^*(A)
=  q^a S_2^*(A) \ +\ \left( q \longrightarrow -\frac{1}{q}\right)
\end{array}
\\
\\
\ee
\be
m=3, \ \ \  {\cal B} = ({\cal R}\otimes I)^{a_1}(I\otimes {\cal R})^{b_1}
({\cal R}\otimes I)^{a_2}(I\otimes {\cal R})^{b_2} \ldots \ \ \ \ \ \ \ \
\ \ \ \ \ \ \ \ \ \ \ \ \ \ \ \ \ \ \ \ \ \ \ \
\ee
$$
H_{[1]}^{(a_1,b_1|a_2,b_2|\ldots)} =
q^{\sum_i (a_i+b_i)} S_3^*(A) + \left(-\frac{1}{q}\right)^{\sum_i(a_i+b_i)} S_{111}^*(A)
+ \Big(\Tr_{2\times 2} \widehat{\cal R}_2^{a_1}U_2 \widehat{\cal R}_2^{b_1}U_2^\dagger
 \widehat{\cal R}_2^{a_2}U_2 \widehat{\cal R}_2^{b_2}U_2^\dagger \ldots\Big) S_{21}^*(A)
$$
\be\label{6}
\\
m=4, \ \ \  {\cal B} = ({\cal R}\otimes I\otimes I)^{a_1}(I\otimes {\cal R}\otimes I)^{b_1}
 ({\cal R}\otimes I\otimes I)^{c_1}
({\cal R}\otimes I\otimes I)^{a_2}(I\otimes {\cal R}\otimes I)^{b_2}
({\cal R}\otimes I\otimes I)^{c_2} \ldots \ \ \ \ \ \ \ \
 \ee
$$
H_{[1]}^{(a_1,b_1,c_1|a_2,b_2,c_2|\ldots)} =
q^{\sum_i (a_i+b_i+c_i)} S_4^*(A) + \left(-\frac{1}{q}\right)^{\sum_i(a_i+b_i+c_i)} S_{1111}^*(A) +
$$
$$
+ \Big(\Tr_{2\times 2} \widehat{\cal R}_2^{a_1}U_2 \widehat{\cal R}_2^{b_1}U_2^\dagger
 \widehat{\cal R}_2^{c_1+a_2}U_2 \widehat{\cal R}_2^{b_2}U_2^\dagger  \widehat{\cal R}_2^{c_2+a_3}
 \ldots\Big) S_{22}^*(A) +
$$
$$
+ \left\{\Big(\Tr_{3\times 3} \widehat{\cal R}_3^{a_1}U_3 \widehat{\cal R}_3^{b_1} V_3U_3\widehat{\cal R}_3^{c_1}
U_3^\dagger V_3^\dagger U_3^\dagger
\widehat{\cal R}_3^{a_2}U_3 \widehat{\cal R}_3^{b_2} V_3U_3\widehat{\cal R}_3^{c_2}
U_3^\dagger V_3^\dagger U_3^\dagger
 \ldots\Big) S_{31}^*(A)\ + \ \left(q \longrightarrow -\frac{1}{q}\right)\right\}
$$
where
\be
\hat{\cal R}_2 = \left(\begin{array}{cc} q & \\ &-\frac{1}{q}\end{array}\right)\ \ \ \ \ \ \ \ \ \
\hat{\cal R}_3 = \left(\begin{array}{ccc} q & & \\ & q &\\ &&-\frac{1}{q}\end{array}\right)\nn
\ee
\be
U_2 = \left(\begin{array}{cc} c_2 & s_2\\ -s_2 &c_2\end{array}\right)\ \ \ \ \ \ \ \
U_3 = \left(\begin{array}{ccc} 1&& \\ &c_2 & s_2\\ &-s_2 &c_2\end{array}\right)\ \ \ \ \ \ \ \
V_3 = \left(\begin{array}{ccc} c_3 & s_3 & \\ -s_3 &c_3 & \\ && 1\end{array}\right)
\label{UVmatr}
\ee
Subscripts refer to the size of the matrices, the entries of rotation matrices $U$ and $V$
are given by
\be
c_k = \frac{1}{[k]}\,, \ \ \ \
s_k = \sqrt{\,1-c_k^2\,} = \frac{\sqrt{\,[k-1]^{\phantom {1^1}}\!\!\![k+1]\,}}{[k]}
\ee
These formulas provide a very transparent and convenient representation for infinitely
many HOMFLY polynomials
and seem to be very useful for any theoretical analysis of their general properties,
from integrability to linear Virasoro like relations
(including $A$-polynomials and spectral curves \cite{Apol,DiFu,3dAGT,Brini},
AMM/EO topological recursion \cite{AMM/EO,GS} etc).
Therefore, further insights are important about the structure of these formulas
and their generalizations
(in \cite{II} the $m=5$ case was also investigated,
and the general formula for the coefficients $h_{[1]}^{[m-1,1]}$ was suggested
for all $m$).

The limitations in \cite{II} are pure technical:
to make the paper readable and the main ideas understandable,
we considered only the implications of the theory of $SU_q(2)$ quantum group:
this allowed us to calculate only contributions of the Young diagrams $Q$
with one and two columns or rows.
Including the diagrams with $l$ columns or rows, which arise when the
number of strands in the braid is $m\geq 5$, requires the similar
use of the $SU_q(l)$ quantum group theory, which is tedious but straightforward,
and will be considered in further publications.
(We emphasize once again that $l$ has nothing to do with $N$ in $A=q^N$,
the relevant $l$ is related to the number $m$ of strands,
and for small $m\leq 4$ the smallest $l=2$ is sufficient to describe
{\it all} the HOMFLY polynomials with $R=[1]$.)

Another restriction in \cite{II} was to $R=[1]$.
It is also partly related to restriction to $SU_q(2)$, but not only.
It is the purpose of the present paper to make a first step
in the direction towards the {\it colored} HOMFLY polynomials with $|R|>1$,
that is, to the symmetric representation $R=[2]$.
{\bf Instead of performing this calculation directly, {\it a la} \cite{II}, we use a shortcut:
we determine the five parameters (angles) of the three orthogonal matrices $\hat U_{[51]}$, $\hat U_{[321]}$
and $\hat U_{[42]}$ by comparison with the answers for torus and composite knots and links
in eq.(\ref{H3n}) below.}
This can be also compared with the results of the long-lasting impressive work by the
Indian group \cite{inds} using the direct evaluation of the Racah coefficients.

Our goal is to find the necessary ingredients for formulas like (\ref{4})-(\ref{6}),
which provide an {\it exhaustive} description of {\it all} braids with a given number
of strands and in a given representation.
The final task would be to find general formulas that depend explicitly on
{\it all} the parameters: the number of strands $m$, the set $(a_{1i},\ldots,a_{mi})$,
specifying the $m$-strand braid (i.e. on the element of the braid group),
and on the Young diagram $R$ labeling the representation.
The formula is going to be for a coefficient in front of a particular character,
like the Schur function $S_Q\{p\}$ (or, alternatively, Hall-Littlewood \cite{HLexp}
or some other element of an appropriate expansion basis).
Now we perform a step of this program: find such {\it parametric} formulas
for given $m=3$ and $R=[2]$. For $R=[1]$ and $m=3,4$ (and partly $m=5$)
they were already derived in \cite{II}.

The formulas of this paper for the mixing matrices, which we obtain here indirectly, can be obtained directly
using the representation theory, like it was done in \cite{II}.
We consider the mixing matrices for a more generic case in
\cite{IV}.




\section{The case of 2 strands, $m=2$}

To determine emerging $Q$ in this case, one suffices to expand the product of two symmetric representations:
\be
\l[2]\times[2] = [4]+[31]+[22],
\ \ \ \
\l[11]\times [11] = [22] + [211] + [1111]
\ee
This decomposition can be easily obtained from the decomposition of the characters.
Indeed, given $S_{[2]} = \frac{1}{2}p_2 + \frac{1}{2}p_1^2$, $S_{[11]} = -\frac{1}{2}p_2 + \frac{1}{2}p_1^2$ and
\be
S_{[4]} = \frac{1}{4}p_4 + \frac{1}{3}p_3p_1 + \frac{1}{8}p_2^2 + \frac{1}{4}p_2p_1^2 + \frac{1}{24}p_1^4,  \\
S_{[31]} = -\frac{1}{4}p_4 - \frac{1}{8}p_2^2 + \frac{1}{4}p_2p_1^2 + \frac{1}{8}p_1^4,  \\
S_{[22]} =  - \frac{1}{3}p_3p_1 + \frac{1}{4}p_2^2   + \frac{1}{12}p_1^4,  \\
S_{[211]} = \frac{1}{4}p_4 -  \frac{1}{8}p_2^2 - \frac{1}{4}p_2p_1^2 + \frac{1}{8}p_1^4,  \\
S_{[1111]} = -\frac{1}{4}p_4 + \frac{1}{3}p_3p_1 + \frac{1}{8}p_2^2 - \frac{1}{4}p_2p_1^2 + \frac{1}{24}p_1^4,
\ee
it is easy to check that
\be
S_{[2]}^2 = S_{[4]} + S_{[31]} + S_{[22]}, \ \ \ \ \ \ S_{[11]}^2 = S_{[22]} + S_{[211]} + S_{[1111]}
\ee
In particular, for the ordinary dimensions $d_Q = S_Q\{p_k = N\}$ of representations $Q$ of
the $SU(N)$ algebra it reads as
$\left(\frac{N(N+1)}{2}\right)^2 = \frac{N(N+1)(N+2)(N+3)}{24} +
\frac{(N-1)N(N+1)(N+2)}{8} + \frac{(N^2-1)N^2}{12} $ etc.
and very similarly for the quantum dimensions $D_Q = S_Q^* = S_Q\{p_k = [N]_q\}$:
\be
\left(\frac{[N][N+1]}{[2]}\right)^2 = \frac{[N][N+1][N+2][N+3]}{[2][3][4]} +
\frac{[N-1][N][N+1][N+2]}{[2][4]} + \frac{[N-1][N]^2[N+1]}{[2]^2[3]}\\
\left({[N-1][N]\over [2]}\right)^2= \frac{[N-1][N]^2[N+1]}{[2]^2[3]}+{[N-2][N-1][N][N+1]\over [2][4]}+
{[N-3][N-2][N-1][N]\over [2][3][4]}
\ee

The HOMFLY polynomials are given by \cite{chi} (see also \cite{DMMSS} for
an extension to superpolynomials \cite{sup})
\be
H^{2,n}_{[1]} = \sum_{Q\in [1]\times [1]} \epsilon_Q^n q^{n\varkappa_Q} S_Q
= q^{n} S_{[2]} \mp q^{-n} S_{[11]}   \\
H^{2,n}_{[2]} = \sum_{Q\in [2]\times [2]} \epsilon_Q^n q^{n\varkappa_Q} S_Q
= q^{6n} S_{[4]} \mp q^{2n} S_{[31]} + S_{[22]}  \\
H^{2,n}_{[11]} = \sum_{Q\in [2]\times [2]} \epsilon_Q^n q^{n\varkappa_Q} S_Q
= q^{-6n} S_{[1111]} \mp q^{-2n} S_{[211]} + S_{[22]}
\label{H2n}
\ee
where\footnote{Similarly to \cite{I,II},
we use here $\varkappa_Q = -\nu_Q + \nu_{Q'}$ with the opposite sign as compared with \cite{DMMSS}.}
$\varkappa_Q = -\nu_Q + \nu_{Q'}$, $\nu_Q=\sum (i-1)Q_i$, $Q'$ is the transposed Young diagram and
the sign factors $\epsilon_Q$ are defined from the Adams rule \cite{chi}: "the initial condition"
$H^{m,0}_{[R]}\{p\} = \widehat{Ad}_mS_R\{p\}$ with
$\widehat{Ad}_m p_k = p_{mk}$ is imposed at $n=0$
\be
\widehat{Ad}_2 S_{[1]}\{p\} = \frac{p_2+p_1^2}{2} = S_{[2]} - S_{[11]}  \\
\widehat{Ad}_2 S_{[2]}\{p\} = \frac{p_4+p_2^2}{2} = S_{[4]} - S_{[31]} + S_{[22]}  \\
\widehat{Ad}_2 S_{[11]}\{p\} = \frac{p_4+p_2^2}{2} = S_{[1111]} - S_{[211]} + S_{[22]}
\label{Ad2knots}
\ee
It should not be mixed with the "physical" initial conditions  for the $n$-evolution
of \cite{DMMSS},
\be
H^{m,n}_R =\ H^{n,m}_R
\ee
imposed at $0<n<m$.

For links one has instead of (\ref{Ad2knots}):
\be
\Big(\widehat{Ad}_1 S_{[1]}\{p\}\Big)^2 = p_1^2 = S_{[2]}+S_{[11]},  \\
\Big(\widehat{Ad}_1 S_{[2]}\{p\}\Big)^2 = \frac{(p_2+p_1^2)^2}{4} = S_{[4]} + S_{[31]} + S_{[22]},  \\
\Big(\widehat{Ad}_1 S_{[11]}\{p\}\Big)^2 = \frac{(-p_2+p_1^2)^2}{4} = S_{[1111]} + S_{[211]} + S_{[22]}
\label{Ad2links}
\ee
Accordingly, the signs $\mp$ at the r.h.s. of (\ref{H2n}) are minuses and pluses
for knots and links ($n$ odd or even) respectively.

In particular, for the unknot with $(m,n)=(2,1)$
\be
H_{[1]}^{2,1} = qS_{[2]}^* - q^{-1}S_{[11]}^* = qS_{[1]}^*\\
H_{[2]}^{2,1} = q^{6}S_{[4]}^* - q^{2}S_{[31]}^* + S_{[22]}^* = A^2q^4S_{[2]}^*  \\
H_{[11]}^{2,1} = q^{-6}S_{[1111]}^* - q^{-2}S_{[211]}^* + S_{[22]}^* = A^2q^4S_{[11]}^*
\ee
for the Hopf link with $(m,n)=(2,2)$
\be
H_{[1]}^{2,2} = q^{2}S_{[2]}^* + q^{-2}S_{[11]}^* = \left((q^2-1+q^{-2})A-A^{-1}\right){S_{[1]}^*\over\{q\}}  \\
H_{[2]}^{2,2} = q^{12}S_{[4]}^* + q^{4}S_{[31]}^* + S_{[22]}^* = \left((q^{12}-q^{10}-q^8+2q^6-q^2+1)A^2
-q^8 +q^4-q^2-1+q^2A^{-2}\right)q{S_{[2]}^*\over\{q\}\{q^2\}} \\
H_{[11]}^{2,2} = q^{-12}S_{[1111]}^* + q^{-4}S_{[211]}^* + S_{[22]}^* =\\= \left((1-q^{-2}+2q^{-6}-q^{-8}-q^{-10}+
q^{-12})A^2-1-q^{-2}+q^{-4}-q^{-8}+q^{-2}A^{-2}\right){S_{[11]}\over q\{q\}\{q^2\}}
\ee
and for the trefoil with $(m,n)=(2,3)$
\be
H_{[1]}^{2,3} = q^{3}S_{[2]}^* - q^{-3}S_{[11]}^* = \left((q^2+q^{-2})A-A^{-1}\right)S_{[1]}^*  \\
H_{[2]}^{2,3} = q^{18}S_{[4]}^* - q^{6}S_{[31]}^* + S_{[22]}^* =\left( (q^{12}+q^6+q^4+1)A^2-q^8-q^6-q^2-1
+q^2A^{-2}\right)q^4S_{[2]}^*  \\
H_{[11]}^{2,3} = q^{-18}S_{[1111]}^* - q^{-6}S_{[211]}^* + S_{[22]}^* = \left((q^{-12}+q^{-6}+q^{-4}+1)A^2-
q^{-8}-q^{-6}-q^{-2}-1+q^{-2}A^{-2}\right){S_{[11]}^*\over q^4}
\ee

\section{ 3 strands, $m=3$}

\subsection{Structure of the answer}

Now
\be
\l[2]\times[2]\times[2] = ([4]+[31]+[22])\times [2] =  \\ =
([6] + [51] + [42]) + ([51] + [42] + [411] + [33] + [321]) + ([42]+[321]+[222])
\ee
For example, for the dimensions of $SU(2)$ representations one has
$3^3=27 = (7+5+3) + (5+3+0+1 + 0) + (3+0+0)$. Again, this decomposition is obtained as the decomposition of
the characters:
\be
S_{[6]} = \frac{1}{6}p_6+\frac{1}{5}p_5p_1+\frac{1}{8}p_4p_2+\frac{1}{8}p_4p_1^2
+\frac{1}{18}p_3^2+\frac{1}{6}p_3p_2p_1+\frac{1}{18}p_3p_1^3+\frac{1}{48}p_2^3+
\frac{1}{16}p_2^2p_1^2+\frac{1}{48}p_2p_1^4+\frac{1}{720}p_1^6,  \\
S_{[51]} = -\frac{1}{6}p_6 -  \frac{1}{8}p_4p_2+\frac{1}{8}p_4p_1^2
-\frac{1}{18}p_3^2 +\frac{1}{9}p_3p_1^3-\frac{1}{48}p_2^3+
\frac{1}{16}p_2^2p_1^2+\frac{1}{16}p_2p_1^4+\frac{1}{144}p_1^6,  \\
S_{[42]} = -\frac{1}{5}p_5p_1+\frac{1}{8}p_4p_2-\frac{1}{8}p_4p_1^2 +\frac{1}{16}p_2^3+
\frac{1}{16}p_2^2p_1^2+\frac{1}{16}p_2p_1^4+\frac{1}{80}p_1^6,  \\
S_{[411]} = \frac{1}{6}p_6
+\frac{1}{18}p_3^2-\frac{1}{6}p_3p_2p_1+\frac{1}{18}p_3p_1^3-\frac{1}{24}p_2^3
-\frac{1}{8}p_2^2p_1^2+\frac{1}{24}p_2p_1^4+\frac{1}{72}p_1^6,  \\
S_{[33]} =  -\frac{1}{8}p_4p_2-\frac{1}{8}p_4p_1^2
+\frac{1}{9}p_3^2+\frac{1}{6}p_3p_2p_1-\frac{1}{18}p_3p_1^3-\frac{1}{16}p_2^3+
\frac{1}{16}p_2^2p_1^2+\frac{1}{48}p_2p_1^4+\frac{1}{144}p_1^6,  \\
S_{[321]} = \frac{1}{5}p_5p_1 - \frac{1}{9}p_3^2 - \frac{1}{9}p_3p_1^3 + \frac{1}{45}p_1^6,  \\
\ldots  \\
S_{[222]} = -\frac{1}{8}p_4p_2+\frac{1}{8}p_4p_1^2
+\frac{1}{9}p_3^2-\frac{1}{6}p_3p_2p_1-\frac{1}{18}p_3p_1^3+\frac{1}{16}p_2^3+
\frac{1}{16}p_2^2p_1^2-\frac{1}{48}p_2p_1^4+\frac{1}{144}p_1^6,  \\
\ldots
\ee

Thus, one needs the $2\times 2$ mixing matrices for representations $[51]$ and $[321]$
and the $3\times 3$ mixing matrix for representation $[42]$.

The answer for the HOMFLY polynomial in the fundamental representation for
the generic $3$-strand knot $(a_1,b_1|a_2,b_2|\ldots)$
has the following form:
\be
H_{[1]}^{a_1,b_1|a_2,b_2|\ldots} =
q^{a_1+b_1+a_2+b_2+\ldots}\,S_{[3]}
+ \left(-\frac{1}{q}\right)^{a_1+b_1+a_2+b_2+\ldots} S_{[111]} +  \\
+ \tr_{2\times 2} \Big\{\hat{\cal R}_{[21]}^{a_1} \hat U_{[21]} \hat{\cal R}_{[21]}^{b_1} \hat U_{[21]}^\dagger
\hat{\cal R}_{[21]}^{a_2} \hat U_{[21]} \hat{\cal R}_{[21]}^{b_2} \hat U_{[21]}^\dagger \ldots\Big\}S_{[21]}
\ee
with
\be
\hat{\cal R}_{[21]}
= \left(\begin{array}{cc} q^{\varkappa_{[2]}} & 0 \\ 0 & -q^{\varkappa_{[11]}} \end{array}\right)
= \left(\begin{array}{cc} q & 0 \\ 0 & -q^{-1} \end{array}\right), \ \ \ \ \ \
\hat U_{[21]} = \left(\begin{array}{cc} c_2 & s_2 \\ -s_2 & c_2 \end{array}\right)
\ee
Likewise, in the symmetric representation, it is going to be
\be
H_{[2]}^{a_1,b_1|a_2,b_2|\ldots} =
(q^6)^{a_1+b_1+a_2+b_2+\ldots}\,S_{[6]}
+ (-q^2)^{a_1+b_1+a_2+b_2+\ldots}\,\Big(S_{[411]}+S_{[33]}\Big) + S_{[222]} +  \\
+ \tr_{2\times 2} \Big\{\hat{\cal R}_{[51]}^{a_1} \hat U_{[51]} \hat{\cal R}_{[51]}^{b_1} \hat U_{[51]}^\dagger
\hat{\cal R}_{[51]}^{a_2} \hat U_{[51]} \hat{\cal R}_{[51]}^{b_1} \hat U_{[51]}^\dagger \ldots\Big\}S_{[51]}
+  \\
+ \tr_{2\times 2} \Big\{\hat{\cal R}_{[321]}^{a_1} \hat U_{[321]} \hat{\cal R}_{[321]}^{b_1} \hat U_{[321]}^\dagger
\hat{\cal R}_{[321]}^{a_2} \hat U_{[321]} \hat{\cal R}_{[321]}^{b_1} \hat U_{[321]}^\dagger \ldots\Big\}S_{[321]}
+  \\
+ \tr_{3\times 3} \Big\{\hat{\cal R}_{[42]}^{a_1} \hat U_{[42]} \hat{\cal R}_{[42]}^{b_1} \hat U_{[42]}^\dagger
\hat{\cal R}_{[42]}^{a_2} \hat U_{[42]} \hat{\cal R}_{[42]}^{b_1} \hat U_{[42]}^\dagger \ldots\Big\}S_{[42]}
\label{H3n}
\ee
Here
\be
\hat{\cal R}_{[6]} = q^{\varkappa_{[4]}} = q^6,\ \ \ \ \ \ \ \
\hat{\cal R}_{[411]} = \hat{\cal R}_{[33]} = -q^{\varkappa_{[31]}} = -q^2, \ \ \ \ \ \ \ \
\hat{\cal R}_{[222]} = q^{\varkappa_{[22]}} = 1,  \\  \\
\hat{\cal R}_{[51]}
= \left(\begin{array}{cc} q^{\varkappa_{[4]}} & 0 \\ 0 & -q^{\varkappa_{[31]}} \end{array}\right)
= \left(\begin{array}{cc} q^6 & 0 \\ 0 & -q^{2} \end{array}\right), \ \ \ \ \ \ \nn
\ee
\be
\hat{\cal R}_{[321]}
= \left(\begin{array}{cc} -q^{\varkappa_{[31]}} & 0 \\ 0 & -q^{\varkappa_{[22]}} \end{array}\right)
= \left(\begin{array}{cc} -q^2 & 0 \\ 0 & 1 \end{array}\right), \ \ \ \ \ \nn
\ee
\be
\hat{\cal R}_{[42]}
= \left(\begin{array}{ccc} q^{\varkappa_{[4]}} & 0  & 0\\ 0 & -q^{\varkappa_{[31]}} & 0 \\
0 & 0 & q^{\varkappa_{[22]}} \end{array}\right)
=\left(\begin{array}{ccc} q^6 & 0 & 0 \\ 0 & -q^{2} & 0 \\ 0 & 0 & 1 \end{array}\right)
\label{R42}
\ee
and the mixing matrices $\hat U_Q$ need to be calculated.

Two of them , those for the double-line diagrams $[51]$ and $[42]$, can be evaluated with the help
of representation theory of $SU_q(2)$, but in the $[321]$ sector at least $SU_q(3)$ would be needed.
Instead of performing this calculation directly, {\it a la} \cite{I,II}, we use a shortcut:
determine the five parameters (angles) of the three orthogonal matrices $\hat U_{[51]}$, $\hat U_{[321]}$
and $\hat U_{[42]}$ by comparison with the answers for torus and composite knots and links.

\subsection{Torus knots}

For torus knots $T[3,n]$ with $a_1=b_1=\ldots=a_n=b_n=1$ one has an alternative decomposition \cite{chi}:
\be
\underline{H}_{[2]}^{3,n} = \sum_{Q\vdash 6} q^{\frac{2n}{3}\varkappa_Q} C_{[2]}^QS_Q
\label{Hbar3[2]}
\ee
where the coefficients are defined from the Adams rule
\be
\widehat{Ad}_3 S_{[2]} = \frac{p_6+p_3^2}{2} = \sum_{Q\vdash 6} C_{[2]}^QS_Q =
S_{[6]} - S_{[5,1]} + \underline{0\cdot S_{[42]}} + S_{[411]}
+ S_{[33]} - S_{[321]} + S_{[222]},  \\
\Big(\widehat{Ad}_1 S_{[2]}\Big)^3 = \frac{(p_2+p_1^2)^3}{8} = \sum_{Q\vdash 6} C_{[2]}^QS_Q =
S_{[6]} + 2S_{[5,1]} + \underline{3\cdot S_{[42]}} + S_{[411]}
+ S_{[33]} + 2S_{[321]} + S_{[222]}
\ee
for knots and links, i.e. for $n=1,2\ ({\rm mod}\ 3)$ and $n = 0\ ({\rm mod}\ 3)$ respectively.

Thus for the knots, $n=1,2\ ({\rm mod}\ 3)$
\be
\underline{H}_{[2]}^{3,n} = q^{10n}S_{[6]} - q^{6n}S_{[5,1]} + \underline{0\cdot q^{10n/3}S_{[42]}} + q^{2n}S_{[411]}
+ q^{2n}S_{[33]} - S_{[321]} + q^{-2n}S_{[222]} =  \\
= q^{-2n}\Big(q^{12n}S_{[6]} - q^{8n}S_{[5,1]} + \underline{0\cdot q^{16n/3}S_{[42]}} + q^{4n}S_{[411]}
+ q^{4n}S_{[33]} - q^{2n}S_{[321]} + S_{[222]}\Big)
\label{H3n[2]kn}
\ee
Note that the only would be contribution with non-integer value of $\frac{1}{3}\varkappa_Q$
(underlined) does not contribute in the case of torus knots: the Adams coefficient
$C_{[2]}^{[42]}=0$.

Looking at the coefficients in front of the fully known "singlet" terms
$S_{[6]}$, $S_{[411]}$, $S_{[33]}$, $S_{[222]}$,
which do not involve yet unknown mixing matrices,
we see that eq.(\ref{Hbar3[2]}) differs from the correct expression by a factor of
\be
\underline{H}_{[2]}^{3,n} = q^{-2n}{H}_{[2]}^{3,n}
\ee
For the generic single-line (symmetric) representations $[p]$ and arbitrary number $m$ of strands
one gets, comparing the coefficients in front of $S_{[pm]}$:
\be\label{chicorr}
\underline{H}_{[p]}^{m,n} = q^{\frac{2n}{m}\varkappa_{[mp]}-n(m-1)\varkappa_{[2p]}}{H}_{[p]}^{m,n}
= q^{-n(m-2)p(p-1)}{H}_{[p]}^{m,n}
\ee
so that there is no discrepancy for either the first fundamental representation $p=1$
or for the case of $m=2$ strands, when all the knots are torus.

For the links, $n=0\ ({\rm mod}\ 3)$
\be
\underline{H}_{[2]}^{3,n} = q^{10n}S_{[6]} +2 q^{6n}S_{[5,1]} + \underline{3\cdot q^{10n/3}S_{[42]}}
+ q^{2n}S_{[411]} + q^{2n}S_{[33]} + 2S_{[321]} + q^{-2n}S_{[222]} =  \\
= q^{-2n}\Big(q^{12n}S_{[6]} + 2q^{8n}S_{[5,1]} + \underline{3\cdot q^{16n/3}S_{[42]}} + q^{4n}S_{[411]}
+ q^{4n}S_{[33]} + 2q^{2n}S_{[321]} + S_{[222]}\Big)
\label{H3n[2]ln}
\ee
This time the underlined terms are non-vanishing, but since for the links $n\vdots 3$,
the power is integer in this case.

Note that the coefficients are the same for knots and links in front of the
terms $S_{[6]}$, $S_{[411]}$, $S_{[33]}$ and $S_{[222]}$, in full accordance with
(\ref{H3n}), because for the torus knots and links $a_i+b_i$ is either $2$ or $0$,
i.e. always even, so that the corresponding signs $\epsilon_Q$ can not affect the answers
in the torus case (however, they affect the answers for the composite knots, see s.\ref{comp} below).

These formulas generalize those for the fundamental representation:
\be
H_{[1]}^{3,n} = q^{2n}S_{[3]} - S_{[21]} + q^{-2n}S_{[111]},\ \ \ \ n=1,2\ ({\rm mod}\ 3)  \\
H_{[1]}^{3,n} = q^{2n}S_{[3]} + 2S_{[21]} + q^{-2n}S_{[111]},\ \ \ \ n=0\ ({\rm mod}\ 3)
\label{H3n[1]}
\ee
considered in \cite{II}.


\subsection{$2\times 2$ matrices $\hat U_{[51]}$ and $\hat U_{[321]}$ from the torus knots}

When mixing matrix is of the size $2\times 2$, it can be parameterized by a single parameter $s$,
sine of the mixing angle, cosine $c$ being related through $c^2+s^2=1$.
Then we have for an elementary building block
\be
\hat{\cal R}^{a} \hat U \hat{\cal R}^{b} \hat U^\dagger =
\left(\begin{array}{cc} \epsilon q^{\varkappa} & 0 \\ 0 & \tilde\epsilon q^{\tilde\varkappa}
\end{array}\right)^{a}
\left(\begin{array}{cc} c & s \\ -s & c
\end{array}\right)
\left(\begin{array}{cc} \epsilon q^{\varkappa} & 0 \\ 0 & \tilde\epsilon q^{\tilde\varkappa}
\end{array}\right)^{b}
\left(\begin{array}{cc} c & -s \\ s & c
\end{array}\right)
=  \\
= \left(\begin{array}{cc}
\epsilon^{a+b} q^{\varkappa(a+b)}c^2 + \epsilon^a\tilde\epsilon^b q^{\varkappa a + \tilde\varkappa b}s^2
& \left( -\epsilon^{a+b} q^{\varkappa(a+b)}
+ \epsilon^a\tilde\epsilon^b q^{\varkappa a + \tilde\varkappa b}\right)cs \\
\left( \tilde\epsilon^{a+b} q^{\tilde\varkappa(a+b)}
- \epsilon^b\tilde\epsilon^a q^{\varkappa b + \tilde\varkappa a}\right)cs
& \tilde\epsilon^{a+b} q^{\tilde\varkappa(a+b)}c^2
+ \epsilon^b\tilde\epsilon^a q^{\varkappa b + \tilde\varkappa a}s^2
\end{array}\right)
\ee
In the case of torus knots and links $a=b=1$ and this reduces to
\be
\hat{\cal R} \hat U \hat{\cal R} \hat U^\dagger =
 \left(\begin{array}{cc}
  q^{2\varkappa}c^2 + (\epsilon\tilde\epsilon) q^{\varkappa + \tilde\varkappa}s^2
& \left( -  q^{2\varkappa}
+ (\epsilon\tilde\epsilon) q^{\varkappa  + \tilde\varkappa }\right)cs \\
\left(   q^{2\tilde\varkappa}
- (\epsilon\tilde\epsilon) q^{\varkappa +\tilde\varkappa }\right)cs
&   q^{2\tilde\varkappa }c^2
+ (\epsilon\tilde\epsilon) q^{\varkappa  \tilde\varkappa }s^2
\end{array}\right)
\ee
and
\be
\Tr_{2\times 2} \hat{\cal R} \hat U \hat{\cal R} \hat U^\dagger =
(q^{2\varkappa}+q^{2\tilde\varkappa })c^2 + 2(\epsilon\tilde\epsilon) q^{\varkappa + \tilde\varkappa}s^2
 = \Big(q^{2\varkappa}+q^{2\tilde\varkappa }\Big)
 - \Big(q^\varkappa - (\epsilon\tilde\epsilon)q^{\tilde\varkappa}\Big)^2s^2
 \label{Tr31}
\ee
Now it remains to substitute the relevant values of $\varkappa$, $\tilde\varkappa$,
$\epsilon$ and $\varepsilon$, and compare this trace with the relevant coefficient of the HOMFLY
polynomial for the torus knot $T[3,1]$
(it is essentially the unknot but realized by a non-simplest braid; since we do not need to restrict
ourselves to the topological locus here, this expression is not the same as $S_{[R]}$).
After that one can calculate the traces of powers of this matrix and check that with the same
value of $s$ they
reproduce the values of the coefficient for all other torus knots and links $T[3,n]$ with different $n$.
This is, in fact, not a problem, because one should just check that with the right value of $s$
the matrix $\hat{\cal R} \hat U \hat{\cal R} \hat U^\dagger$ has appropriate eigenvalues,
proportional to the roots of unity.
Finally, the same value of $s$ determines the coefficient for all other 3-strand braids
$(a_1,b_1,a_2,b_2,\ldots)$.

\bigskip

{\bf The case of $R=[1]$ and the term $S_{[21]}$.}
We start with this already known case, \cite{II} for illustrative purposes. One has to
substitute $\varkappa = \varkappa_{[2]} = 1$, $\tilde\varkappa_{[11]} = -1$,
$\epsilon=1$, $\tilde\epsilon = -1$ and compare (\ref{Tr31}) with the value of the coefficient
in front of $S_{21}$ in (\ref{H3n[1]}) with $n=1$, which is $-1$. This gives:
\be
q^2+q^{-2} - (q+q^{-1})^2s^2 = -1 \ \Longrightarrow \
s = \frac{\sqrt{q^2+1+q^{-2}}}{q+q^{-1}}=\frac{\sqrt{[3]}}{[2]}=s_2, \ \
c = \frac{1}{q+q^{-1}} = \frac{1}{[2]} = c_2
\ee
This reproduces the answer (\ref{UVmatr}) for $U_2$ from \cite{II}.

It is easy to check that, with this values of $s$ and $c$,
\be
{\det}_{2\times 2}\Big(\hat{\cal R} \hat U \hat{\cal R} \hat U^\dagger - \lambda I\Big)
 = \frac{\lambda^3-1}{\lambda-1} = \lambda^2+\lambda+1
\ee
i.e. the two eigenvalues of $\hat{\cal R} \hat U \hat{\cal R} \hat U^\dagger$ are $e^{\pm \frac{2\pi i}{3}}$,
so that
\be
\Tr_{2\times 2} \Big(\hat{\cal R} \hat U \hat{\cal R} \hat U^\dagger\Big)^n
= \left\{\begin{array}{ccc}  -1  & {\rm for} & n=1,2\ ({\rm mod}\ 3) \\
 +2  & {\rm for} & n=0\ ({\rm mod}\ 3)\end{array}\right.
\ee
in full agreement with (\ref{H3n[1]}).

\bigskip

{\bf The case of $R=[2]$ and the term $S_{[51]}$.}
One has to substitute $\varkappa = \varkappa_{[4]} = 6$, $\tilde\varkappa_{[31]} = 2$,
$\epsilon=1$, $\tilde\epsilon = -1$ and compare (\ref{Tr31}) with the value of the coefficient
in front of $S_{[51]}$ in (\ref{H3n[2]kn}) with $n=1$, which is $-q^8$. This gives:
\be
q^{12}+q^{4} - (q^6+q^{2})^2s^2 = -q^8 \ \Longrightarrow \
s = \frac{\sqrt{q^4+1+q^{-4}}}{q^2+q^{-2}} = \frac{\sqrt{[3]_{q^2}}}{[2]_{q^2}}, \ \
c = \frac{1}{q^2+q^{-2}} = \frac{1}{[2]_{q^2}}
\ee
where $[x]_{q^2}\equiv {q^{2x}-q^{-2x}\over q^2-q^{-2}}$.

Again, it is a simple exercise to check that with these values of $s$ and $c$
\be
{\det}_{2\times 2}\Big(\hat{\cal R} \hat U \hat{\cal R} \hat U^\dagger - \lambda I\Big)
= \lambda^2 + q^8\lambda+q^{16}
\ee
i.e. the two eigenvalues of $\hat{\cal R} \hat U \hat{\cal R} \hat U^\dagger$ are
$q^8e^{\pm\frac{2\pi i}{3}}$ and
\be
\Tr_{2\times 2} \Big(\hat{\cal R} \hat U \hat{\cal R} \hat U^\dagger\Big)^n
= \left\{\begin{array}{ccc}  -q^{8n}  & {\rm for} & n=1,2\ ({\rm mod}\ 3) \\
 +2q^{8n}  & {\rm for} & n=0\ ({\rm mod}\ 3)\end{array}\right.
\ee
in full agreement with (\ref{H3n[2]kn}) and (\ref{H3n[2]ln}).

\bigskip

{\bf The case of $R=[2]$ and the term $S_{[321]}$.}
One has to substitute $\varkappa = \varkappa_{[31]} = 2$, $\tilde\varkappa_{[22]} = 0$,
$\epsilon=-1$, $\tilde\epsilon = 1$ and compare (\ref{Tr31}) with the value of the coefficient
in front of $S_{[321]}$ in (\ref{H3n[2]kn}) with $n=1$, which is $-q^2$. This gives:
\be
q^{4}+1 - (q^2+1)^2s^2 = -q^2 \ \Longrightarrow \
s = \frac{\sqrt{q^2+1+q^{-2}}}{q+q^{-1}} = \frac{\sqrt{[3]}}{[2]}, \ \
c = \frac{1}{q+q^{-1}} = \frac{1}{[2]}
\ee

With these values of $s$ and $c$
\be
{\det}_{2\times 2}\Big(\hat{\cal R} \hat U \hat{\cal R} \hat U^\dagger - \lambda I\Big)
= \lambda^2 + q^2\lambda+q^{4}
\ee
i.e. the two eigenvalues of $\hat{\cal R} \hat U \hat{\cal R} \hat U^\dagger$ are
$q^2e^{\pm\frac{2\pi i}{3}}$ and
\be
\Tr_{2\times 2} \Big(\hat{\cal R} \hat U \hat{\cal R} \hat U^\dagger\Big)^n
= \left\{\begin{array}{ccc}  -q^{2n}  & {\rm for} & n=1,2\ ({\rm mod}\ 3) \\
 +2q^{2n}  & {\rm for} & n=0\ ({\rm mod}\ 3)\end{array}\right.
\ee
again in excellent agreement with (\ref{H3n[2]kn}) and (\ref{H3n[2]ln}).

\subsection{Constraining the $3\times 3$ matrix $\hat U_{[42]}$   from the torus knots}

When orthogonal mixing matrix is of the size $3\times 3$, it can be parameterized by three
independent Euler angles, namely by their sines and cosines:
\be
\hat U = \left(\begin{array}{ccc} c_1  & 0 & s_1 \\ 0& 1 & 0\\ -s_1 & 0  & c_1 \end{array}\right)
\left(\begin{array}{ccc} 1&0&0\\0&c_2 & s_2 \\ 0&-s_2 & c_2  \end{array}\right)
\left(\begin{array}{ccc} c_3  & 0 & s_3\\ 0& 1 & 0\\ -s_3  & 0 & c_3  \end{array}\right)
\label{Eude}
\ee
One now needs to perform the same trick: to compare the traces of powers of
$\hat{\cal R} \hat U \hat{\cal R} \hat U^\dagger$, where  $\hat{\cal R}$ is given in (\ref{R42})
with the known coefficients in front of $S_{[42]}$ in (\ref{H3n[2]kn}).
This comparison tells that
\be
\tr_{3\times 3} (\hat{\cal R} \hat U \hat{\cal R} \hat U^\dagger)^n =
\left\{\begin{array}{ccc}  0 & {\rm for} & n = 1,2\ ({\rm mod}\ 3) \\
3q^{16n/3} & {\rm for} & n = 0\ ({\rm mod}\ 3)
\end{array}\right.
\label{probtri}
\ee
for diagonal
\be
\hat{\cal R} = \left(\begin{array}{ccc} q^6 & 0 & 0 \\ 0 & -q^2 & 0 \\ 0 & 0 & 1\end{array}\right)
\ee
The choice of the Euler decomposition in (\ref{Eude}) is obviously adjusted to this form of the
matrix $\hat{\cal R}$.
At $q=1$ a solution is obvious:
\be
\hat{\cal R} = \left(\begin{array}{ccc} 1 & 0 & 0 \\ 0 & -1 & 0 \\ 0 & 0 & 1\end{array}\right)
\ \Longrightarrow \
\hat{U} = \left(\begin{array}{ccc} 1 & 0 & 0 \\ 0 & c & s \\ 0 & -s & c\end{array}\right), \ \ \ \
\hat{\cal R} \hat U \hat{\cal R} \hat U^\dagger =
\left(\begin{array}{ccc} 1 & 0 & 0 \\ 0 & c^2-s^2 & -2cs \\ 0 & 2cs & c^2-s^2 \end{array}\right)
\ee
i.e. one gets the rotation matrix with the doubled angle $\phi$,  $s = \sin\phi$.
Then (\ref{probtri}) means that $6\phi = 2\pi k$ with any integer $k$,
i.e. $\phi = \frac{\pi k}{3}$, and $c = \pm \frac{1}{2}$, $s = \pm\frac{\sqrt{3}}{2}$.
Of course, at $q=1$ there is a huge degeneracy: any rotation involving only the first and the third
lines leaves $\hat{\cal R}(q=1)$ intact, and one can take many other $\hat U$, obtained by such a
rotation, for example,
$\hat{U} = \left(\begin{array}{ccc} c & s & 0 \\ -s & c & 0 \\ 0 & 0 & 1\end{array}\right)$
with the same $c$ and $s$. For $q=1$ only one of the three Euler angles in $U$ is fixed by
conditions (\ref{probtri}).

At $q\neq 1$ conditions (\ref{probtri}) imply that the three eigenvalues of
$\hat{\cal R} \hat U \hat{\cal R} \hat U^\dagger$ are three cubic roots of unity
times $q^{16/3}$, i.e. that
\be
{\det}_{3\times 3} \Big(\hat{\cal R} \hat U \hat{\cal R} \hat U^\dagger - \lambda I\Big)
= q^{16}-\lambda^3
\ee
Clearly, these are only two conditions, so that only two of the three Euler angles
will be fixed by (\ref{probtri}).
One extra condition, not just coming from the 3-strand torus knot and link polynomials, will be needed
to fix $\hat U_{[42]}$ unambiguously.

We impose this condition by making an "educated guess" that $c_3=c_1$ and $s_3=-s_1$.
Then
\be
c_1 = c_3 = \frac{1+q^4}{\sqrt{(2q^4-q^2+2)(1+q^2+q^4)}}\,,
\ \ \ \ \ \ c_2 = -\frac{q^4-q^2+1}{1+q^4} = -\frac{1+q^6}{(1+q^2)(1+q^4)}
\ee
(the sign in $c_2$ is essential).

One can use an alternative parametrization instead of (\ref{Eude}):
\be
\hat U = \left(\begin{array}{ccc} c_1'  & 0 & s_1' \\ 0& 1 & 0\\ -s_1' & 0  & c_1' \end{array}\right)
\left(\begin{array}{ccc} c_2' & s_2' & 0 \\ -s_2' & c_2' &0\\0&0&1 \end{array}\right)
\left(\begin{array}{ccc} c_3'  & 0 & s_3'\\ 0& 1 & 0\\ -s_3'  & 0 & c_3'  \end{array}\right)
\label{Eude'}
\ee
In this case the Euler angles are given by
\be
c_1' = c_3' = \sqrt{\frac{1-q^{10}}{(2q^4-q^2+2)(1-q^6)}}= -s_1=s_3,
\ \ \ \ \ \ c_2' = -\frac{q^4-q^2+1}{1+q^4} = -\frac{1+q^6}{(1+q^2)(1+q^4)}=c_2
\ee
i.e. are "dual" to those for (\ref{Eude}).
In both cases one obtains the same matrix $\hat U$, see eq.(\ref{U42}) below.

It remains an open question, whether a nicer decomposition exists for this
rather sophisticated mixing matrix.

\subsection{The final answer}

Substituting into (\ref{H3n}) the values of the mixing angles,
found in the previous subsections
one finally obtains for arbitrary $3$-strand braid:
\be
H_{[2]}^{a_1,b_1|a_2,b_2|\ldots} =
q^{6(a_1+b_1+a_2+b_2+\ldots)}\,S_{[6]}
+ (-q^2)^{a_1+b_1+a_2+b_2+\ldots}\,\Big(S_{[411]}+S_{[33]}\Big) + S_{[222]} +  \\
+ \tr_{2\times 2}\left\{
\left(\begin{array}{cc} q^6 & 0 \\ 0 & -q^2\end{array}\right)^{a_1}
\left(\begin{array}{cc} -\frac{1}{[2]_{q^2}} & \frac{\sqrt{[3]_{q^2}}}{[2]_{q^2}} \\
-\frac{\sqrt{[3]_{q^2}}}{[2]_{q^2}} & -\frac{1}{[2]_{q^2}}\end{array}\right)
\left(\begin{array}{cc} q^6 & 0 \\ 0 & -q^2\end{array}\right)^{b_1}
\left(\begin{array}{cc} -\frac{1}{[2]_{q^2}} & -\frac{\sqrt{[3]_{q^2}}}{[2]_{q^2}} \\
\frac{\sqrt{[3]_{q^2}}}{[2]_{q^2}} & -\frac{1}{[2]_{q^2}}\end{array}\right)
\ldots\right\}S_{[51]} +  \\
+ \tr_{2\times 2}
\left\{
\left(\begin{array}{cc} -q^2 & 0 \\ 0 & 1\end{array}\right)^{a_1}
\left(\begin{array}{cc} -\frac{1}{[2]_{q}} & \frac{\sqrt{[3]_{q}}}{[2]_{q}} \\
-\frac{\sqrt{[3]_{q}}}{[2]_{q}} & -\frac{1}{[2]_{q}}\end{array}\right)
\left(\begin{array}{cc} -q^2 & 0 \\ 0 & 1\end{array}\right)^{b_1}
\left(\begin{array}{cc} -\frac{1}{[2]_{q}} & -\frac{\sqrt{[3]_{q}}}{[2]_{q}} \\
\frac{\sqrt{[3]_{q}}}{[2]_{q}} & -\frac{1}{[2]_{q}}\end{array}\right)
 \ldots\right\}S_{[321]}
+  \\
+ \tr_{3\times 3}\left\{
\left(\begin{array}{ccc} q^6 &&\\&-q^2 & \\ && 1 \end{array}\right)^{a_1}
U_{[42]}
\left(\begin{array}{ccc} q^6 &&\\&-q^2 & \\ && 1 \end{array}\right)^{b_1}
U_{[42]}^\dagger
\ldots\right\}S_{[42]}\ \ \ \ \ \ \ \ \ \ \
\label{ansH3n}
\ee

The matrix $U_{[42]}$ is equal to:
\be
\left(\begin{array}{ccc}
\frac{q^4}{(q^4+1)(q^4+q^2+1)}
& -\frac{q\sqrt{q^8+q^6+q^4+q^2+1}}{(q^4+1)\sqrt{q^4+q^2+1}}
& -\frac{\sqrt{q^8+q^6+q^4+q^2+1}}{q^4+q^2+1}\\
\frac{q\sqrt{q^8+q^6+q^4+q^2+1}}{(q^4+1)\sqrt{q^4+q^2+1}}&-\frac{q^4-q^2+1}{q^4+1}
&\frac{q}{\sqrt{q^4+q^2+1}}\\
-\frac{\sqrt{q^8+q^6+q^4+q^2+1}}{q^4+q^2+1}&-\frac{q}{\sqrt{q^4+q^2+1}}
&\frac{q^2}{q^4+q^2+1}
\end{array}\right)=\\
= \left(\begin{array}{ccc}
\frac{[2]}{[3][4]}& -\frac{[2]}{[4]}\sqrt{\frac{[5]}{[3]}} & -\frac{\sqrt{[5]}}{[3]} \\
\frac{[2]}{[4]}\sqrt{\frac{[5]}{[3]}} & -\frac{[6]}{[3][4]} & \frac{[1]}{\sqrt{[3]}} \\
-\frac{\sqrt{[5]}}{[3]} & -\frac{1}{\sqrt{[3]}} & \frac{1}{[3]}
\end{array}\right)
\label{U42}
\ee
This is the same matrix as the matrix of the Racah coefficients, (A.20) in \cite{Kaul93}.

\subsection{Composite knots and links: a check
\label{comp}}

In this section we perform further checks, making use of topological equivalence
between different braids, i.e. homotopic equivalence of the corresponding
knots and links.
Accordingly, in this section we can consider only the ordinary HOMFLY polynomials
$H$ reduced to the topological locus
\be
p_k = p_k^* = \frac{A^k-A^{-k}}{q^k-q^{-k}}
\label{tolo2}
\ee

\bigskip

{\bf The case of $b_1=a_2=b_2=\ldots =0$:}
In this simplest example, with only one non-vanishing parameter $a_1$,
the 3-strand knot/link splits into untied a 2-strand knot/link and the unknot.
Accordingly,
\be
H_R^{3,(a,0,0,0,\ldots)} =\ H_R^{2,a}\cdot {H_R^{0}}
\label{H3aaadeco}
\ee
At the same time, in this case (\ref{ansH3n}) is also drastically simplified:
all mixing matrices drop away from the formula and it reduces to just
\be
H_{[2]}^{3,(a,0,0,0,\ldots)} = q^{6a}S_{[6]} + (-q^2)^a\Big(S_{[411]}+S_{[33]}\Big) + S_{[222]} +\\
+ \Big(q^{6a}+(-q^2)^a\Big)S_{[51]} + \Big((-q^2)^a+1\Big)S_{[321]} +
\Big(q^{6a} + (-q^2)^a + 1\Big)S_{[42]}
\label{H3a000}
\ee
Note that, in variance with expressions for the 3-strand {\it torus} knots and links,
this formula is sensitive to the sign of the $R$-matrix eigenvalue $-q^2$.
It remains to reduce (\ref{H3a000}) to the topological locus (\ref{tolo2}), where the Schur functions
turn into the quantum dimensions, and check that this coincides with the r.h.s.
of (\ref{H3aaadeco}) with $R=[2]$, where the unknot polynomial is just $H_R^{0} = S_R^*$
and $H_{[2]}^{2,a}$ is given by the second line of (\ref{H2n}).
Of course, such a relation can {\it not} be lifted to the entire $p$-space: (\ref{H3a000})
does {\it not} coincide with $H_{[2]}^{(2,a)} S_{[2]}$ beyond the topological locus (\ref{tolo2}):
one suffices to note that the former depends on $p_6$, while the latter one does not.

\subsection{Results}

\subsubsection{The figure eight knot $4_1$}

This knot can be realized with the braid
\be
4_1:\ \ \ \  (a_1,b_1|a_2,b_2) = (1,-1|1,-1),
\ee
similar to a possible 3-strand realization of the trefoil,
which is a torus knot $T[2,3]=T[3,2]$
\be
3_1:\ \ \ \ (1,1|1,1)
\ee

In the fundamental representation one had
\be
H^{4_1}_{[1]} = S_{[3]}^* + \Big(q^4 - 2q^2 + 1 - 2q^{-2} + q^{-4}\Big)S^*_{[21]} +S_{[111]}^*
= \Big(A^2 -(q^2-1+q^{-2}) + A^{-2}\Big) S_{[1]}^*
\ee
while
\be
H^{3_1}_{[1]} = q^{4}S_{[3]}^* - S_{[21]}^* + q^{-4}S_{[111]}^* =
\Big((q^2+q^{-2})A^2 - 1 \Big)  S_{[1]}^*
\ee
The second expression is highly asymmetric,
while the formula for $4_1$ is {\it very} symmetric even when expressed in terms of the $A$
variable: this is a specifics of $4_1$.

In the symmetric representation $R=[2]$ the answer is
\be
%
H^{4_1}_{[2]}
= \Big( q^4A^4
-(1+q^2)(1-q^2+q^6)q^{-2}A^2
+ (q^6-q^4+3-q^{-4}+q^{-6}) -  \\
-(1+q^{-2})(1-q^{-2}+q^{-6})q^2A^{-2}
+q^{-4}A^{-4}\Big)S^*_{[2]}
\label{H241}
\ee
This can be compared with the asymmetric formula for the trefoil $(1,1,1,1)$
\be
H^{3_1}_{[2]}
= q^8\Big(A^4(1+q^4+q^6+q^{12}) - A^2(1+q^2)(1+q^6) + q^2\Big)S_{[2]}^*
\ee
Expression (\ref{H241}) certainly coincide with results presented in existing literature,
see e.g. \cite{Rama00}. Moreover, it turns out that in the case of $4_1$ one can get the result for {\it any}
symmetric $[p]$ and antisymmetric $[1^p]$ representation \cite{IMMMfe}.

The HOMFLY polynomials in the symmetric representation for other 3-strand knots with no more than 8 crossings
are collected in the Appendix.

\subsection{Antisymmetric representation}

In order to construct the HOMFLY polynomials in the antisymmetric representation $[1,1]$, one could repeat the
standard machinery of the mixing matrices etc we described above. However, the result can be obtained much simpler
using a symmetry of the HOMFLY polynomials.

Indeed, the character expansions of the HOMFLY polynomials possess a
$Z_2$-symmetry
\be\label{symas}
A,\ q,\ S_R^* \ \ \longleftrightarrow \ A,\ -\frac{1}{q}\,,\ S_{R'}^*
\ee
where $R'$ is a transposition of the Young diagram $R$. This symmetry can be easily understood, since
$S_{R'}\{p_k\} = S_R\big\{(-)^{k-1}p_k\big\}$ and $\kappa_R=-\kappa_{R'}$. At the same time, all the ($SU_q(N)$)
group representation quantities (in particular, the mixing matrices) are also possess this antipodal symmetry. Hence,
one can calculate the HOMFLY polynomials in the antisymmetric representation just making a substitution $q\to -1/q$
in the HOMFLY polynomials for the symmetric representation obtained in the previous sections.

\subsection{Ooguri-Vafa conjecture}

In the paper \cite{OV}, the authors conjectured a connection of the Chern-Simons theory with topological string on the
resolution of the conifold. In fact, they proposed that the generating function $Z$ of average of the Wilson loop
in different representations is associated with the topological string partition function $Z_{str}$.
In accordance with the Ooguri-Vafa result \cite{GV} $Z$ is given by the sum\footnote{
This sum can be obtained as the Chern-Simons average of the Ooguri-Vafa operator $\exp\sum_n {1\over n}
\Tr \left(\oint_K Adx\right)^n\Tr V^n$, where $A$ is the gauge field, $p_k=\Tr V^k$ are external sources and
the traces are taken over the fundamental representation.
}
\be
Z(q,A,K)=\sum_R\chi_R(p)H_R^K(q,A)
\ee
where the sum runs over all the irreducible representation of $SU(N)$ ($A=q^N$). Now the topological nature
of this object implies that the "connected" correlators $f_R(q,A)$ defined by the expansion
\be
\log Z=\sum_{n=0,R}{1\over n}f_R(q^n,A^n)\chi_R(p^{(n)})
\ee
where the set of variables $p^{(n)}_k\equiv p_{nk}$, has the generic structure
\be
f_R(q,A)=\sum_{n,k} \tilde N_{R,n,k}{A^nq^k\over q-q^{-1}}
\ee
$\tilde N_{R,n,k}$ are integer and the parity of $n$ in the sum coincides with the parity of $|R|$ while the parity of $k$ is inverse.
These numbers are related to the Gopakumar-Vafa
integers $n_{\Delta,n,k}$ \cite{GV} by the relation
$$
n_{\Delta,n,k}=\sum_R \Phi_R(\Delta)\tilde N_{R,n,k}
$$
where $\Phi_R(\Delta)$ is the character of the symmetric group $S_{|\Delta|}$. The integers $\tilde N_{R,n,k}$ are
more refined, since their integrality implies that $n_{\Delta,n,k}$ are integer but not vise verse.
In fact, one can consider even more refined integers \cite{LMV}
\be\label{numN}
f_R(q,A)=\sum_{n,k,R_1,R_2} C_{RR_1R_2}\Sigma_{R_1}(q) N_{R_2,n,k}A^n\Big(q^{-1}-q\Big)^{2k-1}
\ee
where
\be
C_{RR_1R_2}=\sum_{\Delta}{\Phi_R(\Delta)\Phi_{R_1}(\Delta)\Phi_{R_2}(\Delta)\over z_{\Delta}}
\ee
the Clebsh-Gordon coefficients of the symmetric group, $z_{\Delta}$ is the standard symmetric factor of the
Young diagram \cite{Fulton} and $\Sigma_R(q)$ is a monomial non-zero only for the corner Young diagrams
$R = [l-d,1^{d}]$ and is equal to
\be
\Sigma_R(q)=(-1)^dq^{2d-l+1}
\ee

First few terms for $f_R$ and $N_{R,n,k}$ are
\be
f_{[1]}(q,A)=H_{[1]}(q,A)\\
f_{[2]}(q,A)=H_{[2]}(q,A)-{1\over 2}\Big( H_{[1]}(q,A)^2+H_{[2]}(q^2,A^2)\Big)\\
f_{[1,1]}(q,A)=H_{[1,1]}(q,A)-{1\over 2}\Big( H_{[1]}(q,A)^2-H_{[2]}(q^2,A^2)\Big)\\
...
\ee
and
\be
f_{[1]}(q,A)=\sum_{n,k}N_{[1],n,k}\Big(q^{-1}-q\Big)^{2k-1}A^n\\
f_{[2]}(q,A)=\sum_{n,k}\Big(q^{-1}N_{[2],n,k}-qN_{[1,1],n,k}\Big)\Big(q^{-1}-q\Big)^{2k-1}A^n\\
f_{[1,1]}(q,A)=\sum_{n,k}\Big(-qN_{[2],n,k}+q^{-1}N_{[1,1],n,k}\Big)\Big(q^{-1}-q\Big)^{2k-1}A^n\\
...
\ee

We calculate both the Ooguri-Vafa polynomials $f_{[2]}(q,A)$ and the numbers $N_{[2],n,k}$
for all 3-strand knots with no more than 8 crossings in the Appendix. The integrality
of these numbers and their using in product formulas is discussed in \cite{Peng}.

\subsection{"Special" polynomials}

The "special" polynomials are defined \cite{DMMSS} as the limit of ratio of the HOMFLY polynomials and
the quantum dimensions as $q\to 1$:
\be
{\mathfrak{H}}_R^{\cal K}(A) = \lim_{q\rightarrow 1} \frac{H_R^{\cal
K}(q,A)}{S_R^*(q,A)}
\label{speHdef}
\ee
Note that the limit is taken with fixed $A$, and both the HOMFLY polynomial
$H_R$
and the quantum dimension $S_R^*$ are singular behaving as
$(q-q^{-1})^{-|R|}$.
Here $|R|$ is the number of boxes in the Young diagram $R$.
Note that in this limit
\be
\lim_{q\rightarrow 1} S_R(A) = d_RS_{[1]}(A)^{|R|}
\ee
where
\be
d_R=S_R\{p\}\Big|_{p_k=\delta_{k,1}}=\prod_{(i,j)\in R}{1\over h_{i,j}}
\ee
and $h_{i,j}$ is the "hook" length.

The conjectured property of the "special" polynomials reads as \cite{DMMSS,IMMMspe}
\be
{\mathfrak{H}}^{\cal K}_R(A) = \Big({\mathfrak{H}}_{[1]}^{\cal
K}(A)\Big)^{|R|}
\label{spepro}
\ee
and is presumably valid for arbitrary ${\cal K}$ and $R$.

For example,
\be
{\mathfrak{H}}_{[2]}^{3_1}(A) = (2A^2-1)^2, \\
{\mathfrak{H}}_{[1]}^{3_1}(A) = 2A^2-1,
\ee
\be
{\mathfrak{H}}_{[2]}^{4_1}(A) = \Big(A^2-1+A^{-2}\Big)^2, \\
{\mathfrak{H}}_{[1]}^{4_1}(A) = A^2-1+A^{-2}
\ee
etc.

This conjecture is an amusing "dual" of a somewhat similar conjecture
\be
{\mathfrak{A}}_{R}^{\cal K}(q) = {\mathfrak{A}}_{[1]}^{\cal
K}\left(q^{|R|}\right)
\label{alepro}
\ee
for the Alexander polynomial
\be
{\mathfrak{A}}_{R}^{\cal K}(q) = \lim_{A\rightarrow 1}
\frac{H_R^{\cal K}(q,A)}{S_R^*}
\ee

We check these two conjectures for the concrete knots in the Appendix.

Similarly, one can consider the "special" limit of $q\to 1$ for other polynomials, e.g. for the Ooguri-Vafa
polynomials $f_R(q,A)$. The special Ooguri-Vafa polynomials
${\mathfrak{f}}_{R}(A)\equiv\lim_{q\rightarrow 1}{f_R(q,A)\over
S_{[R]}^*}$, however, depend on the representation much less trivially than the "special" and Alexander polynomials
(see the Appendix for examples). Note that ${\mathfrak{f}}_{[2]}(A)=-
{\mathfrak{f}}_{[11]}(A)$.

\subsection{Framing factor\label{ff}}

In this text we assume that the ${\cal R}$-matrix is normalized so
that in the channel $Q\in R\otimes R$ its eigenvalue is equal to
$\pm q^{\varkappa_Q}$ and is independent of $R$. This simplifies our
formulas, and this is important for their extension beyond the
topological locus (\ref{tolo}). However, instead this breaks some
properties, important for the knot theory, including topological
(ambient isotopy) invariance. Still, this difference is very easy to
take into account by adding an overall
factor, which is simple, but depends on representation and even on
the rank of the group $SU(N)$. This factor is also important in the
definition of the Ooguri-Vafa polynomials\footnote{For the fully symmetric
knots like
the figure eight $4_1$ with the vanishing writhe number the
factor is unity and the Ooguri-Vafa polynomials can be easily extended
beyond the topological locus \cite{IMMMfe}. For generic knots this
extension needs a separate discussion. } and is ambiguously determined
due to the freedom in choosing the framing \cite{MV}. We choose
the standard, or canonical framing.
Then, the ${\cal R}$-matrix, which
is adequate for knot theory calculations is actually normalized
differently:
\be
{\cal R}_{R\otimes R}^{norm} = A^{-|R|}
q^{-4\varkappa_R} {\cal R}_{R\otimes R} \label{normcon}
\ee
This
means that all our answers for the HOMFLY polynomials should be
multiplied by the additional factor
\be
H_R^{\cal B}\
\longrightarrow\ H_R^{\cal K} =
\Big(A^{|R|}q^{4\varkappa_R}\Big)^{-w^{\cal B}} H^{\cal B}_R\{p^*_k\}
\label{normH}
\ee
where $w^{\cal B}$ is the algebraic number of
intersections in the braid ${\cal B}$ called {\it the writhe
number}. We illustrate the significance of this factor by three
examples. First of them concerns the topological invariance, the second
one the identity (\ref{alepro}) for the Alexander polynomials: in
both cases the additional factor is essential. The third
example demonstrates that (\ref{normH}) is consistent with existing
literature.

\paragraph{Example 1:} If the torus knot $3_1$ is represented
by the $2$-strand braid $[2,3]=(1)^3=(1,1,1)$, then one gets for the
HOMFLY polynomial
\be
\frac{q^3\ {^*\!S}_{[2]} - q^{-3}\
{^*\!S}_{[11]}} { {^*\!S}_{[1]}} = \frac{q^3(Aq-A^{-1}q^{-1}) -
q^{-3}(Aq^{-1}-A^{-1}q)}{q^2-q^{-2}} = A(q^2+q^{-2}) - A^{-1}
\ee
If
the same knot is represented by the $3$-strand braid $[3,2] =
(1,1)^2=(1,1|1,1)$, one gets instead
\be
\frac{q^4\ {^*\!S}_{[3]} -
\tr_{2\times 2}{\Big(\hat{\cal R}\hat U\hat{\cal R}\hat
U^{\dagger}\Big)^2} {^*\!S}_{[21]} + q^{-4}\ {^*\!S}_{[111]} }{
{^*\!S}_{[1]} } = A^2(q^2+q^{-2}) - 1
\ee
Clearly, these two
expressions do not coincide and differ by a factor of $A$, which is
exactly taken into account by the correction factor (\ref{normH}),
because $w^{[3,2]}=4$, while $w^{[2,3]}=3$, and in this example
$\varkappa_{[1]}=0$. For $p=2$ the two HOMFLY polynomials,
calculated in this paper, differ by a factor of $A^2q^4$, which is again
nicely eliminated by (\ref{normH}), because $\varkappa_{[2]}=1$. Note that in the Appendix
we choose the opposite orientation for the trefoil: $(-1,-1,-1)$ in order to better match formulas
from the standard knot tables.

\paragraph{Example 2:} In fact, the Alexander polynomials made from our
extended HOMFLY polynomials,
\be
{\mathfrak A}^{[2,n]}_{[p\,]}(q) =
\lim_{A\rightarrow 1} \frac{q^{p(2p-1)n}\ \SS_{[2p\,]} -
q^{p(2p-3)n}\ \SS_{[2p-1,1]}}{\SS_{[p]}}
\ee
(all other Young
diagrams from the decomposition of  $[p\,]\otimes[p\,]$ do not
contribute at $A=1$), satisfy
\be
{\mathfrak{A}}^{[2,n]}_{[p\,]}(q)\
=\ q^{2p(p-1)n}\ {\mathfrak{A}}^{[2,n]}_{[1]}(q^p)\ = \
q^{4\varkappa_{[p\,]}\cdot w^{[2,n]}}\
{\mathfrak{A}}^{[2,n]}_{[1]}(q^p)
\ee
rather than (\ref{alepro}).
Unwanted factors in this relation are eliminated after the
factor (\ref{normH}) is taken into account.

\paragraph{Example 3:} As we
know from (\ref{chicorr}), the torus polynomial character expansion
of \cite{chi} based on use of the Adams operation, and also
suitable for continuation from the topological locus (\ref{tolo}) to the
entire space of time-variables, differs from ours by a factor of
$q^{-n(m-2)p(p-1)}$. The normalized HOMFLY obtained from ours by
the rule (\ref{normH}) should, therefore, differ from that one by a
factor of $q^{-2n(m-1)p(p-1)}\cdot q^{n(m-2)p(p-1)} = q^{-nmp(p-1)}$
(since $w^{[m,n]} = (m-1)n$ and $\varkappa_{[p\,]}=
\frac{p(p-1)}{2}$):
\be
H_{[p\,]}^{[m,n]} = A^{-(m-1)np}
q^{-mnp(p-1)} \underline{H}^{m,n}_{[p\,]} =
A^{-(m-1)n|R|}q^{-2mn\varkappa_{[p\,]}} \underline{H}^{m,n}_{[p\,]}
\ee
This is exactly the factor used in \cite{chi} for arbitrary
representation $R$. It can deserve noting that $mn$ is {\it not} the
writhe number of the braid associated with the torus knot, and
coefficient $2$ is different from $4$ in (\ref{normH}).

\subsection{Cabling}

The standard way to obtain the colored HOMFLY polynomials is to extract them from
those in the fundamental representation, but for different knots and links.
Namely, if one needs ${\cal H}^{\cal K}_R$, one considers instead
${\cal H}^{{\cal K}^{|R|}}_{[1]}$, where ${\cal K}^{|R|}$ is the {\it cabling}
of the knot ${\cal K}$, obtained by substituting the knot with a set of $|R|$
parallel ones (a "cable"),
i.e. actually a knot ${\cal K}$ is substituted by an $|R|$-component link.
However, to extract information about an arbitrary $R$ of a given size $|R|$
one should also allow additional intertwinings of the wires inside each cable,
which decreases the number of components in the link,
so that ${\cal K}^{|R|}$ is actually a linear combination of several links,
made in this way from the $|R|$-cabled ${\cal K}$.

If the knot ${\cal K}$ is represented by an $m$-strand braid,
the cabling involves $m|R|$ strands.
Since general formulas are known \cite{II}
for arbitrary $r$-strand knots in the fundamental representations,
one can actually demonstrate how the cabling procedure works for arbitrary
$2$-strand knots in symmetric and antisymmetric representations.
For the $3$-strand knots in these representations or for $2$-strand knots
in representations $[3],\ [21],\ [111]$ one needs the knowledge of the
$6$-strand knots in the fundamental representation, which is still not
available in full generality.
Thus, in the rest of this section we rederive ${\cal H}^{[2,n]}_{[2]}$
and ${\cal H}^{[2,n]}_{[11]}$ from ${\cal H}_{[1]}^{[2,n]^2}$.

Cabling is a tedious, but well known and widely used procedure,
we add this subsection for the sake of completeness.

\subsubsection*{Cabling the unknot}

Our first example is actually the $1$-strand knot: the unknot.
The $2$-cabling of a 1-strand braid implies that it is substituted with a $2$-strand one:
the two unlinked unknots and the answer is
\be
{H}_{[1]}^{[2,0]}\{p_k\} = S_{[1]}^2\{p_k\} = S_{[2]}\{p_k\} + S_{[11]}\{p_k\}
= {H}_{[2]}^{[1,0]}\{p_k\} + {H}_{[11]}^{[1,0]}\{p_k\}
\ee
a linear combination of unknot polynomials in two representations of the size $|R|=2$.
Similarly, the untwisted $p$-cabling gives a linear combination
\be
{H}_{[1]}^{[p,0]}\{p_k\} = S_{[1]}^{p}\{p_k\}
= \sum_{R: \ |R|=p} {\cal H}_R^{[1,0]}\{p_k\}
\ee
To extract the individual HOMFLY polynomials for two representations $[2]$ and $[11]$
one needs to consider not only the two non-intersecting strands,
but also to allow one intertwining.
One would naturally assume that one extra intersection just provides
$S_{[2]}-S_{[11]}$, but this is the case only for $q=1$.
If one associates an extra $R$-matrix with this additional intersection,
one gets $q$-dependent factors:
\be
{H}_{[1]}^{[2,1]}\{p_k\} \ \stackrel{(\ref{H2n})}{=}\
q S_{[2]}\{p_k\} - q^{-1} S_{[11]}\{p_k\} =
q{H}_{[2]}^{[1,0]}\{p_k\} - q^{-1}{H}_{[11]}^{[1,0]}\{p_k\}
\ee
and finally the cabling of the unknot implies
\be
{H}_{[2]}^{[1,0]}\{p_k\}  = \frac{1}{1+q^2}{H}_{[1]}^{[2,0]}\{p_k\}
+ \frac{q}{1+q^2}{H}_{[1]}^{[2,1]}\{p_k\}, \\
{H}_{[11]}^{[1,0]}\{p_k\}  = \frac{q^2}{1+q^2}{H}_{[1]}^{[2,0]}\{p_k\}
- \frac{q}{1+q^2}{H}_{[1]}^{[2,1]}\{p_k\}
\label{projt}
\ee
If one restricts the answer to the topological locus {\it and restore the factors},
see s.\ref{ff}, to make contact with the standard calculations, one
would write the same relations as follows:
\be
{H}_{[2]}^{[1,0]}(A|q) =
\frac{1}{1+q^2}  {{H}}_{[1]}^{[2,0]}(A|q)
+ \frac{qA}{1+q^2}  \Big( A^{-1}{H}_{[1]}^{[2,1]}(A|q)\Big), \\
{H}_{[11]}^{[1,0]}(A|q) =
\frac{q^2}{1+q^2}  {{H}}_{[1]}^{[2,0]}(A|q)
- \frac{qA}{1+q^2}  \Big( A^{-1}{H}_{[1]}^{[2,1]}(A|q)\Big)
\label{projA}
\ee
Formulas (\ref{projt}) and (\ref{projA}) actually define {\it the projectors},
specifying the linear combinations of cabled knots with additional twistings \cite{chi},
which select particular representations $[2]$ and $[11]$.
Since they are actually independent of the knot,
the same projectors are used for the same purpose below,
when we switch to a little more interesting examples of $2$-cabling the
$2$-strand knots.

\subsubsection*{2-cabling the 2-strand knots}

A new thing as compared to the previous subsection is that one has
intersections in original 2-strand braid (there were none in the 1-strand case).
Each ${\cal R}$-matrix at the $2$-strand crossing is substituted by
four ${\cal R}$-matrices, after lifting to $4$ strands:
\be
{\cal R} \longrightarrow \Big({\cal R}\otimes I \otimes I\Big)
\Big({\cal R}\otimes{\cal R}\Big) \Big(I\otimes I\otimes {\cal R}\Big)
\ee
so that the $2$-strand braid $[2,n]$ is lifted to a
$4$-strand braid of the type
$(0,1,1|1,1,0)^n$.
Moreover, to separate representations $[2]$ and $[11]$ one also needs to
allow one twisting between the first two and the last two braids,
i.e. to consider the four slightly different links/knots
\be
(0,1,1|1,1,0)^n, \ \ \ (0,1,1|1,1,0)^n (1,0,0), \ \ \
(0,1,1|1,1,0)^n(0,0,1), \ \ \ (0,1,1|1,1,0)^n(1,0,1)
\ee
Making use of projectors (\ref{projt}) and (\ref{projA}), one gets,
in somewhat compressed notation:
\be
q^{-2n}{H}_{[2]}^{[2,n]}\{p_k\}  = \frac{1}{(1+q^2)^2}
\Big({H}_{[1]}^{[4,(000),n]}\{p_k\}
+ q {H}_{[1]}^{[4,(001),n]}\{p_k\} + q {H}_{[1]}^{[4,(100),n]}\{p_k\}
+ q^2 {H}_{[1]}^{[4,(101),n]}\{p_k\}\Big), \\
q^{2n}{H}_{[11]}^{[2,n]}\{p_k\}  = \frac{1}{(1+q^2)^2}
\Big(q^4{H}_{[1]}^{[4,(000),n]}\{p_k\}
- q^3 {H}_{[1]}^{[4,(001),n]}\{p_k\} - q^3 {H}_{[1]}^{[4,(100),n]}\{p_k\}
+ q^2 {H}_{[1]}^{[4,(101),n]}\{p_k\}\Big)
\label{combs4}
\ee
According to \cite{II}, substituting the peculiar braid $(a_1b_1c_1|a_2b_2c_2|a_3b_3c_3|\ldots)
= (\underbrace{011|110|\ldots|011|110}_{n \ {\rm times}})$
into the general formula \cite[eq.(65)]{II} for the 4-strand extended HOMFLY polynomials gives
\be
{H}_{[1]}^{[4,(000),n]} = {H}_{[1]}^{[4,(011|110)^n(000)]}
= q^{4n}S_{[4]} + \tr_{3\times 3} \Big(\hat{\cal R}_{[31]}
\hat V_{[31]}\hat U_{[31]} \hat{\cal R}_{[31]}
\hat U_{[31]}^\dagger \hat V_{[31]}^\dagger\hat U_{[31]}^\dagger
\hat{\cal R}_{[31]}\hat U_{[31]} \hat{\cal R}_{[31]} \Big)^n \cdot S_{[31]} + \\
+ \left(q \longleftrightarrow -\frac{1}{q}\right)
+ \tr_{2\times 2} \Big( \hat{\cal R}_{[22]} \hat U_{22}^\dagger \hat{\cal R}_{[22]}^2
\hat U_{[22]}\hat{\cal R}_{[22]} \Big)^n \cdot S_{[22]}
\label{000}
\ee
and for the other three twisted cablings:
{\footnotesize
\be
{H}_{[1]}^{[4,(100),n]}
= q^{4n+1}S_{[4]} + \tr_{3\times 3} \left\{\hat U_{[31]} \Big(\hat{\cal R}_{[31]}
\hat V_{[31]}\hat U_{[31]} \hat{\cal R}_{[31]}
\hat U_{[31]}^\dagger \hat V_{[31]}^\dagger\hat U_{[31]}^\dagger
\hat{\cal R}_{[31]}\hat U_{[31]} \hat{\cal R}_{[31]} \Big)^n
\hat U_{[31]}^\dagger\hat{\cal R}_{[31]}
\right\} \cdot S_{[31]} + \\
+ \left(q \longleftrightarrow -\frac{1}{q}\right)
+ \tr_{2\times 2}\left\{ \Big( \hat{\cal R}_{[22]} \hat U_{22}^\dagger \hat{\cal R}_{[22]}^2
\hat U_{[22]}\hat{\cal R}_{[22]} \Big)^n
\hat U_{[22]}^\dagger\hat{\cal R}_{[22]}\hat U_{[22]}\right\}\cdot S_{[22]},  \\
{H}_{[1]}^{[4,(001),n]}
= q^{4n+1}S_{[4]} + \tr_{3\times 3}\left\{ \Big(\hat{\cal R}_{[31]}
\hat V_{[31]}\hat U_{[31]} \hat{\cal R}_{[31]}
\hat U_{[31]}^\dagger \hat V_{[31]}^\dagger\hat U_{[31]}^\dagger
\hat{\cal R}_{[31]}\hat U_{[31]} \hat{\cal R}_{[31]} \Big)^n
\hat V_{[31]}\hat U_{[31]}\hat{\cal R}_{[31]}\hat U_{[31]}^\dagger\hat V_{[31]}^\dagger
\right\} \cdot S_{[31]} + \\
+ \left(q \longleftrightarrow -\frac{1}{q}\right)
+ \tr_{2\times 2}\left\{ \Big( \hat{\cal R}_{[22]} \hat U_{22}^\dagger \hat{\cal R}_{[22]}^2
\hat U_{[22]}\hat{\cal R}_{[22]} \Big)^n \hat U_{[22]}^\dagger
\hat{\cal R}_{[22]}\hat U_{[22]}
\right\} \cdot S_{[22]},  \\
\!\!\!\!\!\!\!\!\!\!\!\!\!\!\!\!\!\!\!\!
{H}_{[1]}^{[4,(101),n]}
= q^{4n+2}S_{[4]} + \tr_{3\times 3} \left\{\Big(\hat{\cal R}_{[31]}
\hat V_{[31]}\hat U_{[31]} \hat{\cal R}_{[31]}
\hat U_{[31]}^\dagger \hat V_{[31]}^\dagger\hat U_{[31]}^\dagger
\hat{\cal R}_{[31]}\hat U_{[31]} \hat{\cal R}_{[31]} \Big)^n
\hat U_{[31]}^\dagger\hat{\cal R}_{[31]}\hat U_{[31]}
\hat V_{[31]}\hat U_{[31]}\hat{\cal R}_{[31]}\hat U_{[31]}^\dagger\hat V_{[31]}^\dagger
\right\}\cdot S_{[31]} + \\
+ \left(q \longleftrightarrow -\frac{1}{q}\right)
+ \tr_{2\times 2}\left\{ \Big( \hat{\cal R}_{[22]} \hat U_{22}^\dagger \hat{\cal R}_{[22]}^2
\hat U_{[22]}\hat{\cal R}_{[22]} \Big)^n
\hat U_{[22]}^\dagger\hat{\cal R}_{[22]}^2\hat U_{[22]}
\right\}
\cdot S_{[22]}
\label{111}
\ee
}
\!\!Now let us look at the coefficient in front of $S_{[4]}$.
The two linear combinations, corresponding to (\ref{combs4})
for this coefficient give just
\be
\frac{q^{4n} + 2q\cdot q^{4n+1} + q^2\cdot q^{4n+2}}{(1+q^2)^2} = q^{4n},  \\
\frac{q^4\cdot q^{4n} - 2q^3\cdot q^{4n+1} + q^2\cdot q^{4n+2}}{(1+q^2)^2} = 0
\ee
Similarly, for two linear combinations in front of $S_{[22]}$ one has
\be
\frac{ (q^{2n}+q^{-2n}) + 2q\cdot(-q^{2n-1}+q^{1-2n})
+ q^2\cdot(q^{2n-2}+q^{2-2n})}{(1+q^2)^2} = q^{-2n}, \\
\frac{ q^4\cdot(q^{2n}+q^{-2n}) - 2q^3\cdot(-q^{2n-1}+q^{1-2n})
+ q^2\cdot(q^{2n-2}+q^{2-2n})}{(1+q^2)^2} = q^{2n}
\ee
and for those in front of $S_{[31]}$ the intermediate expressions are
different for knots and links:
for $n$ odd
\be
\frac{ -1 + 2q\cdot(-q )
+ q^2\cdot(-q^2)}{(1+q^2)^2} = -1 \\
\frac{ q^4\cdot(-1) - 2q^3\cdot(-q )
+ q^2\cdot(-q^2)}{(1+q^2)^2} = 0
\ee
while for $n$ even
\be
\frac{ (2q^{2n}+1) + 2q\cdot(q^{2n+1}-q^{2n-1}+q)
+ q^2\cdot(q^2-2q^{2n})}{(1+q^2)^2} = 1 \\
\frac{ q^4\cdot(2q^{2n}+1) - 2q^3\cdot(q^{2n+1}-q^{2n-1}+q)
+ q^2\cdot(q^2-2q^{2n})}{(1+q^2)^2} = 0
\ee
Thus, one finally obtains
\be
q^{-2n} H^{2,n}_{[2]}
= q^{4n} S_{[4]} \mp  S_{[31]} + q^{-2n}S_{[22]}  \\
q^{2n}H^{2,n}_{[11]}
= q^{-4n} S_{[1111]} \mp   S_{[211]} + q^{2n}S_{[22]}
\label{H2H11}
\ee
which coincides with (\ref{H2n}).

\section{Summary}

In this paper we continued our program of constructing simple matrix expressions for the colored HOMFLY
polynomials of arbitrary knots/links started in \cite{I,II}. In practice, we always deal with braid representations
of knots. Here we considered the symmetric and antisymmetric representations $[2]$ and $[1,1]$ for 3-strand braids.
One can construct the result inductively, using the representation group theory, however, in this paper we used an
indirect way of using the known answers for the torus knot/link polynomials in order to restore all the necessary
ingredients (in particular, the mixing matrices) for the generic answer. We return to using the group theory
approach elsewhere \cite{IV}.

Using the formulas, obtained in this paper (we listed various knot polynomials for the knots that can be described
by 3-strand braids with no more than 8 crossings in the Appendix) we tested various conjectures, from the Ooguri-Vafa
conjecture \cite{OV} and its generalization \cite{LMV} to the conjecture of the representation dependence of the
"special" polynomials. The HOMFLY polynomials calculated in the paper were partly obtained earlier in a series of papers
by the Indian group \cite{Rama00,inds} within a different though close approach. In these cases our results confirm
these earlier calculations.

The results presented here are substantially extended in \cite{IV} to include higher symmetric representations
but this requires a deeper insight into the structure of the mixing matrices and, hence, is beyond the scope of the
present paper.

\section*{Note added}

After this paper was published there appeared a paper \cite{Rama}
with calculations of the HOMFLY polynomials in the first symmetric representation and of the corresponding
Ooguri-Vafa polynomials for various knots and links. Their results for the 3-strand knots coincide with formulas
of this paper for the only exception of the HOMFLY polynomial for knot $7_5$ where we made a misprint (the Ooguri-Vafa
polynomial was written in our paper correctly). We are grateful to the authors of \cite{Rama} for the correction.

\section*{Acknowledgements}

Our work is partly supported by Ministry of Education and Science of
the Russian Federation under contract 14.740.11.0081, by NSh-3349.2012.2,
by RFBR grants 10-02-00509 (A.Mir.), 10-02-00499 (A.Mor.), 11-02-01220 (And.Mor.) and
by joint grants 11-02-90453-Ukr, 12-02-91000-ANF,
11-01-92612-Royal Society.
The research of H.~I.~ and A.Mir.
is supported in part by the Grant-in-Aid for Scientific Research (23540316)
from the Ministry of Education, Science and Culture, Japan, and that of A.Mor. by
by JSPS Invitation Fellowship Program for Research in Japan (S-11137).
Support from JSPS/RFBR bilateral collaboration "Synthesis of integrabilities
 arising from gauge-string duality" (FY2010-2011: 12-02-92108-Yaf-a) is gratefully appreciated.

\newpage

\section*{Appendix. Tables of polynomials}

In this Appendix we list the HOMFLY polynomials and related quantities for all 3-strand knots in the
fundamental, symmetric $[2]$ and antisymmetric $[1,1]$ representations. Namely, for each knot, besides HOMFLY, we
write down expressions for the Jones ($A=q^2$), Alexander ($A=1$), "special" ($q\to 1$) polynomials,
the Ooguri-Vafa polynomials $f_R(q,A)$ and their "special" limit $q\to 1$, and for the numbers $N_{R,n,k}$ in
(\ref{numN}). Note that all the expressions are listed with the factors (\ref{normH}) taken into account.
We also use the notation $\{x\}\equiv x-x^{-1}$.

\section*{\fbox{Knot $3_1$}}

{\large $(-1,-1|-1,-1)$}

\bigskip

We use here the $(-1,-1|-1,-1)$ representation of the trefoil (in contrast with $(1,1|1,1)$ used throughout
the main body of the paper) in order to match the standard knot tables.

\begin{footnotesize}
\paragraph{HOMFLY polynomials}
\be \frac{H_{[1]}}{\
^*{S_{[1]}}}=A^4\Big((q^{2}+q^{-2})A^{-2}-1\Big)=\frac{A^4}{\
^*{S_{[1]}}}\Big( \ ^*{S_{[3]}}q^{-4}- \ ^*{S_{[2,1]}}+ \
^*{S_{[1,1,1]}}q^{4}\Big) \ee \be \frac{H_{[1,1]}}{\
^*{S_{[1,1]}}}=A^{8}q^{-16}\Big((q^{20}+q^{14}+q^{12}+q^{8})A^{-4}+(-q^{16}-q^{14}-q^{10}-q^{8})A^{-2}+q^{10}
\Big)=
\\
=\frac{A^{8}q^{-16}}{\ ^*{S_{[1,1]}}}\Big(\ ^*{S_{[3,3]}}- \
^*{S_{[3,2,1]}}q^{4}+ \ ^*{S_{[3,1,1,1]}}q^{8}+ \
^*{S_{[2,2,2]}}q^{8}- \ ^*{S_{[2,1,1,1,1]}}q^{16}+ \
^*{S_{[1,1,1,1,1,1]}}q^{24}\Big) \ee \be \frac{H_{[2]}}{\
^*{S_{[2]}}}=A^{8}q^{16}\Big((q^{-8}+q^{-12}+q^{-14}+q^{-20})A^{-4}+(-q^{-8}-q^{-10}-q^{-14}-q^{-16})A^{-2}+q^{-10}
\Big)=
\\
=\frac{A^{8}q^{16}}{\ ^*{S_{[2]}}}\Big(\ ^*{S_{[6]}}q^{-24}- \
^*{S_{[5,1]}}q^{-16}+ \ ^*{S_{[4,1,1]}}q^{-8}+ \
^*{S_{[3,3]}}q^{-8}- \ ^*{S_{[3,2,1]}}q^{-4}+ \ ^*{S_{[2,2,2]}}\Big)
\ee

\paragraph{Alexander polynomials}

\be \mathfrak{A}_{[1]}=q^{2}-1+q^{-2} \ee \be \mathfrak{A}_{[1,
1]}=\mathfrak{A}_{[2]}=\mathfrak{A}_{[1]}(q^2)=q^{4}-1+q^{-4} \ee

\paragraph{Jones polynomials}

\be J_{[1]}=-q^{8}+q^{6}+q^{2} \ee \be J_{[1, 1]}=1 \ee \be
J_{[2]}=q^{22}-q^{20}-q^{18}+q^{16}-q^{14}+q^{10}+q^{4} \ee

\paragraph{Special polynomials}

\be \mathfrak{H}_{[1]}=2A^{2}-A^{4} \ee \be \mathfrak{H}_{[1,
1]}=\mathfrak{H}_{[2]}=(\mathfrak{H}_{[1]})^2=4A^{4}-4A^{6}+A^{8}
\ee

\paragraph{Ooguri-Vafa polynomials}

\be f_{[1, 1]}=-\frac{A^6\{A\}^2\{A/q\}\{Aq\}(q^2+q^{-2})
}{q^3\{q\}} \ee \be
f_{[2]}=\frac{q^3A^6\{A\}^2\{A/q\}\{Aq\}(q^2+q^{-2}) }{\{q\}} \ee

\paragraph{Special Ooguri-Vafa polynomials}

\be {\mathfrak f}_{[2]}=-{\mathfrak f}_{[1,1]}=2A^6(A-A^{-1})^3 \ee

\subsubsection*{Numbers $N_{R,n,k}$}

\begin{tabular}{cccccc}
$N_{[1]}:$&
\begin{tabular}{|c|ccc|}
\hline
&&& \\
$k \backslash n$=& 1 & 3 & 5\\
&&& \\
\hline
&&& \\
0 & -2 & 3 & -1 \\
&&& \\
1 & -1 & 1 & 0 \\
&&& \\
\hline
\end{tabular}
& $N_{[1,1]}:$ &
\begin{tabular}{|c|ccccc|}
\hline
&&&&& \\
$k \backslash n=$& 2 & 4 & 6 & 8 & 10 \\
&&&&& \\
\hline
&&&&& \\
0 & 4 & -16 & 24 & -16 & 4 \\
&&&&& \\
1 & 4 & -20 & 32 & -20 & 4 \\
&&&&& \\
2 & 1 & -8 & 14 & -8 & 1 \\
&&&&& \\
3 & 0 & -1 & 2 & -1 & 0 \\
&&&&& \\
\hline
\end{tabular}
& $N_{[2]}:$ &
\begin{tabular}{|c|ccccc|}
\hline
&&&&& \\
$k \backslash n=$& 2 & 4 & 6 & 8 & 10 \\
&&&&& \\
\hline
&&&&& \\
0 & 2 & -8 & 12 & -8 & 2 \\
&&&&& \\
1 & 1 & -6 & 10 & -6 & 1 \\
&&&&& \\
2 & 0 & -1 & 2 & -1 & 0 \\
&&&&& \\
\hline
\end{tabular}
\end{tabular}

\bigskip


\section*{\fbox{Knot $4_1$}}

{\large $(1,-1|1,-1)$}

\bigskip

\paragraph{HOMFLY polynomials}

\be \frac{H_{[1]}}{\ ^*{S_{[1]}}}=A^{-2}+(-q^{2}+1-q^{-2})+A^{2}=
\frac{1}{\ ^*{S_{[1]}}}\Big(\ ^*{S_{[3]}}+ \
^*{S_{[2,1]}}(q^{4}-2q^{2}+1-2q^{-2}+q^{-4})+ \ ^*{S_{[1,1,1]}}\Big)
\ee \be \frac{H_{[1,1]}}{\
^*{S_{[1,1]}}}=q^{4}A^{-4}+(-q^{6}-q^{4}+q^{2}-q^{-2})A^{-2}+(q^{6}-q^{4}+3-q^{-4}+q^{-6})+(-q^{2}+q^{-2}-q^{-4}-q^{-6})A^{2}+q^{-4}A^{4}=
\\
=\frac{1}{\ ^*{S_{[1,1]}}}\Big(\ ^*{S_{[3,3]}}+ \
^*{S_{[3,2,1]}}(q^{4}-2q^{2}+1-2q^{-2}+q^{-4})+ \ ^*{S_{[3,1,1,1]}}+
\ ^*{S_{[2,2,2]}}+
\\
+ \ ^*{S_{[2,2,1,1]}}(q^{12}-2q^{10}-q^{8}+4q^{6}-3q^{4}-2q^{2}+6-2q^{-2}-3q^{-4}+4q^{-6}-q^{-8}-2q^{-10}+q^{-12})+\\
+ \ ^*{S_{[2,1,1,1,1]}}(q^{8}-2q^{4}+1-2q^{-4}+q^{-8})+ \
^*{S_{[1,1,1,1,1,1]}}\Big) \ee \be \frac{H_{[2]}}{\
^*{S_{[2]}}}=q^{-4}A^{-4}+(-q^{2}+q^{-2}-q^{-4}-q^{-6})A^{-2}+(q^{6}-q^{4}+3-q^{-4}+q^{-6})+(-q^{6}-q^{4}+q^{2}-q^{-2})A^{2}+q^{4}A^{4}=
\\
=\frac{1}{\ ^*{S_{[2]}}}\Big(\ ^*{S_{[6]}}+ \
^*{S_{[5,1]}}(q^{8}-2q^{4}+1-2q^{-4}+q^{-8})+ \\+
^*{S_{[4,2]}}(q^{12}-2q^{10}-q^{8}+4q^{6}-3q^{4}-2q^{2}+6-2q^{-2}-3q^{-4}+4q^{-6}-q^{-8}-2q^{-10}+q^{-12})+
\\
+\ ^*{S_{[4,1,1]}}+ \ ^*{S_{[3,3]}}+ \
^*{S_{[3,2,1]}}(q^{4}-2q^{2}+1-2q^{-2}+q^{-4})+ \
^*{S_{[2,2,2]}}\Big) \ee

\paragraph{Alexander polynomials}

\be \mathfrak{A}_{[1]}=-q^{2}+3-q^{-2} \ee \be \mathfrak{A}_{[1,
1]}=\mathfrak{A}_{[2]}=\mathfrak{A}_{[1]}(q^2)=-q^{4}+3-q^{-4} \ee

\paragraph{Jones polynomials}

\be J_{[1]}=q^{4}-q^{2}+1-q^{-2}+q^{-4} \ee \be J_{[1, 1]}=1 \ee \be
J_{[2]}=q^{12}-q^{10}-q^{8}+2q^{6}-q^{4}-q^{2}+3-q^{-2}-q^{-4}+2q^{-6}-q^{-8}-q^{-10}+q^{-12}
\ee

\paragraph{Special polynomials}

\be \mathfrak{H}_{[1]}=A^{-2}-1+A^{2} \ee \be \mathfrak{H}_{[1,
1]}=\mathfrak{H}_{[2]}=(\mathfrak{H}_{[1]})^2=A^{-4}-2A^{-2}+3-2A^{2}+A^{4}
\ee

\paragraph{Ooguri-Vafa polynomials}

\be f_{[1, 1]}=-\frac{\{A\}^2\{Aq\}\{Aq^{-2}\}\{A^2q^{-2}\}}{\{q\}}
\ee \be f_{[2]}=\frac{\{A\}^2\{Aq^2\}\{A^2q^{2}\}\{Aq^{-1}\}}{\{q\}}
\ee

\paragraph{Special Ooguri-Vafa polynomials}

\be {\mathfrak f}_{[2]}=-{\mathfrak
f}_{[1,1]}=(A^2-A^{-2})(A-A^{-1})^3 \ee

\subsubsection*{Numbers $N_{R,n,k}$}

\noindent
\begin{tabular}{cc}
$N_{[1]}:$ &
\begin{tabular}{|c|cccc|}
\hline
&&&& \\
$k \backslash n=$& -3 & -1 & 1 & 3 \\
&&&& \\
\hline
&&&& \\
0 & -1 & 2 & -2 & 1 \\
&&&& \\
1 & 0 & 1 & -1 & 0 \\
&&&& \\
\hline
\end{tabular}
\end{tabular}

\bigskip

\noindent
\begin{tabular}{cccc}
$N_{[1,1]}:$ &
\begin{tabular}{|c|ccccccc|}
\hline
&&&&&&& \\
$k \backslash n=$& -6 & -4 & -2 & 0 & 2 & 4 & 6 \\
&&&&&&& \\
\hline
&&&&&&& \\
0 & 1 & -3 & 4 & -6 & 9 & -7 & 2 \\
&&&&&&& \\
1 & 0 & -1 & 2 & -5 & 9 & -6 & 1 \\
&&&&&&& \\
2 & 0 & 0 & 0 & -1 & 2 & -1 & 0 \\
&&&&&&& \\
\hline
\end{tabular}
& $N_{[2]}:$ &
\begin{tabular}{|c|ccccccc|}
\hline
&&&&&&& \\
$k \backslash n=$& -6 & -4 & -2 & 0 & 2 & 4 & 6 \\
&&&&&&& \\
\hline
&&&&&&& \\
0 & 2 & -7 & 9 & -6 & 4 & -3 & 1 \\
&&&&&&& \\
1 & 1 & -6 & 9 & -5 & 2 & -1 & 0 \\
&&&&&&& \\
2 & 0 & -1 & 2 & -1 & 0 & 0 & 0 \\
&&&&&&& \\
\hline
\end{tabular}
\end{tabular}

\bigskip

\section*{\fbox{Knot $5_2$}}

{\large $(1,-1|1,3)$}

\bigskip

\end{footnotesize}

The choice of representation for this knot is different in
\cite{katlas} and in \cite{ZR}. We here follow the convention of
\cite{ZR}.

\begin{footnotesize}

\paragraph{HOMFLY polynomials}
\be \frac{H_{[1]}}{\
^*{S_{[1]}}}=A^{-4}\Big((q^{2}-1+q^{-2})A^{-2}+(q^{2}-1+q^{-2})-A^{2}
\Big)=
\\
=\frac{A^{-4}}{\ ^*{S_{[1]}}}\Big(\ \ ^*{S_{[3]}}q^{4}+ \
^*{S_{[2,1]}}(-q^{4}+q^{2}-1+q^{-2}-q^{-4})+ \
^*{S_{[1,1,1]}}q^{-4}\Big) \ee \be \frac{H_{[1,1]}}{\
^*{S_{[1,1]}}}=q^{16}A^{-8}\Big((q^{20}-q^{18}-q^{16}+2q^{14}-q^{10}+q^{8})A^{-4}+(q^{20}+q^{18}-2q^{16}+3q^{12}-q^{10}-q^{8}+q^{6})A^{-2}+
\\
+(-2q^{14}+2q^{10}-q^{8}-q^{6}+q^{4})+(-q^{12}+q^{8}-q^{6}-q^{4})A^{2}+q^{6}A^{4}
\Big)=
\\
=\frac{q^{16}A^{-8}}{\ ^*{S_{[1,1]}}}\Big(\ ^*{S_{[3,3]}}1+ \
^*{S_{[3,2,1]}}(-1+q^{-2}-q^{-4}+q^{-6}-q^{-8})+ \
^*{S_{[3,1,1,1]}}q^{-8}+ \ ^*{S_{[2,2,2]}}q^{-8}+
\\
+\
^*{S_{[2,2,1,1]}}(q^{2}-1-q^{-2}+2q^{-4}-2q^{-8}+q^{-10}+q^{-12}-2q^{-14}+q^{-16}-q^{-20}+q^{-22})+
\\
+\ ^*{S_{[2,1,1,1,1]}}(-q^{-8}+q^{-12}-q^{-16}+q^{-20}-q^{-24})+ \
^*{S_{[1,1,1,1,1,1]}}q^{-24}\Big) \ee \be \frac{H_{[2]}}{\
^*{S_{[2]}}}=q^{-16}A^{-8}\Big((q^{-8}-q^{-10}+2q^{-14}-q^{-16}-q^{-18}+q^{-20})A^{-4}+
\\
+(q^{-6}-q^{-8}-q^{-10}+3q^{-12}-2q^{-16}+q^{-18}+q^{-20})A^{-2}+
\\
+(q^{-4}-q^{-6}-q^{-8}+2q^{-10}-2q^{-14})+(-q^{-4}-q^{-6}+q^{-8}-q^{-12})A^{2}+q^{-6}A^{4}
\Big)=
\\
=\frac{q^{-16}A^{-8}}{\ ^*{S_{[2]}}}\Big(\ ^*{S_{[6]}}q^{24}+ \
^*{S_{[5,1]}}(-q^{24}+q^{20}-q^{16}+q^{12}-q^{8})+
\\
+\
^*{S_{[4,2]}}(q^{22}-q^{20}+q^{16}-2q^{14}+q^{12}+q^{10}-2q^{8}+2q^{4}-q^{2}-1+q^{-2})+
\\
+ \ ^*{S_{[4,1,1]}}q^{8}+ \ ^*{S_{[3,3]}}q^{8}+ \
^*{S_{[3,2,1]}}(-q^{8}+q^{6}-q^{4}+q^{2}-1)+ \ ^*{S_{[2,2,2]}}\Big)
\ee

\paragraph{Alexander polynomials}

\be \mathfrak{A}_{[1]}=2q^{2}-3+2q^{-2} \ee \be \mathfrak{A}_{[1,
1]}=\mathfrak{A}_{[2]}=\mathfrak{A}_{[1]}(q^2)=2q^{4}-3+2q^{-4} \ee

\paragraph{Jones polynomials}

\be J_{[1]}=q^{-2}-q^{-4}+2q^{-6}-q^{-8}+q^{-10}-q^{-12} \ee \be
J_{[1, 1]}=1 \ee \be
J_{[2]}=q^{-4}-q^{-6}+3q^{-10}-2q^{-12}-q^{-14}+4q^{-16}-3q^{-18}-q^{-20}+3q^{-22}-2q^{-24}-q^{-26}+2q^{-28}-q^{-30}-q^{-32}+q^{-34}
\ee

\paragraph{Special polynomials}

\be \mathfrak{H}_{[1]}=-A^{-6}+A^{-4}+A^{-2} \ee \be
\mathfrak{H}_{[1,1]}=\mathfrak{H}_{[2]}=(\mathfrak{H}_{[1]})^2=A^{-12}-2A^{-10}-A^{-8}+2A^{-6}+A^{-4}
\ee

\paragraph{Ooguri-Vafa polynomials}

\be
f_{[1,1]}=\frac{\{A\}^2\{A/q\}\{Aq\}(q^2-1+q^{-2})\Big((q^{9}+q^{7}+q^{5})A^{-10}+(q^{7}+q^{5})A^{-8}+q^{5}A^{-6}\Big)}{\{q\}}
\ee \be
f_{[2]}=\frac{\{A\}^2\{A/q\}\{Aq\}(q^2-1+q^{-2})\Big((-q^{-1}-q^{-5}-q^{-9})A^{-10}+(-q^{-1}-q^{-7})A^{-8}+(-q^{-1}+q^{-3}-q^{-5})A^{-6}\Big)}{\{q\}}
\ee

\paragraph{Special Ooguri-Vafa polynomials}

\be {\mathfrak f}_{[2]}=-{\mathfrak
f}_{[1,1]}=-\frac{(A^4+2A^2+3)(A-A^{-1})^3}{A^{10}} \ee

\subsubsection*{Numbers $N_{R,n,k}$}

\noindent
\begin{tabular}{cc}
$N_{[1]}:$ &
\begin{tabular}{|c|cccc|}
\hline
&&&& \\
$k \backslash n=$& -7 & -5 & -3 & -1 \\
&&&& \\
\hline
&&&& \\
0 & 1 & -2 & 0 & 1 \\
&&&& \\
1 & 0 & -1 & 0 & 1 \\
&&&& \\
\hline
\end{tabular}
\end{tabular}

\bigskip

\noindent
\begin{tabular}{cccc}
$N_{[1,1]}:$ &
\begin{tabular}{|c|ccccccc|}
\hline
&&&&&&& \\
$k \backslash n=$& -14 & -12 & -10 & -8 & -6 & -4 & -2 \\
&&&&&&& \\
\hline
&&&&&&& \\
0 & 6 & -21 & 25 & -10 & 0 & -1 & 1 \\
&&&&&&& \\
1 & 11 & -46 & 60 & -25 & 0 & -1 & 1 \\
&&&&&&& \\
2 & 6 & -34 & 50 & -22 & 0 & 0 & 0 \\
&&&&&&& \\
3 & 1 & -10 & 17 & -8 & 0 & 0 & 0 \\
&&&&&&& \\
4 & 0 & -1 & 2 & -1 & 0 & 0 & 0 \\
&&&&&&& \\
\hline
\end{tabular}
& $N_{[2]}:$ &
\begin{tabular}{|c|ccccccc|}
\hline
&&&&&&& \\
$k \backslash n=$& -14 & -12 & -10 & -8 & -6 & -4 & -2 \\
&&&&&&& \\
\hline
&&&&&&& \\
0 & 9 & -31 & 36 & -14 & 0 & -3 & 2 \\
&&&&&&& \\
1 & 24 & -95 & 120 & -49 & 0 & -4 & 3 \\
&&&&&&& \\
2 & 22 & -106 & 147 & -63 & 0 & -1 & 1 \\
&&&&&&& \\
3 & 8 & -53 & 82 & -37 & 0 & 0 & 0 \\
&&&&&&& \\
4 & 1 & -12 & 21 & -10 & 0 & 0 & 0 \\
&&&&&&& \\
5 & 0 & -1 & 2 & -1 & 0 & 0 & 0 \\
&&&&&&& \\
\hline
\end{tabular}
\end{tabular}

\bigskip

\section*{\fbox{Knot $6_2$}}

{\large $(1,-1|1,-3)$}

\bigskip

\paragraph{HOMFLY polynomials}
\be \frac{H_{[1]}}{\
^*{S_{[1]}}}=A^2\Big((q^{2}+q^{-2})A^{-2}+(-q^{4}+q^{2}-2+q^{-2}-q^{-4})+(q^{2}-1+q^{-2})A^{2}
\Big)=
\\
=\frac{A^2}{\ ^*{S_{[1]}}}\Big(\ ^*{S_{[3]}}q^{-2}+ \
^*{S_{[2,1]}}(q^{6}-2q^{4}+2q^{2}-3+2q^{-2}-2q^{-4}+q^{-6})+ \
^*{S_{[1,1,1]}}q^{2}\Big) \ee \be \frac{H_{[1,1]}}{\
^*{S_{[1,1]}}}=A^{4}q^{-8}\Big((q^{16}+q^{10}+q^{8}+q^{4})A^{-4}+(-q^{18}-q^{16}+q^{14}-q^{12}-3q^{10}-2q^{4}-q^{-2})A^{-2}+
\\
+(q^{18}-q^{16}+4q^{12}-3q^{8}+4q^{6}+2q^{4}-2q^{2}+2+q^{-2}-q^{-4}+q^{-6})+
\\
+(-q^{14}+2q^{10}-2q^{8}-3q^{6}+3q^{4}-3+q^{-2}-q^{-6})A^{2}+(q^{8}-q^{6}-q^{4}+2q^{2}-q^{-2}+q^{-4})A^{4}
\Big)=
\\
=\frac{A^{4}q^{-8}}{\ ^*{S_{[1,1]}}}\Big(\ ^*{S_{[3,3]}}+ \
^*{S_{[3,2,1]}}(q^{8}-2q^{6}+2q^{4}-3q^{2}+2-2q^{-2}+q^{-4})+ \
^*{S_{[3,1,1,1]}}q^{4}+ \ ^*{S_{[2,2,2]}}q^{4}+
\\
+\
^*{S_{[2,2,1,1]}}(q^{24}-2q^{22}-q^{20}+5q^{18}-3q^{16}-5q^{14}+8q^{12}-9q^{8}+
\\
+7q^{6}+3q^{4}-8q^{2}+5+2q^{-2}-5q^{-4}+3q^{-6}-2q^{-10}+q^{-12})+
\\
+\
^*{S_{[2,1,1,1,1]}}(q^{20}-2q^{16}+2q^{12}-3q^{8}+2q^{4}-2+q^{-4})+
\ ^*{S_{[1,1,1,1,1,1]}}q^{12}\Big) \ee \be \frac{H_{[2]}}{\
^*{S_{[2]}}}=A^{4}q^8\Big((q^{-4}+q^{-8}+q^{-10}+q^{-16})A^{-4}+(-q^{2}-2q^{-4}-3q^{-10}-q^{-12}+q^{-14}-q^{-16}-q^{-18})A^{-2}+
\\
+(q^{6}-q^{4}+q^{2}+2-2q^{-2}+2q^{-4}+4q^{-6}-3q^{-8}+4q^{-12}-q^{-16}+q^{-18})+
\\
+(-q^{6}+q^{2}-3+3q^{-4}-3q^{-6}-2q^{-8}+2q^{-10}-q^{-14})A^{2}+(q^{4}-q^{2}+2q^{-2}-q^{-4}-q^{-6}+q^{-8})A^{4}
\Big)=
\\
=\frac{A^{4}q^8}{\ ^*{S_{[2]}}}\Big(\ ^*{S_{[6]}}q^{-12}+ \
^*{S_{[5,1]}}(q^{4}-2+2q^{-4}-3q^{-8}+2q^{-12}-2q^{-16}+q^{-20})+
\\
+\
^*{S_{[4,2]}}(q^{12}-2q^{10}+3q^{6}-5q^{4}+2q^{2}+5-8q^{-2}s+3q^{-4}+
\\
+7q^{-6}-9q^{-8}+8q^{-12}-5q^{-14}-3q^{-16}+5q^{-18}-q^{-20}-2q^{-22}+q^{-24})+
\\
+\ ^*{S_{[4,1,1]}}q^{-4}+ \ ^*{S_{[3,3]}}q^{-4}+ \
^*{S_{[3,2,1]}}(q^{4}-2q^{2}+2-3q^{-2}+2q^{-4}-2q^{-6}+q^{-8})+ \
^*{S_{[2,2,2]}}\Big) \ee

\paragraph{Alexander polynomials}

\be \mathfrak{A}_{[1]}=-q^{4}+3q^{2}-3+3q^{-2}-q^{-4} \ee \be
\mathfrak{A}_{[1,
1]}=\mathfrak{A}_{[2]}=\mathfrak{A}_{[1]}(q^2)=-q^{8}+3q^{4}-3+3q^{-4}-q^{-8}
\ee

\paragraph{Jones polynomials}

\be J_{[1]}=q^{10}-2q^{8}+2q^{6}-2q^{4}+2q^{2}-1+q^{-2} \ee \be
J_{[1, 1]}=1 \ee \be
J_{[2]}=q^{28}-2q^{26}+4q^{22}-5q^{20}+6q^{16}-6q^{14}+6q^{10}-5q^{8}-q^{6}+5q^{4}-3q^{2}-1+3q^{-2}-q^{-4}-q^{-6}+q^{-8}
\ee

\paragraph{Special polynomials}

\be \mathfrak{H}_{[1]}=2-2A^{2}+A^{4} \ee \be \mathfrak{H}_{[1,
1]}=\mathfrak{H}_{[2]}=(\mathfrak{H}_{[1]})^2=4-8A^{2}+8A^{4}-4A^{6}+A^{8}
\ee

\paragraph{Ooguri-Vafa polynomials}

\be f_{[1,
1]}=\frac{\{A\}^2\{A/q\}\{Aq\}(q^2-1+q^{-2})((q^{7}+q^{5})A^{2}+(q^{5}+q^{-3})A^{4}+(-q^{-1}-q^{-7})A^{6})}{\{q\}}
\ee \be
f_{[2]}=\frac{\{A\}^2\{A/q\}\{Aq\}(q^2-1+q^{-2})((-q^{-1}-q^{-3})A^{2}+(-q^{7}-q^{-1})A^{4}+(q^{11}+q^{5})A^{6})}{\{q\}}
\ee

\paragraph{Special Ooguri-Vafa polynomials}

\be {\mathfrak f}_{[2]}=-{\mathfrak
f}_{[1,1]}=2A^2(A^4-A^2-1)(A-A^{-1})^3 \ee

\subsubsection*{Numbers $N_{R,n,k}$}

\noindent
\begin{tabular}{cc}
$N_{[1]}:$ &
\begin{tabular}{|c|cccc|}
\hline
&&&& \\
$k \backslash n=$& -1 & 1 & 3 & 5 \\
&&&& \\
\hline
&&&& \\
0 & -2 & 4 & -3 & 1 \\
&&&& \\
1 & -1 & 4 & -4 & 1 \\
&&&& \\
2 & 0 & 1 & -1 & 0 \\
&&&& \\
\hline
\end{tabular}
\end{tabular}

\bigskip

\noindent
\begin{tabular}{cccc}
$N_{[1,1]}:$ &
\begin{tabular}{|c|ccccccc|}
\hline
&&&&&&& \\
$k \backslash n=$& -2 & 0 & 2 & 4 & 6 & 8 & 10 \\
&&&&&&& \\
\hline
&&&&&&& \\
0 & 3 & -14 & 33 & -52 & 53 & -30 & 7 \\
&&&&&&& \\
1 & 4 & -25 & 84 & -178 & 212 & -125 & 28 \\
&&&&&&& \\
2 & 1 & -13 & 82 & -246 & 335 & -201 & 42 \\
&&&&&&& \\
3 & 0 & -2 & 40 & -175 & 267 & -159 & 29 \\
&&&&&&& \\
4 & 0 & 0 & 10 & -67 & 113 & -65 & 9 \\
&&&&&&& \\
5 & 0 & 0 & 1 & -13 & 24 & -13 & 1 \\
&&&&&&& \\
6 & 0 & 0 & 0 & -1 & 2 & -1 & 0 \\
&&&&&&& \\
\hline
\end{tabular}
& $N_{[2]}:$ &
\begin{tabular}{|c|ccccccc|}
\hline
&&&&&&& \\
$k \backslash n=$& -2 & 0 & 2 & 4 & 6 & 8 & 10 \\
&&&&&&& \\
\hline
&&&&&&& \\
0 & 5 & -20 & 35 & -40 & 35 & -20 & 5 \\
&&&&&&& \\
1 & 10 & -45 & 85 & -110 & 110 & -65 & 15 \\
&&&&&&& \\
2 & 6 & -34 & 72 & -113 & 132 & -79 & 16 \\
&&&&&&& \\
3 & 1 & -10 & 25 & -54 & 75 & -44 & 7 \\
&&&&&&& \\
4 & 0 & -1 & 3 & -12 & 20 & -11 & 1 \\
&&&&&&& \\
5 & 0 & 0 & 0 & -1 & 2 & -1 & 0 \\
&&&&&&& \\
\hline
\end{tabular}
\end{tabular}

\bigskip

\section*{\fbox{Knot $6_3$}}

{\large $(2,-1|1,-2)$}

\bigskip

\paragraph{HOMFLY polynomials}
\be \frac{H_{[1]}}{\
^*{S_{[1]}}}=(-q^{2}+1-q^{-2})A^{-2}+(q^{4}-q^{2}+3-q^{-2}+q^{-4})+(-q^{2}+1-q^{-2})A^{2}=
\\
=\frac{1}{\ ^*{S_{[1]}}}\Big(\ ^*{S_{[3]}}+
^*{S_{[2,1]}}(-q^{6}+2q^{4}-3q^{2}+3-3q^{-2}+2q^{-4}-q^{-6})+
^*{S_{[1,1,1]}}\Big) \ee \be \frac{H_{[1,1]}}{\
^*{S_{[1,1]}}}=(q^{10}-q^{8}-q^{6}+2q^{4}-1+q^{-2})A^{-4}+(-q^{12}+2q^{8}-3q^{6}-3q^{4}+4q^{2}-1-4q^{-2}+q^{-4}-q^{-8})A^{-2}+
\\
+(q^{12}-q^{10}+q^{8}+4q^{6}-3q^{4}-q^{2}+9-q^{-2}-3q^{-4}+4q^{-6}+q^{-8}-q^{-10}+q^{-12})+
\\
+(-q^{8}+q^{4}-4q^{2}-1+4q^{-2}-3q^{-4}-3q^{-6}+2q^{-8}-q^{-12})A^{2}+(q^{2}-1+2q^{-4}-q^{-6}-q^{-8}+q^{-10})A^{4}=
\\
=\frac{1}{\ ^*{S_{[1,1]}}}\Big(\ ^*{S_{[3,3]}}+\
^*{S_{[3,2,1]}}(-q^{6}+2q^{4}-3q^{2}+3-3q^{-2}+2q^{-4}-q^{-6})+\
^*{S_{[3,1,1,1]}}+\ ^*{S_{[2,2,2]}}+
\\
+\
^*{S_{[2,2,1,1]}}(q^{18}-2q^{16}+5q^{12}-6q^{10}-2q^{8}+12q^{6}-9q^{4}-7q^{2}+
\\
+16-7q^{-2}-9q^{-4}+12q^{-6}-2q^{-8}-6q^{-10}+5q^{-12}-2q^{-16}+q^{-18})+
\\
+\
^*{S_{[2,1,1,1,1]}}(-q^{12}+2q^{8}-3q^{4}+3-3q^{-4}+2q^{-8}-q^{-12})+\
^*{S_{[1,1,1,1,1,1]}}\Big)
\\
\frac{H_{[2]}}{\
^*{S_{[2]}}}=q^{2}-1+2q^{-4}-q^{-6}-q^{-8}+q^{-10})A^{-4}+(-q^{8}+q^{4}-4q^{2}-1+4q^{-2}-3q^{-4}-3q^{-6}+2q^{-8}-q^{-12})A^{-2}+
\\
+(q^{12}-q^{10}+q^{8}+4q^{6}-3q^{4}-q^{2}+9-q^{-2}-3q^{-4}+4q^{-6}+q^{-8}-q^{-10}+q^{-12})+
\\
+(-q^{12}+2q^{8}-3q^{6}-3q^{4}+4q^{2}-1-4q^{-2}+q^{-4}-q^{-8})A^{2}+(q^{10}-q^{8}-q^{6}+2q^{4}-1+q^{-2})A^{4}=
\\
=\frac{1}{\ ^*{S_{[2]}}}\Big(\ ^*{S_{[6]}}+\
^*{S_{[5,1]}}(-q^{12}+2q^{8}-3q^{4}+3-3q^{-4}+2q^{-8}-q^{-12})+
\\
+\
^*{S_{[4,2]}}(q^{18}-2q^{16}+5q^{12}-6q^{10}-2q^{8}+12q^{6}-9q^{4}-7q^{2}+
\\
+16-7q^{-2}-9q^{-4}+12q^{-6}-2q^{-8}-6q^{-10}+5q^{-12}-2q^{-16}+q^{-18})+
\\
+\ ^*{S_{[4,1,1]}}+\ ^*{S_{[3,3]}}+
\\
+\ ^*{S_{[3,2,1]}}(-q^{6}+2q^{4}-3q^{2}+3-3q^{-2}+2q^{-4}-q^{-6})+\
^*{S_{[2,2,2]}}\Big) \ee

\paragraph{Alexander polynomials}

\be \mathfrak{A}_{[1]}=q^{4}-3q^{2}+5-3q^{-2}+q^{-4} \ee \be
\mathfrak{A}_{[1,
1]}=\mathfrak{A}_{[2]}=\mathfrak{A}_{[1]}(q^2)=q^{8}-3q^{4}+5-3q^{-4}+q^{-8}
\ee

\paragraph{Jones polynomials}

\be J_{[1]}=-q^{6}+2q^{4}-2q^{2}+3-2q^{-2}+2q^{-4}-q^{-6} \ee \be
J_{[1, 1]}=1 \ee \be
J_{[2]}=q^{18}-2q^{16}-q^{14}+5q^{12}-4q^{10}-3q^{8}+9q^{6}-5q^{4}-5q^{2}+
\\
+11-5q^{-2}-5q^{-4}+9q^{-6}-3q^{-8}-4q^{-10}+5q^{-12}-q^{-14}-2q^{-16}+q^{-18}
\ee

\paragraph{Special polynomials}

\be \mathfrak{H}_{[1]}=-A^{-2}+3-A^{2} \ee \be \mathfrak{H}_{[1,
1]}=\mathfrak{H}_{[2]}=(\mathfrak{H}_{[1]})^2=A^{-4}-6A^{-2}+11-6A^{2}+A^{4}
\ee

\paragraph{Ooguri-Vafa polynomials}

\be f_{[1,
1]}=\frac{\{A\}^2\{A/q\}\{Aq\}(q^2-1+q^{-2})\Big((-q^{7}+q^{3}-q^{-1})A^{-2}+(q^{3}+2q^{1}-2q^{-1}-q^{-3})+(q^{1}-q^{-3}+q^{-7})A^{2}\Big)}{\{q\}}
\ee \be
f_{[2]}=\frac{\{A\}^2\{A/q\}\{Aq\}(q^2-1+q^{-2})\Big((-q^{1}+q^{-3}-q^{-7})A^{-2}+(-q^{3}-2q^{1}+2q^{-1}+q^{-3})+(q^{7}-q^{3}+q^{-1})A^{2}\Big)}{\{q\}}
\ee

\paragraph{Special Ooguri-Vafa polynomials}

\be {\mathfrak f}_{[2]}=-{\mathfrak
f}_{[1,1]}=(A^2-A^{-2})(A-A^{-1})^3 \ee

\subsubsection*{Numbers $N_{R,n,k}$}

\noindent
\begin{tabular}{cc}
$N_{[1]}:$ &
\begin{tabular}{|c|cccc|}
\hline
&&&& \\
$k \backslash n=$& -3 & -1 & 1 & 3 \\
&&&& \\
\hline
&&&& \\
0 & 1 & -4 & 4 & -1 \\
&&&& \\
1 & 1 & -4 & 4 & -1 \\
&&&& \\
2 & 0 & -1 & 1 & 0 \\
&&&& \\
\hline
\end{tabular}
\end{tabular}

\bigskip

\noindent
\begin{tabular}{cccc}
$N_{[1,1]}:$ &
\begin{tabular}{|c|ccccccc|}
\hline
&&&&&&& \\
$k \backslash n=$& -6 & -4 & -2 & 0 & 2 & 4 & 6 \\
&&&&&&& \\
\hline
&&&&&&& \\
0 & 1 & -9 & 28 & -42 & 33 & -13 & 2 \\
&&&&&&& \\
1 & 5 & -27 & 72 & -113 & 104 & -52 & 11 \\
&&&&&&& \\
2 & 5 & -26 & 65 & -114 & 127 & -72 & 15 \\
&&&&&&& \\
3 & 1 & -9 & 24 & -54 & 74 & -43 & 7 \\
&&&&&&& \\
4 & 0 & -1 & 3 & -12 & 20 & -11 & 1 \\
&&&&&&& \\
5 & 0 & 0 & 0 & -1 & 2 & -1 & 0 \\
&&&&&&& \\
\hline
\end{tabular}
& $N_{[2]}:$ &
\begin{tabular}{|c|ccccccc|}
\hline
&&&&&&& \\
$k \backslash n=$& -6 & -4 & -2 & 0 & 2 & 4 & 6 \\
&&&&&&& \\
\hline
&&&&&&& \\
0 & 2 & -13 & 33 & -42 & 28 & -9 & 1 \\
&&&&&&& \\
1 & 11 & -52 & 104 & -113 & 72 & -27 & 5 \\
&&&&&&& \\
2 & 15 & -72 & 127 & -114 & 65 & -26 & 5 \\
&&&&&&& \\
3 & 7 & -43 & 74 & -54 & 24 & -9 & 1 \\
&&&&&&& \\
4 & 1 & -11 & 20 & -12 & 3 & -1 & 0 \\
&&&&&&& \\
5 & 0 & -1 & 2 & -1 & 0 & 0 & 0 \\
&&&&&&& \\
\hline
\end{tabular}
\end{tabular}

\bigskip

\section*{\fbox{Knot $7_3$}}

{\large $(1,-1|1,5)$}

\bigskip

\paragraph{HOMFLY polynomials}
\be \frac{H_{[1]}}{\
^*{S_{[1]}}}=A^{-6}\Big((-q^{2}-q^{-2})A^{-2}+(q^{4}-q^{2}+2-q^{-2}+q^{-4})+(q^{4}-q^{2}+1-q^{-2}+q^{-4})A^{2}
\Big)=
\\
=\frac{A^{-6}}{\ ^*{S_{[1]}}}\Big(\ ^*{S_{[3]}}q^{6}+ \
^*{S_{[2,1]}}(-q^{6}+q^{4}-2q^{2}+3-2q^{-2}+q^{-4}-q^{-6})+ \
^*{S_{[1,1,1]}}q^{-6}\Big) \ee \be \frac{H_{[1,1]}}{\
^*{S_{[1,1]}}}=A^{-12}q^{24}\Big((q^{-6}+q^{-10}+q^{-12}+q^{-18})A^{-4}+
\\
+(-q^{-4}-q^{-6}-2q^{-10}-q^{-12}+q^{-14}-2q^{-16}-2q^{-18}+q^{-20}-q^{-24})A^{-2}+
\\
+(q^{-4}-q^{-6}+2q^{-10}-2q^{-12}+4q^{-16}-3q^{-18}-2q^{-20}+3q^{-22}-2q^{-26})+
\\
+(q^{-6}-q^{-8}+2q^{-12}-2q^{-14}+q^{-16}+3q^{-18}-3q^{-20}-q^{-22}+4q^{-24}-2q^{-28}+q^{-30}+q^{-32})A^{2}+
\\
+(q^{-8}-q^{-10}+q^{-14}-q^{-16}+q^{-18}+q^{-20}-2q^{-22}+2q^{-26}-q^{-28}-q^{-30}+q^{-32})A^{4}
\Big)=
\\
=\frac{A^{-12}q^{24}}{\ ^*{S_{[1,1]}}}\Big(\ ^*{S_{[3,3]}}+ \
^*{S_{[3,2,1]}}(-1+q^{-2}-2q^{-4}+3q^{-6}-2q^{-8}+q^{-10}-q^{-12})+
\ ^*{S_{[3,1,1,1]}}q^{-12}+ \ ^*{S_{[2,2,2]}}q^{-12}+
\\
+\
^*{S_{[2,2,1,1]}}(q^{2}-1+2q^{-4}-3q^{-6}-q^{-8}+6q^{-10}-5q^{-12}-4q^{-14}+
\\
+10q^{-16}-4q^{-18}-5q^{-20}+6q^{-22}-q^{-24}-3q^{-26}+2q^{-28}-q^{-32}+q^{-34})+
\\
+\
^*{S_{[2,1,1,1,1]}}(-q^{-12}+q^{-16}-2q^{-20}+3q^{-24}-2q^{-28}+q^{-32}-q^{-36})+
\ ^*{S_{[1,1,1,1,1,1]}}q^{-36}\Big) \ee \be \frac{H_{[2]}}{\
^*{S_{[2]}}}=A^{-12}q^{-24}\Big((q^{18}+q^{12}+q^{10}+q^{6})A^{-4}+
\\
+(-q^{24}+q^{20}-2q^{18}-2q^{16}+q^{14}-q^{12}-2q^{10}-q^{6}-q^{4})A^{-2}+
\\
+(-2q^{26}+3q^{22}-2q^{20}-3q^{18}+4q^{16}-2q^{12}+2q^{10}-q^{6}+q^{4})+
\\
+(q^{32}+q^{30}-2q^{28}+4q^{24}-q^{22}-3q^{20}+3q^{18}+q^{16}-2q^{14}+2q^{12}-q^{8}+q^{6})A^{2}+
\\
+(q^{32}-q^{30}-q^{28}+2q^{26}-2q^{22}+q^{20}+q^{18}-q^{16}+q^{14}-q^{10}+q^{8})A^{4}
\Big)=
\\
=\frac{A^{-12}q^{-24}}{\ ^*{S_{[2]}}}\Big(\ ^*{S_{[6]}}q^{36}+ \
^*{S_{[5,1]}}(-q^{36}+q^{32}-2q^{28}+3q^{24}-2q^{20}+q^{16}-q^{12})+
\\
+\
^*{S_{[4,2]}}(q^{34}-q^{32}+2q^{28}-3q^{26}-q^{24}+6q^{22}-5q^{20}-
\\
-4q^{18}+10q^{16}-4q^{14}-5q^{12}+6q^{10}-q^{8}-3q^{6}+2q^{4}-1+q^{-2})+
\\
+\ ^*{S_{[4,1,1]}}q^{12}+ \ ^*{S_{[3,3]}}q^{12}+ \
^*{S_{[3,2,1]}}(-q^{12}+q^{10}-2q^{8}+3q^{6}-2q^{4}+q^{2}-1)+ \
^*{S_{[2,2,2]}}\Big) \ee

\paragraph{Alexander polynomials}

\be \mathfrak{A}_{[1]}=2q^{4}-3q^{2}+3-3q^{-2}+2q^{-4} \ee \be
\mathfrak{A}_{[1,
1]}=\mathfrak{A}_{[2]}=\mathfrak{A}_{[1]}(q^2)=2q^{8}-3q^{4}+3-3q^{-4}+2q^{-8}
\ee

\paragraph{Jones polynomials}

\be
J_{[1]}=q^{-4}-q^{-6}+2q^{-8}-2q^{-10}+3q^{-12}-2q^{-14}+q^{-16}-q^{-18}
\ee \be J_{[1, 1]}=1 \ee \be
J_{[2]}=q^{-8}-q^{-10}+3q^{-14}-2q^{-16}-2q^{-18}+5q^{-20}-2q^{-22}-4q^{-24}+7q^{-26}-2q^{-28}-6q^{-30}+
\\
+8q^{-32}-2q^{-34}-5q^{-36}+5q^{-38}-q^{-40}-3q^{-42}+2q^{-44}-q^{-48}+q^{-50}
\ee

\paragraph{Special polynomials}

\be \mathfrak{H}_{[1]}=-2A^{-8}+2A^{-6}+A^{-4} \ee \be
\mathfrak{H}_{[1,
1]}=\mathfrak{H}_{[2]}=(\mathfrak{H}_{[1]})^2=4A^{-16}-8A^{-14}+4A^{-10}+A^{-8}
\ee

\paragraph{Ooguri-Vafa polynomials}

\be f_{[1,
1]}=\frac{\{A\}^2\{A/q\}\{Aq\}}{\{q\}}\Big((q^{17}+2q^{13}+q^{11}+2q^{9}+q^{7}+3q^{5}+q^{3}+2q^{1}+q^{-3})A^{-14}+
\\
+(q^{15}+q^{11}+q^{9}+q^{5}+q^{3}+q^{1}-q^{-1}+q^{-3})A^{-12}+(q^{13}-q^{11}+q^{9}+2q^{1}-2q^{-1}+q^{-3})A^{-10}\Big)
\ee \be
f_{[2]}=\frac{\{A\}^2\{A/q\}\{Aq\}}{\{q\}}\Big((-q^{3}-2q^{-1}-q^{-3}-3q^{-5}-q^{-7}-2q^{-9}-q^{-11}-2q^{-13}-q^{-17})A^{-14}+
\\
+(-q^{3}+q^{1}-q^{-1}-q^{-3}-q^{-5}-q^{-9}-q^{-11}-q^{-15})A^{-12}+(-q^{3}+2q^{1}-2q^{-1}-q^{-9}+q^{-11}-q^{-13})A^{-10}\Big)
\ee

\paragraph{Special Ooguri-Vafa polynomials}

\be {\mathfrak f}_{[2]}=-{\mathfrak
f}_{[1,1]}=-\frac{2(A^4+3A^2+7)(A-A^{-1})^3}{A^{14}} \ee

\subsubsection*{Numbers $N_{R,n,k}$}

\vspace{-1cm}

\noindent
\begin{tabular}{cccc}
$N_{[1]}:$ &
\begin{tabular}{|c|cccc|}
\hline
&&&& \\
$k \backslash n=$& -9 & -7 & -5 & -3 \\
&&&& \\
\hline
&&&& \\
0 & 2 & -4 & 1 & 1 \\
&&&& \\
1 & 1 & -4 & 0 & 3 \\
&&&& \\
2 & 0 & -1 & 0 & 1 \\
&&&& \\
\hline
\end{tabular}
& $N_{[1,1]}:$ &
\begin{tabular}{|c|ccccccc|}
\hline
&&&&&&& \\
$k \backslash n=$& -18 & -16 & -14 & -12 & -10 & -8 & -6 \\
&&&&&&& \\
\hline
&&&&&&& \\
0 & 41 & -146 & 179 & -76 & -1 & -2 & 5 \\
&&&&&&& \\
1 & 200 & -755 & 944 & -390 & -8 & -15 & 24 \\
&&&&&&& \\
2 & 398 & -1639 & 2131 & -890 & -6 & -35 & 41 \\
&&&&&&& \\
3 & 412 & -1917 & 2628 & -1123 & -1 & -28 & 29 \\
&&&&&&& \\
4 & 241 & -1320 & 1926 & -847 & 0 & -9 & 9 \\
&&&&&&& \\
5 & 80 & -549 & 859 & -390 & 0 & -1 & 1 \\
&&&&&&& \\
6 & 14 & -135 & 228 & -107 & 0 & 0 & 0 \\
&&&&&&& \\
7 & 1 & -18 & 33 & -16 & 0 & 0 & 0 \\
&&&&&&& \\
8 & 0 & -1 & 2 & -1 & 0 & 0 & 0 \\
&&&&&&& \\
\hline
\end{tabular}
\end{tabular}

\bigskip

\noindent
\begin{tabular}{cc}
$N_{[2]}:$ &
\begin{tabular}{|c|ccccccc|}
\hline
&&&&&&& \\
$k \backslash n=$& -18 & -16 & -14 & -12 & -10 & -8 & -6 \\
&&&&&&& \\
\hline
&&&&&&& \\
0 & 55 & -196 & 241 & -104 & 1 & -4 & 7 \\
&&&&&&& \\
1 & 330 & -1231 & 1531 & -634 & -10 & -27 & 41 \\
&&&&&&& \\
2 & 821 & -3284 & 4213 & -1751 & -15 & -75 & 91 \\
&&&&&&& \\
3 & 1085 & -4787 & 6418 & -2716 & -7 & -85 & 92 \\
&&&&&&& \\
4 & 837 & -4202 & 5940 & -2575 & -1 & -45 & 46 \\
&&&&&&& \\
5 & 389 & -2314 & 3472 & -1547 & 0 & -11 & 11 \\
&&&&&&& \\
6 & 107 & -803 & 1286 & -590 & 0 & -1 & 1 \\
&&&&&&& \\
7 & 16 & -170 & 292 & -138 & 0 & 0 & 0 \\
&&&&&&& \\
8 & 1 & -20 & 37 & -18 & 0 & 0 & 0 \\
&&&&&&& \\
9 & 0 & -1 & 2 & -1 & 0 & 0 & 0 \\
&&&&&&& \\
\hline
\end{tabular}
\end{tabular}

\bigskip

\section*{\fbox{Knot $7_5$}}

{\large $(-2,1|-1,-4)$}

\bigskip

\paragraph{HOMFLY polynomials}
\be \frac{H_{[1]}}{\
^*{S_{[1]}}}=A^6\Big((q^{4}-q^{2}+2-q^{-2}+q^{-4})A^{-2}+(q^{4}-2q^{2}+2-2q^{-2}+q^{-4})+(-q^{2}+1-q^{-2})A^{2}
\Big)=
\\
=\frac{A^6}{\ ^*{S_{[1]}}}\Big(\ ^*{S_{[3]}}q^{-6}+ \
^*{S_{[2,1]}}(-q^{6}+2q^{4}-3q^{2}+3-3q^{-2}+2q^{-4}-q^{-6})+ \
^*{S_{[1,1,1]}}q^{6}\Big) \ee \be \frac{H_{[1,1]}}{\
^*{S_{[1,1]}}}=A^{12}q^{-24}\Big((q^{32}-q^{30}-q^{28}+3q^{26}-3q^{22}+3q^{20}
+q^{18}-2q^{16}+2q^{14}+q^{12}-q^{10}+q^{8})A^{-4}+
\\+(q^{32}+q^{30}-3q^{28}+6q^{24}-3q^{22}-5q^{20}+6q^{18}-5q^{14}+3q^{12}-2q^{8}+q^{6})A^{-2}+
\\+(-3q^{26}+5q^{22}-5q^{20}-5q^{18}+7q^{16}-q^{14}-5q^{12}+4q^{10}-2q^{6}+q^{4})+
\\+(-q^{24}+q^{22}+3q^{20}-3q^{18}-2q^{16}+5q^{14}-q^{12}-3q^{10}+2q^{8}-q^{4})A^{2}+
\\+(q^{18}-q^{16}-q^{14}+2q^{12}-q^{8}+q^{6})A^{4}
\Big)=
\\
=\frac{A^{12}q^{-24}}{\ ^*{S_{[1,1]}}}\Big(\ ^*{S_{[3,3]}}+ \
^*{S_{[3,2,1]}}(-q^{12}+2q^{10}-3q^{8}+3q^{6}-3q^{4}+2q^{2}-1)+ \
^*{S_{[3,1,1,1]}}q^{12}+ \ ^*{S_{[2,2,2]}}q^{12}+
\\
+\
^*{S_{[2,2,1,1]}}(q^{34}-2q^{32}+5q^{28}-6q^{26}-2q^{24}+12q^{22}-9q^{20}-
\\
-7q^{18}+16q^{16}-7q^{14}-9q^{12}+12q^{10}-2q^{8}-6q^{6}+5q^{4}-2+q^{-2})+
\\
+\
^*{S_{[2,1,1,1,1]}}(-q^{36}+2q^{32}-3q^{28}+3q^{24}-3q^{20}+2q^{16}-q^{12})+
\ ^*{S_{[1,1,1,1,1,1]}}q^{36}\Big) \ee \be \frac{H_{[2]}}{\
^*{S_{[2]}}}=A^{12}q^{24}\Big((q^{-8}-q^{-10}+q^{-12}+2q^{-14}-2q^{-16}+q^{-18}
+3q^{-20}-3q^{-22}+3q^{-26}-q^{-28}-q^{-30}+q^{-32})A^{-4}+
\\+(q^{-6}-2q^{-8}+3q^{-12}-5q^{-14}+6q^{-18}-5q^{-20}-3q^{-22}
+6q^{-24}-3q^{-28}+q^{-30}+q^{-32})A^{-2}+
\\+(q^{-4}-2q^{-6}+4q^{-10}-5q^{-12}-q^{-14}+7q^{-16}-5q^{-18}-5q^{-20}+5q^{-22}-3q^{-26})+
\\+(-q^{-4}+2q^{-8}-3q^{-10}-q^{-12}+5q^{-14}-2q^{-16}-3q^{-18}+3q^{-20}+q^{-22}-q^{-24})A^{2}+
\\+(q^{-6}-q^{-8}+2q^{-12}-q^{-14}-q^{-16}+q^{-18})A^{4}
\Big)=
\\
=\frac{A^{12}q^{24}}{\ ^*{S_{[2]}}}\Big(\ ^*{S_{[6]}}q^{-36}+ \
^*{S_{[5,1]}}(-q^{-12}+2q^{-16}-3q^{-20}+3q^{-24}-3q^{-28}+2q^{-32}-q^{-36})+
\\
+\
^*{S_{[4,2]}}(q^{2}-2+5q^{-4}-6q^{-6}-2q^{-8}+12q^{-10}-9q^{-12}-7q^{-14}+
\\
+16q^{-16}-7q^{-18}-9q^{-20}+12q^{-22}-2q^{-24}-6q^{-26}+5q^{-28}-2q^{-32}+q^{-34})+
\\
+\ ^*{S_{[4,1,1]}}q^{-12}+ \ ^*{S_{[3,3]}}q^{-12}+ \
^*{S_{[3,2,1]}}(-1+2q^{-2}-3q^{-4}+3q^{-6}-3q^{-8}+2q^{-10}-q^{-12})+
\ ^*{S_{[2,2,2]}}\Big) \ee

\paragraph{Alexander polynomials}

\be \mathfrak{A}_{[1]}=2q^{4}-4q^{2}+5-4q^{-2}+2q^{-4} \ee \be
\mathfrak{A}_{[1,
1]}=\mathfrak{A}_{[2]}=\mathfrak{A}_{[1]}(q^2)=2q^{8}-4q^{4}+5-4q^{-4}+2q^{-8}
\ee

\paragraph{Jones polynomials}

\be
J_{[1]}=-q^{18}+2q^{16}-3q^{14}+3q^{12}-3q^{10}+3q^{8}-q^{6}+q^{4}
\ee \be J_{[1, 1]}=1 \ee \be
J_{[2]}=q^{50}-2q^{48}+5q^{44}-6q^{42}-2q^{40}+11q^{38}-9q^{36}-4q^{34}+14q^{32}-
\\
-10q^{30}-5q^{28}+13q^{26}-7q^{24}-5q^{22}+9q^{20}-3q^{18}-3q^{16}+4q^{14}-q^{10}+q^{8}
\ee

\paragraph{Special polynomials}

\be \mathfrak{H}_{[1]}=2A^{4}-A^{8} \ee \be \mathfrak{H}_{[1,
1]}=\mathfrak{H}_{[2]}=(\mathfrak{H}_{[1]})^2=4A^{8}-4A^{12}+A^{16}
\ee

\paragraph{Ooguri-Vafa polynomials}

\be f_{[1,
1]}=\frac{\{A\}^2\{A/q\}\{Aq\}(q^2-1+q^{-2})}{\{q\}}\Big((-2q^{3}+q^{1}-q^{-1}-2q^{-3}-q^{-5}-q^{-9})A^{10}+
\\
+(-2q^{3}-q^{-3}-2q^{-5}-q^{-11})A^{12}+(-q^{3}-q^{-5}-q^{-7}-q^{-13})A^{14}\Big)
\ee \be
f_{[2]}=\frac{\{A\}^2\{A/q\}\{Aq\}(q^2-1+q^{-2})}{\{q\}}\Big((q^{13}+q^{9}+2q^{7}+q^{5}-q^{3}+2q^{1})A^{10}+
\\
+(q^{15}+2q^{9}+q^{7}+2q^{1})A^{12}+(q^{17}+q^{11}+q^{9}+q^{1})A^{14}\Big)
\ee

\paragraph{Special Ooguri-Vafa polynomials}

\be {\mathfrak f}_{[2]}=-{\mathfrak
f}_{[1,1]}=2A^{10}(2A^4+3A^2+3)(A-A^{-1})^3 \ee

\subsubsection*{Numbers $N_{R,n,k}$}

\noindent
\begin{tabular}{cc}
$N_{[1]}:$ &
\begin{tabular}{|c|cccc|}
\hline
&&&& \\
$k \backslash n=$& 3 & 5 & 7 & 9 \\
&&&& \\
\hline
&&&& \\
0 & -2 & 2 & 1 & -1 \\
&&&& \\
1 & -3 & 1 & 3 & -1 \\
&&&& \\
2 & -1 & 0 & 1 & 0 \\
&&&& \\
\hline
\end{tabular}
\end{tabular}

\bigskip

\noindent \hspace{-1cm}\begin{tabular}{cccc} $N_{[1,1]}:$ &
\begin{tabular}{|c|ccccccc|}
\hline
&&&&&&& \\
$k \backslash n=$& 6 & 8 & 10 & 12 & 14 & 16 & 18 \\
&&&&&&& \\
\hline
&&&&&&& \\
0 & 17 & -50 & 47 & -28 & 47 & -50 & 17 \\
&&&&&&& \\
1 & 71 & -203 & 181 & -218 & 481 & -443 & 131 \\
&&&&&&& \\
2 & 118 & -324 & 269 & -736 & 1764 & -1484 & 393 \\
&&&&&&& \\
3 & 101 & -261 & 197 & -1383 & 3332 & -2604 & 618 \\
&&&&&&& \\
4 & 47 & -112 & 75 & -1563 & 3674 & -2681 & 560 \\
&&&&&&& \\
5 & 11 & -24 & 14 & -1091 & 2480 & -1689 & 299 \\
&&&&&&& \\
6 & 1 & -2 & 1 & -470 & 1032 & -654 & 92 \\
&&&&&&& \\
7 & 0 & 0 & 0 & -121 & 257 & -151 & 15 \\
&&&&&&& \\
8 & 0 & 0 & 0 & -17 & 35 & -19 & 1 \\
&&&&&&& \\
9 & 0 & 0 & 0 & -1 & 2 & -1 & 0 \\
&&&&&&& \\
\hline
\end{tabular}
& $N_{[2]}:$ &
\begin{tabular}{|c|ccccccc|}
\hline
&&&&&&& \\
$k \backslash n=$& 6 & 8 & 10 & 12 & 14 & 16 & 18 \\
&&&&&&& \\
\hline
&&&&&&& \\
0 & 11 & -32 & 31 & -24 & 41 & -40 & 13 \\
&&&&&&& \\
1 & 37 & -103 & 91 & -144 & 325 & -289 & 83 \\
&&&&&&& \\
2 & 48 & -127 & 101 & -400 & 968 & -793 & 203 \\
&&&&&&& \\
3 & 30 & -74 & 52 & -619 & 1482 & -1125 & 254 \\
&&&&&&& \\
4 & 9 & -20 & 12 & -560 & 1295 & -912 & 176 \\
&&&&&&& \\
5 & 1 & -2 & 1 & -299 & 665 & -433 & 67 \\
&&&&&&& \\
6 & 0 & 0 & 0 & -92 & 197 & -118 & 13 \\
&&&&&&& \\
7 & 0 & 0 & 0 & -15 & 31 & -17 & 1 \\
&&&&&&& \\
8 & 0 & 0 & 0 & -1 & 2 & -1 & 0 \\
&&&&&&& \\
\hline
\end{tabular}
\end{tabular}

\bigskip

\section*{\fbox{Knot $8_2$}}

{\large $(1,-1|1,-5)$}

\bigskip

\paragraph{HOMFLY polynomials}
\be \frac{H_{[1]}}{\
^*{S_{[1]}}}=A^4\Big((q^{4}+1+q^{-4})A^{-2}+(-q^{6}+q^{4}-2q^{2}+1-2q^{-2}+q^{-4}-q^{-6})+(q^{4}-q^{2}+1-q^{-2}+q^{-4})A^{2}
\Big)=
\\
=\frac{A^4}{\ ^*{S_{[1]}}}\Big(\ ^*{S_{[3]}}q^{-4}+ \
^*{S_{[2,1]}}(q^{8}-2q^{6}+2q^{4}-3q^{2}+3-3q^{-2}+2q^{-4}-2q^{-6}+q^{-8})+
\ ^*{S_{[1,1,1]}}q^{4}\Big) \ee \be \frac{H_{[1,1]}}{\
^*{S_{[1,1]}}}=A^8q^{-16}\Big((q^{28}+q^{22}+q^{20}+q^{16}+q^{14}+q^{12}+q^{10}+q^{8}+q^{4})A^{-4}+
\\
+(-q^{30}-q^{28}+q^{26}-q^{24}-3q^{22}-3q^{16}-2q^{14}-q^{12}-2q^{10}-q^{8}-q^{6}-2q^{4}-q^{-2})A^{-2}+
\\
+(q^{30}-q^{28}+4q^{24}-3q^{20}+4q^{18}+3q^{16}-2q^{14}+q^{12}+3q^{10}+2q^{6}+q^{4}-q^{2}+2+q^{-2}-q^{-4}+q^{-6})+
\\
+(-q^{26}+2q^{22}-2q^{20}-3q^{18}+3q^{16}+q^{14}-4q^{12}+2q^{8}-2q^{6}-2+q^{-2}-q^{-6})A^{2}+
\\
+(q^{20}-q^{18}-q^{16}+2q^{14}-2q^{10}+q^{8}+q^{6}-q^{4}+q^{2}-q^{-2}+q^{-4})A^{4}
\Big)=
\\
=\frac{A^8q^{-16}}{\ ^*{S_{[1,1]}}}\Big(\ ^*{S_{[3,3]}}+ \
^*{S_{[3,2,1]}}(q^{12}-2q^{10}+2q^{8}-3q^{6}+3q^{4}-3q^{2}+2-2q^{-2}+q^{-4})+
\ ^*{S_{[3,1,1,1]}}q^{8}+ \ ^*{S_{[2,2,2]}}q^{8}+
\\
+\
^*{S_{[2,2,1,1]}}(q^{36}-2q^{34}-q^{32}+5q^{30}-3q^{28}-5q^{26}+8q^{24}+q^{22}-9q^{20}+5q^{18}+4q^{16}-7q^{14}+
\\
+2q^{12}+3q^{10}-4q^{8}+2q^{6}+2q^{4}-4q^{2}+3+q^{-2}-4q^{-4}+3q^{-6}-2q^{-10}+q^{-12})+
\\
+\
^*{S_{[2,1,1,1,1]}}(q^{32}-2q^{28}+2q^{24}-3q^{20}+3q^{16}-3q^{12}+2q^{8}-2q^{4}+1)+
\ ^*{S_{[1,1,1,1,1,1]}}q^{24}\Big) \ee \be \frac{H_{[2]}}{\
^*{S_{[2]}}}=A^8q^{16}\Big((q^{-4}+q^{-8}+q^{-10}+q^{-12}+q^{-14}+q^{-16}+q^{-20}+q^{-22}+q^{-28})A^{-4}+
\\
+(-q^{2}-2q^{-4}-q^{-6}-q^{-8}-2q^{-10}-q^{-12}-2q^{-14}-3q^{-16}-3q^{-22}-q^{-24}+q^{-26}-q^{-28}-q^{-30})A^{-2}+
\\
+(q^{6}-q^{4}+q^{2}+2-q^{-2}+q^{-4}+2q^{-6}+3q^{-10}+q^{-12}-2q^{-14}+3q^{-16}+4q^{-18}-3q^{-20}+4q^{-24}-q^{-28}+q^{-30})+
\\
+(-q^{6}+q^{2}-2-2q^{-6}+2q^{-8}-4q^{-12}+q^{-14}+3q^{-16}-3q^{-18}-2q^{-20}+2q^{-22}-q^{-26})A^{2}+
\\
+(q^{4}-q^{2}+q^{-2}-q^{-4}+q^{-6}+q^{-8}-2q^{-10}+2q^{-14}-q^{-16}-q^{-18}+q^{-20})A^{4}
\Big)=
\\
=\frac{A^8q^{16}}{\ ^*{S_{[2]}}}\Big(\ ^*{S_{[6]}}q^{-24}+ \
^*{S_{[5,1]}}(1-2q^{-4}+2q^{-8}-3q^{-12}+3q^{-16}-3q^{-20}+2q^{-24}-2q^{-28}+q^{-32})+
\\
+\
^*{S_{[4,2]}}(q^{12}-2q^{10}+3q^{6}-4q^{4}+q^{2}+3-4q^{-2}+2q^{-4}+2q^{-6}-4q^{-8}+3q^{-10}+
\\
+2q^{-12}-7q^{-14}+4q^{-16}+5q^{-18}-9q^{-20}+q^{-22}+8q^{-24}-5q^{-26}-3q^{-28}+5q^{-30}-q^{-32}-2q^{-34}+q^{-36})+
\\
+\ ^*{S_{[4,1,1]}}q^{-8}+ \ ^*{S_{[3,3]}}q^{-8}+ \
^*{S_{[3,2,1]}}(q^{4}-2q^{2}+2-3q^{-2}+3q^{-4}-3q^{-6}+2q^{-8}-2q^{-10}+q^{-12})+
\ ^*{S_{[2,2,2]}}\Big) \ee

\paragraph{Alexander polynomials}

\be \mathfrak{A}_{[1]}=-q^{6}+3q^{4}-3q^{2}+3-3q^{-2}+3q^{-4}-q^{-6}
\ee \be \mathfrak{A}_{[1,
1]}=\mathfrak{A}_{[2]}=\mathfrak{A}_{[1]}(q^2)=-q^{12}+3q^{8}-3q^{4}+3-3q^{-4}+3q^{-8}-q^{-12}
\ee

\paragraph{Jones polynomials}

\be
J_{[1]}=q^{16}-2q^{14}+2q^{12}-3q^{10}+3q^{8}-2q^{6}+2q^{4}-q^{2}+1
\ee \be J_{[1, 1]}=1 \ee \be
J_{[2]}=q^{44}-2q^{42}+3q^{38}-4q^{36}+2q^{34}+3q^{32}-6q^{30}+3q^{28}+4q^{26}-7q^{24}+2q^{22}+
\\
+5q^{20}-7q^{18}+q^{16}+5q^{14}-5q^{12}+5q^{8}-3q^{6}-q^{4}+3q^{2}-1-q^{-2}+q^{-4}
\ee

\paragraph{Special polynomials}

\be \mathfrak{H}_{[1]}=3A^{2}-3A^4+A^{6} \ee \be
\mathfrak{H}_{[1,1]}=\mathfrak{H}_{[2]}=(\mathfrak{H}_{[1]})^2=9A^{4}-18A^{6}+15-6A^{10}+A^{12}
\ee

\paragraph{Ooguri-Vafa polynomials}

\be f_{[1,
1]}=\frac{\{A\}^2\{A/q\}\{Aq\}(q^2-1+q^{-2})}{\{q\}}\Big((q^{11}+q^{9}-q^{-1}-q^{-3}-q^{-5})A^{6}+
\\
+(q^{9}+2q^{1}+q^{-7}+q^{-9}+q^{-11})A^8+(-q^{3}+q^{1}-q^{-3}-q^{-5}-q^{-15})A^{10}\Big)
\ee \be f_{[2]}=\frac{\{A\}^2\{A/q\}\{Aq\}(q^2-1+q^{-2})
((q^{9}+q^{7}+q^{5}-q^{-5}-q^{-7})A^{6}+(-q^{15}-q^{13}-q^{11}-2q^{3}-q^{-5})A^8+(q^{19}+q^{9}+q^{7}-q^{3}+q^{1})A^{10})
}{\{q\}} \ee

\paragraph{Special Ooguri-Vafa polynomials}

\be {\mathfrak f}_{[2]}=-{\mathfrak
f}_{[1,1]}=A^6(3A^4-6A^2+1)(A-A^{-1})^3 \ee

\subsubsection*{Numbers $N_{R,n,k}$}

\noindent
\begin{tabular}{cccc}
$N_{[1]}:$ &
\begin{tabular}{|c|cccc|}
\hline
&&&& \\
$k \backslash n=$& 1 & 3 & 5 & 7 \\
&&&& \\
\hline
&&&& \\
0 & -3 & 6 & -4 & 1 \\
&&&& \\
1 & -4 & 11 & -10 & 3 \\
&&&& \\
2 & -1 & 6 & -6 & 1 \\
&&&& \\
3 & 0 & 1 & -1 & 0 \\
&&&& \\
\hline
\end{tabular}
& $N_{[1,1]}:$ &
\begin{tabular}{|c|ccccccc|}
\hline
&&&&&&& \\
$k \backslash n=$& 2 & 4 & 6 & 8 & 10 & 12 & 14 \\
&&&&&&& \\
\hline
&&&&&&& \\
0 & 16 & -81 & 179 & -226 & 174 & -77 & 15 \\
&&&&&&& \\
1 & 45 & -320 & 964 & -1585 & 1482 & -735 & 149 \\
&&&&&&& \\
2 & 43 & -526 & 2311 & -4880 & 5319 & -2870 & 603 \\
&&&&&&& \\
3 & 16 & -473 & 3217 & -8572 & 10567 & -6017 & 1262 \\
&&&&&&& \\
4 & 2 & -254 & 2838 & -9430 & 12860 & -7524 & 1508 \\
&&&&&&& \\
5 & 0 & -81 & 1629 & -6760 & 10036 & -5903 & 1079 \\
&&&&&&& \\
6 & 0 & -14 & 604 & -3197 & 5107 & -2969 & 469 \\
&&&&&&& \\
7 & 0 & -1 & 139 & -987 & 1682 & -954 & 121 \\
&&&&&&& \\
8 & 0 & 0 & 18 & -191 & 345 & -189 & 17 \\
&&&&&&& \\
9 & 0 & 0 & 1 & -21 & 40 & -21 & 1 \\
&&&&&&& \\
10 & 0 & 0 & 0 & -1 & 2 & -1 & 0 \\
&&&&&&& \\
\hline
\end{tabular}
\end{tabular}

\bigskip

\noindent
\begin{tabular}{cc}
$N_{[2]}:$ &
\begin{tabular}{|c|ccccccc|}
\hline
&&&&&&& \\
$k \backslash n=$& 2 & 4 & 6 & 8 & 10 & 12 & 14 \\
&&&&&&& \\
\hline
&&&&&&& \\
0 & 15 & -71 & 146 & -174 & 131 & -59 & 12 \\
&&&&&&& \\
1 & 50 & -281 & 706 & -1049 & 955 & -482 & 101 \\
&&&&&&& \\
2 & 63 & -434 & 1414 & -2695 & 2911 & -1601 & 342 \\
&&&&&&& \\
3 & 37 & -332 & 1547 & -3875 & 4835 & -2795 & 583 \\
&&&&&&& \\
4 & 10 & -132 & 1025 & -3432 & 4817 & -2838 & 550 \\
&&&&&&& \\
5 & 1 & -26 & 427 & -1938 & 2992 & -1754 & 298 \\
&&&&&&& \\
6 & 0 & -2 & 110 & -697 & 1164 & -667 & 92 \\
&&&&&&& \\
7 & 0 & 0 & 16 & -154 & 275 & -152 & 15 \\
&&&&&&& \\
8 & 0 & 0 & 1 & -19 & 36 & -19 & 1 \\
&&&&&&& \\
9 & 0 & 0 & 0 & -1 & 2 & -1 & 0 \\
&&&&&&& \\
\hline
\end{tabular}
\end{tabular}

\bigskip

\section*{\fbox{Knot $8_5$}}

{\large $(-1,3|-1,3)$}

\bigskip

\paragraph{HOMFLY polynomials}
\be \frac{H_{[1]}}{\
^*{S_{[1]}}}=A^{-4}\Big((q^{4}-q^{2}+2-q^{-2}+q^{-4})A^{-2}+(-q^{6}+q^{4}-3q^{2}+1-3q^{-2}+q^{-4}-q^{-6})+(q^{4}+2+q^{-4})A^{2}
\Big)=
\\
=\frac{A^{-4}}{\ ^*{S_{[1]}}}\Big(\ ^*{S_{[3]}}q^{4}+ \
^*{S_{[2,1]}}(q^{8}-2q^{6}+3q^{4}-4q^{2}+3-4q^{-2}+3q^{-4}-2q^{-6}+q^{-8})+
\ ^*{S_{[1,1,1]}}q^{-4}\Big) \ee \be \frac{H_{[1,1]}}{\
^*{S_{[1,1]}}}=A^{-8}q^{16}\Big((q^{4}-q^{2}+1+q^{-2}-2q^{-4}+2q^{-6}+2q^{-8}-3q^{-10}+q^{-12}+3q^{-14}-q^{-16}-q^{-18}+q^{-20})A^{-4}+
\\
+(-q^{6}-3+q^{-2}-q^{-4}-5q^{-6}+2q^{-8}-q^{-10}-8q^{-12}+3q^{-16}-5q^{-18}-3q^{-20}+2q^{-22}-q^{-26})A^{-2}+
\\
+(q^{6}-q^{4}+2q^{2}+2-q^{-2}+4q^{-4}+4q^{-6}+8q^{-10}+5q^{-12}-
\\
-q^{-14}+7q^{-16}+8q^{-18}-3q^{-20}+q^{-22}+5q^{-24}-q^{-28}+q^{-30})+
\\
+(-q^{2}-q^{-2}-3q^{-4}-q^{-6}-3q^{-8}-6q^{-10}-3q^{-12}-4q^{-14}-
\\
-7q^{-16}-2q^{-18}-q^{-20}-5q^{-22}-2q^{-24}+q^{-26}-q^{-28}-q^{-30})A^{2}+
\\
+(q^{-4}+2q^{-8}+2q^{-10}+q^{-12}+2q^{-14}+3q^{-16}+2q^{-20}+2q^{-22}+q^{-28})A^{4}
\Big)=
\\
=\frac{A^{-8}q^{16}}{\ ^*{S_{[1,1]}}}\Big(\ ^*{S_{[3,3]}}+ \
^*{S_{[3,2,1]}}(q^{4}-2q^{2}+3-4q^{-2}+3q^{-4}-4q^{-6}+3q^{-8}-2q^{-10}+q^{-12})+
\ ^*{S_{[3,1,1,1]}}q^{-8}+
\\
+\ ^*{S_{[2,2,2]}}q^{-8}+\
^*{S_{[2,2,1,1]}}(q^{12}-2q^{10}+q^{8}+2q^{6}-6q^{4}+4q^{2}+3-8q^{-2}+6q^{-4}+2q^{-6}-9q^{-8}+8q^{-10}+
\\
+3q^{-12}-12q^{-14}+8q^{-16}+8q^{-18}-14q^{-20}+2q^{-22}+10q^{-24}-8q^{-26}-3q^{-28}+6q^{-30}-q^{-32}-2q^{-34}+q^{-36})+
\\
+\
^*{S_{[2,1,1,1,1]}}(1-2q^{-4}+3q^{-8}-4q^{-12}+3q^{-16}-4q^{-20}+3q^{-24}-2q^{-28}+q^{-32})+
\ ^*{S_{[1,1,1,1,1,1]}}q^{-24}\Big) \ee \be \frac{H_{[2]}}{\
^*{S_{[2]}}}=A^{-8}q^{-16}\Big((q^{20}-q^{18}-q^{16}+3q^{14}+q^{12}-3q^{10}+2q^{8}+2q^{6}-2q^{4}+q^{2}+1-q^{-2}+q^{-4})A^{-4}+
\\
+(-q^{26}+2q^{22}-3q^{20}-5q^{18}+3q^{16}-8q^{12}-q^{10}+2q^{8}-5q^{6}-q^{4}+q^{2}-3-q^{-6})A^{-2}+
\\
+(q^{30}-q^{28}+5q^{24}+q^{22}-3q^{20}+8q^{18}+7q^{16}-q^{14}+5q^{12}+8q^{10}+4q^{6}+4q^{4}-q^{2}+2+2q^{-2}-q^{-4}+q^{-6})+
\\
+(-q^{30}-q^{28}+q^{26}-2q^{24}-5q^{22}-q^{20}-2q^{18}-7q^{16}-4q^{14}-3q^{12}-6q^{10}-3q^{8}-q^{6}-3q^{4}-q^{2}-q^{-2})A^{2}+
\\
+(q^{28}+2q^{22}+2q^{20}+3q^{16}+2q^{14}+q^{12}+2q^{10}+2q^{8}+q^{4})A^{4}
\Big)=
\\
=\frac{A^{-8}q^{-16}}{\ ^*{S_{[2]}}}\Big(\ ^*{S_{[6]}}q^{24}+ \
^*{S_{[5,1]}}(q^{32}-2q^{28}+3q^{24}-4q^{20}+3q^{16}-4q^{12}+3q^{8}-2q^{4}+1)+
\\
+\
^*{S_{[4,2]}}(q^{36}-2q^{34}-q^{32}+6q^{30}-3q^{28}-8q^{26}+10q^{24}+2q^{22}-14q^{20}+8q^{18}+8q^{16}-12q^{14}+
\\
+3q^{12}+8q^{10}-9q^{8}+2q^{6}+6q^{4}-8q^{2}+3+4q^{-2}-6q^{-4}+2q^{-6}+q^{-8}-2q^{-10}+q^{-12})+
\\
+\ ^*{S_{[4,1,1]}}q^{8}+ \ ^*{S_{[3,3]}}q^{8}+ \
^*{S_{[3,2,1]}}(q^{12}-2q^{10}+3q^{8}-4q^{6}+3q^{4}-4q^{2}+3-2q^{-2}+q^{-4})+
\ ^*{S_{[2,2,2]}}\Big) \ee

\paragraph{Alexander polynomials}

\be \mathfrak{A}_{[1]}=-q^{6}+3q^{4}-4q^{2}+5-4q^{-2}+3q^{-4}-q^{-6}
\ee \be \mathfrak{A}_{[1,
1]}=\mathfrak{A}_{[2]}=\mathfrak{A}_{[1]}(q^2)=-q^{12}+3q^{8}-4q^{4}+5-4q^{-4}+3q^{-8}-q^{-12}
\ee

\paragraph{Jones polynomials}

\be
J_{[1]}=1-q^{-2}+3q^{-4}-3q^{-6}+3q^{-8}-4q^{-10}+3q^{-12}-2q^{-14}+q^{-16}
\ee \be J_{[1, 1]}=1 \ee \be
J_{[2]}=q^{4}-q^{2}-1+4q^{-2}-q^{-4}-5q^{-6}+7q^{-8}-9q^{-12}+8q^{-14}+3q^{-16}-12q^{-18}+7q^{-20}+6q^{-22}-12q^{-24}+
\\
+5q^{-26}+7q^{-28}-10q^{-30}+3q^{-32}+5q^{-34}-6q^{-36}+2q^{-38}+q^{-40}-2q^{-42}+q^{-44}
\ee

\paragraph{Special polynomials}

\be \mathfrak{H}_{[1]}=2A^{-6}-5A^{-4}+4A^{-2} \ee \be
\mathfrak{H}_{[1,1]}=\mathfrak{H}_{[2]}=(\mathfrak{H}_{[1]})^2=4A^{-12}-20A^{-10}+41A^{-8}-40A^{-6}+16A^{-4}
\ee

\paragraph{Ooguri-Vafa polynomials}

\be f_{[1, 1]}=\frac{\{A\}^2\{A/q\}\{Aq\}}{\{q\}}
\Big((q^{19}-q^{17}+2q^{15}+2q^{9}+q^{7}+q^{3}+3q^{1}-2q^{-1}+2q^{-3})A^{-10}+
\\
+(-q^{15}-2q^{11}-q^{9}-2q^{7}-q^{5}-5q^{3}+2q^{1}-4q^{-1}-2q^{-5}+q^{-7}-q^{-9})A^{-8}+(q^{9}+2q^{5}+q^{1}-2q^{-5}-q^{-11})A^{-6}\Big)
\ee \be f_{[2]}=\frac{\{A\}^2\{A/q\}\{Aq\}}{\{q\}}
\Big((-2q^{3}+2q^{1}-3q^{-1}-q^{-3}-q^{-7}-2q^{-9}-2q^{-15}+q^{-17}-q^{-19})A^{-10}+
\\
+(q^{9}-q^{7}+2q^{5}+4q^{1}-2q^{-1}+5q^{-3}+q^{-5}+2q^{-7}+q^{-9}+2q^{-11}+q^{-15})A^{-8}+(q^{11}+2q^{5}-q^{-1}-2q^{-5}-q^{-9})A^{-6}\Big)
\ee

\paragraph{Special Ooguri-Vafa polynomials}

\be {\mathfrak f}_{[2]}=-{\mathfrak
f}_{[1,1]}=-\frac{(A^4-16A^2+9)(A-A^{-1})^3}{A^{10}} \ee

\subsubsection*{Numbers $N_{R,n,k}$}

\noindent
\begin{tabular}{cc} $N_{[1]}:$ &
\begin{tabular}{|c|cccc|}
\hline
&&&& \\
$k \backslash n=$& -7 & -5 & -3 & -1 \\
&&&& \\
\hline
&&&& \\
0 & -2 & 7 & -9 & 4 \\
&&&& \\
1 & -3 & 11 & -12 & 4 \\
&&&& \\
2 & -1 & 6 & -6 & 1 \\
&&&& \\
3 & 0 & 1 & -1 & 0 \\
&&&& \\
\hline
\end{tabular}
\end{tabular}

\bigskip

\noindent \hspace{-1.5cm}\begin{tabular}{cccc} $N_{[1,1]}:$ &
\begin{tabular}{|c|ccccccc|}
\hline
&&&&&&& \\
$k \backslash n=$& -14 & -12 & -10 & -8 & -6 & -4 & -2 \\
&&&&&&& \\
\hline
&&&&&&& \\
0 & 25 & -123 & 262 & -318 & 237 & -103 & 20 \\
&&&&&&&\\
1 & 170 & -802 & 1536 & -1573 & 951 & -337 & 55 \\
&&&&&&&\\
2 & 475 & -2229 & 3987 & -3529 & 1702 & -470 & 64 \\
&&&&&&&\\
3 & 704 & -3421 & 5890 & -4601 & 1733 & -342 & 37 \\
&&&&&&&\\
4 & 605 & -3180 & 5402 & -3797 & 1093 & -133 & 10 \\
&&&&&&&\\
5 & 310 & -1857 & 3175 & -2043 & 440 & -26 & 1 \\
&&&&&&&\\
6 & 93 & -683 & 1194 & -713 & 111 & -2 & 0 \\
&&&&&&&\\
7 & 15 & -153 & 277 & -155 & 16 & 0 & 0 \\
&&&&&&&\\
8 & 1 & -19 & 36 & -19 & 1 & 0 & 0 \\
&&&&&&&\\
9 & 0 & -1 & 2 & -1 & 0 & 0 & 0 \\
&&&&&&&\\
\hline
\end{tabular}
& $N_{[2]}:$ &
\begin{tabular}{|c|ccccccc|}
\hline
&&&&&&& \\
$k \backslash n=$& -14 & -12 & -10 & -8 & -6 & -4 & -2 \\
&&&&&&& \\
\hline
&&&&&&& \\
0 & 34 & -175 & 381 & -454 & 316 & -123 & 21 \\
&&&&&&&\\
1 & 264 & -1285 & 2521 & -2559 & 1425 & -416 & 50 \\
&&&&&&&\\
2 & 882 & -4185 & 7631 & -6766 & 3013 & -619 & 44 \\
&&&&&&&\\
3 & 1601 & -7697 & 13423 & -10646 & 3822 & -519 & 16 \\
&&&&&&&\\
4 & 1729 & -8758 & 14953 & -10809 & 3148 & -265 & 2 \\
&&&&&&&\\
5 & 1157 & -6437 & 10961 & -7321 & 1722 & -82 & 0 \\
&&&&&&&\\
6 & 483 & -3103 & 5348 & -3333 & 619 & -14 & 0 \\
&&&&&&&\\
7 & 122 & -972 & 1716 & -1005 & 140 & -1 & 0 \\
&&&&&&&\\
8 & 17 & -190 & 347 & -192 & 18 & 0 & 0 \\
&&&&&&&\\
9 & 1 & -21 & 40 & -21 & 1 & 0 & 0 \\
&&&&&&&\\
10 & 0 & -1 & 2 & -1 & 0 & 0 & 0 \\
&&&&&&&\\
\hline
\end{tabular}
\end{tabular}

\bigskip

\section*{\fbox{Knot $8_7$}}

{\large $(-2,1|-1,4)$}

\bigskip

\paragraph{HOMFLY polynomials}
\be \frac{H_{[1]}}{\
^*{S_{[1]}}}=A^{-2}\Big((-q^{4}+q^{2}-2+q^{-2}-q^{-4})A^{-2}+
\\
+(q^{6}-q^{4}+3q^{2}-2+3q^{-2}-q^{-4}+q^{-6})+(-q^{4}+q^{2}-1+q^{-2}-q^{-4})A^{2}
\Big)=
\\
=\frac{A^{-2}}{\ ^*{S_{[1]}}}\Big(\ ^*{S_{[3]}}q^{2}+ \
^*{S_{[2,1]}}(-q^{8}+2q^{6}-3q^{4}+4q^{2}-5+4q^{-2}-3q^{-4}+2q^{-6}-q^{-8})+
\ ^*{S_{[1,1,1]}}q^{-2}\Big) \ee \be \frac{H_{[1,1]}}{\
^*{S_{[1,1]}}}=A^{-4}q^{8}\Big((q^{10}-q^{8}+2q^{4}-2q^{2}+4q^{-2}-2q^{-4}-q^{-6}+3q^{-8}-q^{-12}+q^{-14})A^{-4}+
\\
+(-q^{12}+q^{8}-3q^{6}+2q^{2}-7+5q^{-4}-8q^{-6}-4q^{-8}+
\\
+5q^{-10}-2q^{-12}-4q^{-14}+q^{-16}-q^{-20})A^{-2}+
\\
+(q^{12}-q^{10}+q^{8}+3q^{6}-2q^{4}+2q^{2}+5-5q^{-2}+5q^{-4}+8q^{-6}-
\\
-7q^{-8}+10q^{-12}-2q^{-14}-3q^{-16}+4q^{-18}+q^{-20}-q^{-22}+q^{-24})+
\\
+(-q^{8}+q^{4}-3q^{2}+q^{-2}-4q^{-4}+2q^{-6}+2q^{-8}-6q^{-10}+5q^{-14}-3q^{-16}-3q^{-18}+2q^{-20}-q^{-24})A^{2}+
\\
+(q^{2}-1+q^{-4}-q^{-6}+q^{-8}+q^{-10}-2q^{-12}+2q^{-16}-q^{-18}-q^{-20}+q^{-22})A^{4}
\Big)=
\\
=\frac{A^{-4}q^{8}}{\ ^*{S_{[1,1]}}}\Big(\ ^*{S_{[3,3]}}+ \
^*{S_{[3,2,1]}}(-q^{6}+2q^{4}-3q^{2}+4-5q^{-2}+4q^{-4}-3q^{-6}+2q^{-8}-q^{-10})+
\ ^*{S_{[3,1,1,1]}}q^{-4}+
\\ +\ ^*{S_{[2,2,2]}}q^{-4}+ \ ^*{S_{[2,2,1,1]}}(q^{18}-2q^{16}+4q^{12}-5q^{10}+q^{8}+7q^{6}-11q^{4}+4q^{2}+11-19q^{-2}+5q^{-4}+
\\
+17q^{-6}-20q^{-8}-q^{-10}+19q^{-12}-12q^{-14}-7q^{-16}+13q^{-18}-3q^{-20}-6q^{-22}+5q^{-24}-2q^{-28}+q^{-30})+
\\
+ \
^*{S_{[2,1,1,1,1]}}(-q^{8}+2q^{4}-3+4q^{-4}-5q^{-8}+4q^{-12}-3q^{-16}+2q^{-20}-q^{-24})+
\ ^*{S_{[1,1,1,1,1,1]}}q^{-12}\Big) \ee \be \frac{H_{[2]}}{\
^*{S_{[2]}}}=A^{-4}q^{-8}\Big((q^{14}-q^{12}+3q^{8}-q^{6}-2q^{4}+4q^{2}-2q^{-2}+2q^{-4}-q^{-8}+q^{-10})A^{-4}+
\\
+(-q^{20}+q^{16}-4q^{14}-2q^{12}+5q^{10}-4q^{8}-8q^{6}+5q^{4}-7+2q^{-2}-3q^{-6}+q^{-8}-q^{-12})A^{-2}+
\\
+(q^{24}-q^{22}+q^{20}+4q^{18}-3q^{16}-2q^{14}+10q^{12}-7q^{8}+8q^{6}+5q^{4}-5q^{2}+5+2q^{-2}-2q^{-4}+3q^{-6}+q^{-8}-q^{-10}+q^{-12})+
\\
+(-q^{24}+2q^{20}-3q^{18}-3q^{16}+5q^{14}-6q^{10}+2q^{8}+2q^{6}-4q^{4}+q^{2}-3q^{-2}+q^{-4}-q^{-8})A^{2}+
\\
+(q^{22}-q^{20}-q^{18}+2q^{16}-2q^{12}+q^{10}+q^{8}-q^{6}+q^{4}-1+q^{-2})A^{4}
\Big)=
\\
=\frac{A^{-4}q^{-8}}{\ ^*{S_{[2]}}}\Big(\ ^*{S_{[6]}}q^{12}+ \
^*{S_{[5,1]}}(-q^{24}+2q^{20}-3q^{16}+4q^{12}-5q^{8}+4q^{4}-3+2q^{-4}-q^{-8})+
\\
+\
^*{S_{[4,2]}}(q^{30}-2q^{28}+5q^{24}-6q^{22}-3q^{20}+13q^{18}-7q^{16}-12q^{14}+19q^{12}-q^{10}-20q^{8}+
\\
+17q^{6}+5q^{4}-19q^{2}+11+4q^{-2}-11q^{-4}+7q^{-6}+q^{-8}-5q^{-10}+4q^{-12}-2q^{-16}+q^{-18})+
\\
+\ ^*{S_{[4,1,1]}}q^{4}+ \ ^*{S_{[3,3]}}q^{4}+ \
^*{S_{[3,2,1]}}(-q^{10}+2q^{8}-3q^{6}+4q^{4}-5q^{2}+4-3q^{-2}+2q^{-4}-q^{-6})+
\ ^*{S_{[2,2,2]}}\Big) \ee

\paragraph{Alexander polynomials}

\be \mathfrak{A}_{[1]}=q^{6}-3q^{4}+5q^{2}-5+5q^{-2}-3q^{-4}+q^{-6}
\ee \be \mathfrak{A}_{[1,
1]}=\mathfrak{A}_{[2]}=\mathfrak{A}_{[1]}(q^2)=q^{12}-3q^{8}+5q^{4}-5+5q^{-4}-3q^{-8}+q^{-12}
\ee

\paragraph{Jones polynomials}

\be
J_{[1]}=-q^{4}+2q^{2}-2+4q^{-2}-4q^{-4}+4q^{-6}-3q^{-8}+2q^{-10}-q^{-12}
\ee \be J_{[1, 1]}=1 \ee \be
J_{[2]}=q^{14}-2q^{12}-q^{10}+5q^{8}-4q^{6}-4q^{4}+10q^{2}-3-9q^{-2}+14q^{-4}-2q^{-6}-14q^{-8}+
\\
+16q^{-10}-16q^{-14}+14q^{-16}+q^{-18}-12q^{-20}+9q^{-22}+q^{-24}-6q^{-26}+4q^{-28}-2q^{-32}+q^{-34}
\ee

\paragraph{Special polynomials}

\be \mathfrak{H}_{[1]}=-2A^{-4}+4A^{-2}-1 \ee \be \mathfrak{H}_{[1,
1]}=\mathfrak{H}_{[2]}=(\mathfrak{H}_{[1]})^2=4A^{-8}-16A^{-6}+20A^{-4}-8A^{-2}+1
\ee

\paragraph{Ooguri-Vafa polynomials}
\be f_{[1, 1]}=\frac{\{A\}^2\{A/q\}\{Aq\}(q^2-1+q^{-2})}{\{q\}}
\Big((q^{17}+q^{11}+3q^{5}+q^{-3})A^{-6}+
\\
+(-q^{13}-q^{11}-2q^{7}+q^{5}-2q^{-1}+2q^{-3}+q^{-5})A^{-4}+(q^{1}-q^{-1}+q^{-5}-q^{-9})A^{-2}\Big)
\ee \be f_{[2]}=\frac{\{A\}^2\{A/q\}\{Aq\}(q^2-1+q^{-2})}{\{q\}}
\Big((-q^{7}-3q^{-1}-q^{-7}-q^{-13})A^{-6}+
\\
+(-q^{9}-2q^{7}+2q^{5}-q^{-1}+2q^{-3}+q^{-7}+q^{-9})A^{-4}+(q^{13}-q^{9}+q^{5}-q^{3})A^{-2}\Big)
\ee

\paragraph{Special Ooguri-Vafa polynomials}

\be {\mathfrak f}_{[2]}=-{\mathfrak
f}_{[1,1]}=-\frac{2(A^3-3)(A-A^{-1})^3}{A^6} \ee

\subsubsection*{Numbers $N_{R,n,k}$}

\noindent
\begin{tabular}{cccc}
$N_{[1]}:$ &
\begin{tabular}{|c|cccc|}
\hline
&&&& \\
$k \backslash n=$& -1 & 1 & 3 & 5 \\
&&&& \\
\hline
&&&& \\
0 & 2 & -6 & 5 & -1 \\
&&&&\\
1 & 3 & -11 & 11 & -3 \\
&&&&\\
2 & 1 & -6 & 6 & -1 \\
&&&&\\
3 & 0 & -1 & 1 & 0 \\
&&&&\\
\hline
\end{tabular}
& $N_{[1,1]}:$ &
\begin{tabular}{|c|ccccccc|}
\hline
&&&&&&& \\
$k \backslash n=$& -2 & 0 & 2 & 4 & 6 & 8 & 10 \\
&&&&&&& \\
\hline
&&&&&&& \\
0 & 39 & -286 & 785 & -1028 & 665 & -194 & 19 \\
&&&&&&&\\
1 & 275 & -2553 & 8365 & -12718 & 9477 & -3227 & 381 \\
&&&&&&&\\
2 & 853 & -10436 & 41492 & -73336 & 62457 & -24578 & 3548 \\
&&&&&&&\\
3 & 1497 & -25234 & 123381 & -252783 & 242378 & -107162 & 17923 \\
&&&&&&&\\
4 & 1617 & -39707 & 242296 & -573983 & 610752 & -294896 & 53921 \\
&&&&&&&\\
5 & 1103 & -42639 & 331047 & -906126 & 1057291 & -544430 & 103754 \\
&&&&&&&\\
6 & 471 & -31998 & 324752 & -1028553 & 1303219 & -701974 & 134083 \\
&&&&&&&\\
7 & 121 & -16898 & 232948 & -857227 & 1169635 & -648615 & 120036 \\
&&&&&&&\\
8 & 17 & -6230 & 123125 & -530576 & 773886 & -435958 & 75736 \\
&&&&&&&\\
9 & 1 & -1564 & 47840 & -244654 & 378959 & -214440 & 33858 \\
&&&&&&&\\
10 & 0 & -254 & 13479 & -83606 & 136710 & -76977 & 10648 \\
&&&&&&&\\
11 & 0 & -24 & 2675 & -20851 & 35800 & -19901 & 2301 \\
&&&&&&&\\
12 & 0 & -1 & 354 & -3683 & 6608 & -3603 & 325 \\
&&&&&&&\\
13 & 0 & 0 & 28 & -436 & 814 & -433 & 27 \\
&&&&&&&\\
14 & 0 & 0 & 1 & -31 & 60 & -31 & 1 \\
&&&&&&&\\
15 & 0 & 0 & 0 & -1 & 2 & -1 & 0 \\
&&&&&&&\\
\hline
\end{tabular}
\end{tabular}

\bigskip

\noindent
\begin{tabular}{cc}
$N_{[2]}:$ &
\begin{tabular}{|c|ccccccc|}
\hline
&&&&&&& \\
$k \backslash n=$& -2 & 0 & 2 & 4 & 6 & 8 & 10 \\
&&&&&&& \\
\hline
&&&&&&& \\
0 & 29 & -216 & 605 & -808 & 531 & -156 & 15 \\
&&&&&&&\\
1 & 173 & -1685 & 5733 & -8988 & 6875 & -2397 & 289 \\
&&&&&&&\\
2 & 451 & -6003 & 25170 & -46202 & 40551 & -16425 & 2458 \\
&&&&&&&\\
3 & 654 & -12559 & 65901 & -141221 & 139859 & -63619 & 10985 \\
&&&&&&&\\
4 & 570 & -16938 & 113269 & -282887 & 311420 & -154255 & 28821 \\
&&&&&&&\\
5 & 300 & -15385 & 134426 & -391395 & 473113 & -248899 & 47840 \\
&&&&&&&\\
6 & 92 & -9578 & 113396 & -386023 & 507192 & -277756 & 52677 \\
&&&&&&&\\
7 & 15 & -4078 & 68991 & -276387 & 391286 & -219388 & 39561 \\
&&&&&&&\\
8 & 1 & -1162 & 30345 & -144784 & 219153 & -124072 & 20519 \\
&&&&&&&\\
9 & 0 & -211 & 9549 & -55385 & 89010 & -50298 & 7335 \\
&&&&&&&\\
10 & 0 & -22 & 2093 & -15273 & 25898 & -14468 & 1772 \\
&&&&&&&\\
11 & 0 & -1 & 303 & -2952 & 5252 & -2878 & 276 \\
&&&&&&&\\
12 & 0 & 0 & 26 & -379 & 704 & -376 & 25 \\
&&&&&&&\\
13 & 0 & 0 & 1 & -29 & 56 & -29 & 1 \\
&&&&&&&\\
14 & 0 & 0 & 0 & -1 & 2 & -1 & 0 \\
&&&&&&&\\
\hline
\end{tabular}
\end{tabular}

\section*{\fbox{Knot $8_9$}}

{\large $(3,-1|1,-3)$}

\bigskip

\paragraph{HOMFLY polynomials}
\be \frac{H_{[1]}}{\
^*{S_{[1]}}}=(q^{4}-q^{2}+2-q^{-2}+q^{-4})A^{-2}+(-q^{6}+q^{4}-3q^{2}+3-3q^{-2}+q^{-4}-q^{-6})+(q^{4}-q^{2}+2-q^{-2}+q^{-4})A^{2}=
\\
=\frac{1}{\ ^*{S_{[1]}}}\Big(\ ^*{S_{[3]}}+ \
^*{S_{[2,1]}}(q^{8}-2q^{6}+3q^{4}-5q^{2}+5-5q^{-2}+3q^{-4}-2q^{-6}+q^{-8})+
\ ^*{S_{[1,1,1]}}\Big) \ee \be \frac{H_{[1,1]}}{\
^*{S_{[1,1]}}}=(q^{16}-q^{14}+3q^{10}-q^{8}-2q^{6}+4q^{4}-2+2q^{-2}-q^{-6}+q^{-8})A^{-4}+
\\
+(-q^{18}+q^{14}-4q^{12}-q^{10}+5q^{8}-6q^{6}-6q^{4}+8q^{2}-2-7q^{-2}+4q^{-4}-3q^{-8}+q^{-10}-q^{-14})A^{-2}+
\\
+(q^{18}-q^{16}+q^{14}+3q^{12}-3q^{10}+q^{8}+9q^{6}-7q^{4}-3q^{2}+15-
\\
-3q^{-2}-7q^{-4}+9q^{-6}+q^{-8}-3q^{-10}+3q^{-12}+q^{-14}-q^{-16}+q^{-18})+
\\
+(-q^{14}+q^{10}-3q^{8}+4q^{4}-7q^{2}-2+8q^{-2}-6q^{-4}-6q^{-6}+5q^{-8}-q^{-10}-4q^{-12}+q^{-14}-q^{-18})A^{2}+
\\
+(q^{8}-q^{6}+2q^{2}-2+4q^{-4}-2q^{-6}-q^{-8}+3q^{-10}-q^{-14}+q^{-16})A^{4}=
\\
=\frac{1}{\ ^*{S_{[1,1]}}}\Big(\ ^*{S_{[3,3]}}+ \
^*{S_{[3,2,1]}}(q^{8}-2q^{6}+3q^{4}-5q^{2}+5-5q^{-2}+3q^{-4}-2q^{-6}+q^{-8})+
\ ^*{S_{[3,1,1,1]}}+ \ ^*{S_{[2,2,2]}}+
\\
+\
^*{S_{[2,2,1,1]}}(q^{24}-2q^{22}+4q^{18}-6q^{16}+11q^{12}-13q^{10}-4q^{8}+24q^{6}-18q^{4}-13q^{2}+
\\
+32-13q^{-2}-18q^{-4}+24q^{-6}-4q^{-8}-13q^{-10}+11q^{-12}-6q^{-16}+4q^{-18}-2q^{-22}+q^{-24})+
\\
+\
^*{S_{[2,1,1,1,1]}}(q^{16}-2q^{12}+3q^{8}-5q^{4}+5-5q^{-4}+3q^{-8}-2q^{-12}+q^{-16})+
\ ^*{S_{[1,1,1,1,1,1]}}\Big) \ee \be \frac{H_{[2]}}{\
^*{S_{[2]}}}=(q^{8}-q^{6}+2q^{2}-2+4q^{-4}-2q^{-6}-q^{-8}+3q^{-10}-q^{-14}+q^{-16})A^{-4}+
\\
+(-q^{14}+q^{10}-3q^{8}+4q^{4}-7q^{2}-2+8q^{-2}-6q^{-4}-6q^{-6}+5q^{-8}-q^{-10}-4q^{-12}+q^{-14}-q^{-18})A^{-2}+
\\
+(q^{18}-q^{16}+q^{14}+3q^{12}-3q^{10}+q^{8}+9q^{6}-7q^{4}-3q^{2}+15-
\\
-3q^{-2}-7q^{-4}+9q^{-6}+q^{-8}-3q^{-10}+3q^{-12}+q^{-14}-q^{-16}+q^{-18})+
\\
+(-q^{18}+q^{14}-4q^{12}-q^{10}+5q^{8}-6q^{6}-6q^{4}+8q^{2}-2-7q^{-2}+4q^{-4}-3q^{-8}+q^{-10}-q^{-14})A^{2}+
\\
+(q^{16}-q^{14}+3q^{10}-q^{8}-2q^{6}+4q^{4}-2+2q^{-2}-q^{-6}+q^{-8})A^{4}=
\\
=\frac{1}{\ ^*{S_{[2]}}}\Big(\ ^*{S_{[6]}}+ \
^*{S_{[5,1]}}(q^{16}-2q^{12}+3q^{8}-5q^{4}+5-5q^{-4}+3q^{-8}-2q^{-12}+q^{-16})+
\\
+\
^*{S_{[4,2]}}(q^{24}-2q^{22}+4q^{18}-6q^{16}+11q^{12}-13q^{10}-4q^{8}+24q^{6}-18q^{4}-13q^{2}+
\\
+32-13q^{-2}-18q^{-4}+24q^{-6}-4q^{-8}-13q^{-10}+11q^{-12}-6q^{-16}+4q^{-18}-2q^{-22}+q^{-24})+
\\
+\ ^*{S_{[4,1,1]}}+ \ ^*{S_{[3,3]}}+ \
^*{S_{[3,2,1]}}(q^{8}-2q^{6}+3q^{4}-5q^{2}+5-5q^{-2}+3q^{-4}-2q^{-6}+q^{-8})+
\ ^*{S_{[2,2,2]}}\Big) \ee

\paragraph{Alexander polynomials}

\be \mathfrak{A}_{[1]}=-q^{6}+3q^{4}-5q^{2}+7-5q^{-2}+3q^{-4}-q^{-6}
\ee \be
\mathfrak{A}_{[1,1]}=\mathfrak{A}_{[2]}=\mathfrak{A}_{[1]}(q^2)=-q^{12}+3q^{8}-5q^{4}+7-5q^{-4}+3q^{-8}-q^{-12}
\ee

\paragraph{Jones polynomials}

\be
J_{[1]}=q^{8}-2q^{6}+3q^{4}-4q^{2}+5-4q^{-2}+3q^{-4}-2q^{-6}+q^{-8}
\ee \be J_{[1, 1]}=1 \ee \be
J_{[2]}=q^{24}-2q^{22}+5q^{18}-6q^{16}-2q^{14}+12q^{12}-10q^{10}-7q^{8}+20q^{6}-12q^{4}-11q^{2}+
\\
+25-11q^{-2}-12q^{-4}+20q^{-6}-7q^{-8}-10q^{-10}+12q^{-12}-2q^{-14}-6q^{-16}+5q^{-18}-2q^{-22}+q^{-24}
\ee

\paragraph{Special polynomials}

\be \mathfrak{H}_{[1]}=2A^{-2}-3+2A^{2} \ee \be \mathfrak{H}_{[1,
1]}=\mathfrak{H}_{[2]}=(\mathfrak{H}_{[1]})^2=4A^{-4}-12A^{-2}+17-12A^{2}+4A^{4}
\ee

\paragraph{Ooguri-Vafa polynomials}
\be f_{[1, 1]}=\frac{\{A\}^2\{A/q\}\{Aq\}}{\{q\}}
\Big((q^{15}+2q^{9}+q^{7}+q^{3}+2q^{1}-q^{-1})A^{-2}+
\\
+(-q^{11}+q^{7}+q^{5}-3q^{3}+3q^{1}-q^{-1}-q^{-3}+q^{-7})+(q^{5}-2q^{3}-q^{1}-q^{-3}-2q^{-5}-q^{-11})A^{2}\Big)
\ee \be f_{[2]}=\frac{\{A\}^2\{A/q\}\{Aq\}}{\{q\}}
\Big((q^{5}-2q^{3}-q^{1}-q^{-3}-2q^{-5}-q^{-11})A^{-2}+
\\
+(-q^{11}+q^{7}+q^{5}-3q^{3}+3q^{1}-q^{-1}-q^{-3}+q^{-7})+(q^{15}+2q^{9}+q^{7}+q^{3}+2q^{1}-q^{-1})A^{2}\Big)
\ee

\paragraph{Special Ooguri-Vafa polynomials}

\be {\mathfrak f}_{[2]}=-{\mathfrak
f}_{[1,1]}=6(A^2-A^{-2})(A-A^{-1})^3 \ee

\subsubsection*{Numbers $N_{R,n,k}$}

\noindent
\begin{tabular}{cccc}
$N_{[1]}:$ &
\begin{tabular}{|c|cccc|}
\hline
&&&& \\
$k \backslash n=$& -3 & -1 & 1 & 3 \\
&&&& \\
\hline
&&&& \\
0 & -2 & 5 & -5 & 2 \\
&&&&\\
1 & -3 & 11 & -11 & 3 \\
&&&&\\
2 & -1 & 6 & -6 & 1 \\
&&&&\\
3 & 0 & 1 & -1 & 0 \\
&&&&\\
\hline
\end{tabular}
& $N_{[1,1]}:$ &
\begin{tabular}{|c|ccccccc|}
\hline
&&&&&&& \\
$k \backslash n=$& -6 & -4 & -2 & 0 & 2 & 4 & 6 \\
&&&&&&& \\
\hline
&&&&&&& \\
0 & 14 & -60 & 120 & -160 & 150 & -84 & 20 \\
&&&&&&&\\
1 & 59 & -278 & 598 & -846 & 815 & -448 & 100 \\
&&&&&&&\\
2 & 103 & -521 & 1183 & -1803 & 1845 & -1026 & 219 \\
&&&&&&&\\
3 & 94 & -514 & 1221 & -2052 & 2288 & -1298 & 261 \\
&&&&&&&\\
4 & 46 & -288 & 716 & -1377 & 1705 & -979 & 177 \\
&&&&&&&\\
5 & 11 & -91 & 239 & -561 & 781 & -446 & 67 \\
&&&&&&&\\
6 & 1 & -15 & 42 & -136 & 214 & -119 & 13 \\
&&&&&&&\\
7 & 0 & -1 & 3 & -18 & 32 & -17 & 1 \\
&&&&&&&\\
8 & 0 & 0 & 0 & -1 & 2 & -1 & 0 \\
&&&&&&&\\
\hline
\end{tabular}
\end{tabular}

\bigskip

\noindent
\begin{tabular}{cc}
$N_{[2]}:$ &
\begin{tabular}{|c|ccccccc|}
\hline
&&&&&&& \\
$k \backslash n=$& -6 & -4 & -2 & 0 & 2 & 4 & 6 \\
&&&&&&& \\
\hline
&&&&&&& \\
0 & 20 & -84 & 150 & -160 & 120 & -60 & 14 \\
&&&&&&&\\
1 & 100 & -448 & 815 & -846 & 598 & -278 & 59 \\
&&&&&&&\\
2 & 219 & -1026 & 1845 & -1803 & 1183 & -521 & 103 \\
&&&&&&&\\
3 & 261 & -1298 & 2288 & -2052 & 1221 & -514 & 94 \\
&&&&&&&\\
4 & 177 & -979 & 1705 & -1377 & 716 & -288 & 46 \\
&&&&&&&\\
5 & 67 & -446 & 781 & -561 & 239 & -91 & 11 \\
&&&&&&&\\
6 & 13 & -119 & 214 & -136 & 42 & -15 & 1 \\
&&&&&&&\\
7 & 1 & -17 & 32 & -18 & 3 & -1 & 0 \\
&&&&&&&\\
8 & 0 & -1 & 2 & -1 & 0 & 0 & 0 \\
&&&&&&&\\
\hline
\end{tabular}
\end{tabular}

\section*{\fbox{Knot $8_{10}$}}

{\large $(-2,2|-1,3)$}

\bigskip

\paragraph{HOMFLY polynomials}
\be \frac{H_{[1]}}{\
^*{S_{[1]}}}=A^{-2}\Big((-q^{4}+q^{2}-3+q^{-2}-q^{-4})A^{-2}+
\\
+(q^{6}-q^{4}+4q^{2}-2+4q^{-2}-q^{-4}+q^{-6})+(-q^{4}+q^{2}-2+q^{-2}-q^{-4})A^{2}
\Big)=
\\
=\frac{A^{-2}}{\ ^*{S_{[1]}}}\Big(\ ^*{S_{[3]}}q^{2}+ \
^*{S_{[2,1]}}(-q^{8}+2q^{6}-4q^{4}+5q^{2}-5+5q^{-2}-4q^{-4}+2q^{-6}-q^{-8})+
\ ^*{S_{[1,1,1]}}q^{-2}\Big) \ee \be \frac{H_{[1,1]}}{\
^*{S_{[1,1]}}}=A^{-4}q^{8}\Big((q^{10}-q^{8}+q^{6}+3q^{4}-3q^{2}+1+7q^{-2}-3q^{-4}-q^{-6}+4q^{-8}-q^{-12}+q^{-14})A^{-4}+
\\
+(-q^{12}-5q^{6}+q^{2}-13-2q^{-2}+6q^{-4}-13q^{-6}-7q^{-8}+6q^{-10}-3q^{-12}-5q^{-14}+q^{-16}-q^{-20})A^{-2}+
\\
+(q^{12}-q^{10}+2q^{8}+4q^{6}-2q^{4}+6q^{2}+10-6q^{-2}+10q^{-4}+15q^{-6}-
\\
-8q^{-8}+q^{-10}+15q^{-12}-2q^{-14}-3q^{-16}+5q^{-18}+q^{-20}-q^{-22}+q^{-24})+
\\
+(-q^{8}-5q^{2}-9q^{-4}+q^{-6}+3q^{-8}-10q^{-10}-2q^{-12}+6q^{-14}-4q^{-16}-4q^{-18}+2q^{-20}-q^{-24})A^{2}+
\\
+(q^{2}-1+q^{-2}+2q^{-4}-2q^{-6}+q^{-8}+3q^{-10}-3q^{-12}+3q^{-16}-q^{-18}-q^{-20}+q^{-22})A^{4}
\Big)=
\\
=\frac{A^{-4}q^{8}}{\ ^*{S_{[1,1]}}}\Big(\ ^*{S_{[3,3]}}+ \
^*{S_{[3,2,1]}}(-q^{6}+2q^{4}-4q^{2}+5-5q^{-2}+5q^{-4}-4q^{-6}+2q^{-8}-q^{-10})+
\ ^*{S_{[3,1,1,1]}}q^{-4}+
\\
+\ ^*{S_{[2,2,2]}}q^{-4}+\
^*{S_{[2,2,1,1]}}(q^{18}-2q^{16}+q^{14}+4q^{12}-8q^{10}+3q^{8}+10q^{6}-18q^{4}+7q^{2}+17-28q^{-2}+8q^{-4}+
\\
+24q^{-6}-29q^{-8}-q^{-10}+27q^{-12}-18q^{-14}-9q^{-16}+18q^{-18}-5q^{-20}-7q^{-22}+6q^{-24}-2q^{-28}+q^{-30})+
\\
+\
^*{S_{[2,1,1,1,1]}}(-q^{8}+2q^{4}-4+5q^{-4}-5q^{-8}+5q^{-12}-4q^{-16}+2q^{-20}-q^{-24})+
\ ^*{S_{[1,1,1,1,1,1]}}q^{-12}\Big) \ee \be \frac{H_{[2]}}{\
^*{S_{[2]}}}=A^{-4}q^{-8}\Big((q^{14}-q^{12}+4q^{8}-q^{6}-3q^{4}+7q^{2}+1-3q^{-2}+3q^{-4}+q^{-6}-q^{-8}+q^{-10})A^{-4}+
\\
+(-q^{20}+q^{16}-5q^{14}-3q^{12}+6q^{10}-7q^{8}-13q^{6}+6q^{4}-2q^{2}-13+q^{-2}-5q^{-6}-q^{-12})A^{-2}+
\\
+(q^{24}-q^{22}+q^{20}+5q^{18}-3q^{16}-2q^{14}+15q^{12}+q^{10}-8q^{8}+
\\
+15q^{6}+10q^{4}-6q^{2}+10+6q^{-2}-2q^{-4}+4q^{-6}+2q^{-8}-q^{-10}+q^{-12})+
\\
+(-q^{24}+2q^{20}-4q^{18}-4q^{16}+6q^{14}-2q^{12}-10q^{10}+3q^{8}+q^{6}-9q^{4}-5q^{-2}-q^{-8})A^{2}+
\\
+(q^{22}-q^{20}-q^{18}+3q^{16}-3q^{12}+3q^{10}+q^{8}-2q^{6}+2q^{4}+q^{2}-1+q^{-2})A^{4}
\Big)=
\\
=\frac{A^{-4}q^{-8}}{\ ^*{S_{[2]}}}\Big(\ ^*{S_{[6]}}q^{12}+ \
^*{S_{[5,1]}}(-q^{24}+2q^{20}-4q^{16}+5q^{12}-5q^{8}+5q^{4}-4+2q^{-4}-q^{-8})+
\\
+\
^*{S_{[4,2]}}(q^{30}-2q^{28}+6q^{24}-7q^{22}-5q^{20}+18q^{18}-9q^{16}-18q^{14}+27q^{12}-q^{10}-29q^{8}+
\\
+24q^{6}+8q^{4}-28q^{2}+17+7q^{-2}-18q^{-4}+10q^{-6}+3q^{-8}-8q^{-10}+4q^{-12}+q^{-14}-2q^{-16}+q^{-18})+
\\
+\ ^*{S_{[4,1,1]}}q^{4}+ \ ^*{S_{[3,3]}}q^{4}+ \
^*{S_{[3,2,1]}}(-q^{10}+2q^{8}-4q^{6}+5q^{4}-5q^{2}+5-4q^{-2}+2q^{-4}-q^{-6})+
\ ^*{S_{[2,2,2]}}\Big) \ee

\paragraph{Alexander polynomials}

\be \mathfrak{A}_{[1]}=q^{6}-3q^{4}+6q^{2}-7+6q^{-2}-3q^{-4}+q^{-6}
\ee \be \mathfrak{A}_{[1,
1]}=\mathfrak{A}_{[2]}=\mathfrak{A}_{[1]}(q^2)=q^{12}-3q^{8}+6q^{4}-7+6q^{-4}-3q^{-8}+q^{-12}
\ee

\paragraph{Jones polynomials}

\be
J_{[1]}=-q^{4}+2q^{2}-3+5q^{-2}-4q^{-4}+5q^{-6}-4q^{-8}+2q^{-10}-q^{-12}
\ee \be J_{[1, 1]}=1 \ee \be
J_{[2]}=q^{14}-2q^{12}-q^{10}+6q^{8}-5q^{6}-6q^{4}+14q^{2}-5-14q^{-2}+21q^{-4}-2q^{-6}-21q^{-8}+23q^{-10}+
\ee \be
+2q^{-12}-24q^{-14}+20q^{-16}+3q^{-18}-19q^{-20}+12q^{-22}+3q^{-24}-9q^{-26}+4q^{-28}+q^{-30}-2q^{-32}+q^{-34}
\ee

\paragraph{Special polynomials}

\be \mathfrak{H}_{[1]}=-3A^{-4}+6A^{-2}-2 \ee \be \mathfrak{H}_{[1,
1]}=\mathfrak{H}_{[2]}=(\mathfrak{H}_{[1]})^2=9A^{-8}-36A^{-6}+48A^{-4}-24A^{-2}+4
\ee

\paragraph{Ooguri-Vafa polynomials}

\be f_{[1, 1]}=\frac{\{A\}^2\{A/q\}\{Aq\}}{\{q\}}
\Big((q^{17}-q^{15}+2q^{13}+2q^{11}-q^{9}+3q^{7}+6q^{5}-3q^{3}+6q^{1}+2q^{-3}-q^{-5}+q^{-7})A^{-6}+
\\
+(-q^{13}-q^{9}-5q^{7}+3q^{5}-5q^{3}-4q^{-1}+5q^{-3}-4q^{-5}+q^{-7}+q^{-9})A^{-4}+
\\
+(2q^{1}-4q^{-1}+3q^{-3}-2q^{-7}+q^{-11}-q^{-13})A^{-2}\Big) \ee \be
f_{[2]}=\frac{\{A\}^2\{A/q\}\{Aq\}}{\{q\}}
\Big((-q^{7}+q^{5}-2q^{3}-6q^{-1}+3q^{-3}-6q^{-5}-3q^{-7}+q^{-9}-2q^{-11}-2q^{-13}+q^{-15}-q^{-17})A^{-6}+
\\
+(-q^{9}-q^{7}+4q^{5}-5q^{3}+4q^{1}+5q^{-3}-3q^{-5}+5q^{-7}+q^{-9}+q^{-13})A^{-4}+(q^{13}-q^{11}+2q^{7}-3q^{3}+4q^{1}-2q^{-1})A^{-2}\Big)
\ee

\paragraph{Special Ooguri-Vafa polynomials}

\be {\mathfrak f}_{[2]}=-{\mathfrak
f}_{[1,1]}=\frac{6(A^4+10A^2-17)(A-A^{-1})^3}{A^6} \ee

\subsubsection*{Numbers $N_{R,n,k}$}

\noindent
\begin{tabular}{cccc}
$N_{[1]}:$ &
\begin{tabular}{|c|cccc|}
\hline
&&&& \\
$k \backslash n=$& -5 & -3 & -1 & 1 \\
&&&& \\
\hline
&&&& \\
0 & 3 & -9 & 8 & -2 \\
&&&&\\
1 & 3 & -12 & 12 & -3 \\
&&&&\\
2 & 1 & -6 & 6 & -1 \\
&&&&\\
3 & 0 & -1 & 1 & 0 \\
&&&&\\
\hline
\end{tabular}
& $N_{[1,1]}:$ &
\begin{tabular}{|c|ccccccc|}
\hline
&&&&&&& \\
$k \backslash n=$& -10 & -8 & -6 & -4 & -2 & 0 & 2 \\
&&&&&&& \\
\hline
&&&&&&& \\
0 & 32 & -155 & 307 & -318 & 182 & -55 & 7 \\
&&&&&&&\\
1 & 138 & -665 & 1281 & -1283 & 731 & -240 & 38 \\
&&&&&&&\\
2 & 272 & -1316 & 2431 & -2294 & 1281 & -456 & 82 \\
&&&&&&&\\
3 & 296 & -1502 & 2680 & -2338 & 1250 & -472 & 86 \\
&&&&&&&\\
4 & 187 & -1055 & 1849 & -1468 & 719 & -277 & 45 \\
&&&&&&&\\
5 & 68 & -460 & 808 & -576 & 239 & -90 & 11 \\
&&&&&&&\\
6 & 13 & -120 & 216 & -137 & 42 & -15 & 1 \\
&&&&&&&\\
7 & 1 & -17 & 32 & -18 & 3 & -1 & 0 \\
&&&&&&&\\
8 & 0 & -1 & 2 & -1 & 0 & 0 & 0 \\
&&&&&&&\\
\hline
\end{tabular}
\end{tabular}

\bigskip

\noindent
\begin{tabular}{cc}
$N_{[2]}:$ &
\begin{tabular}{|c|ccccccc|}
\hline
&&&&&&& \\
$k \backslash n=$& -10 & -8 & -6 & -4 & -2 & 0 & 2 \\
&&&&&&& \\
\hline
&&&&&&& \\
0 & 49 & -233 & 448 & -442 & 233 & -61 & 6 \\
&&&&&&&\\
1 & 246 & -1162 & 2150 & -1979 & 949 & -227 & 23 \\
&&&&&&&\\
2 & 576 & -2731 & 4838 & -4065 & 1697 & -352 & 37 \\
&&&&&&&\\
3 & 769 & -3786 & 6511 & -4967 & 1728 & -283 & 28 \\
&&&&&&&\\
4 & 624 & -3321 & 5649 & -3933 & 1092 & -120 & 9 \\
&&&&&&&\\
5 & 312 & -1884 & 3225 & -2069 & 440 & -25 & 1 \\
&&&&&&&\\
6 & 93 & -685 & 1198 & -715 & 111 & -2 & 0 \\
&&&&&&&\\
7 & 15 & -153 & 277 & -155 & 16 & 0 & 0 \\
&&&&&&&\\
8 & 1 & -19 & 36 & -19 & 1 & 0 & 0 \\
&&&&&&&\\
9 & 0 & -1 & 2 & -1 & 0 & 0 & 0 \\
&&&&&&&\\
\hline
\end{tabular}
\end{tabular}

\bigskip

\section*{\fbox{Knot $8_{16}$}}

{\large $(1,-1|1,-2|1,-2)$}

\bigskip

\paragraph{HOMFLY polynomials}
\be \frac{H_{[1]}}{\
^*{S_{[1]}}}=A^2\Big((-q^{4}+2q^{2}-2+2q^{-2}-q^{-4})A^{-2}+
\\
+(q^{6}-2q^{4}+4q^{2}-4+4q^{-2}-2q^{-4}+q^{-6})+(-q^{4}+2q^{2}-3+2q^{-2}-q^{-4})A^{2}
\Big)=
\\
=\frac{A^2}{\ ^*{S_{[1]}}}\Big(\ ^*{S_{[3]}}q^{-2}+ \
^*{S_{[2,1]}}(-q^{8}+3q^{6}-5q^{4}+6q^{2}-7+6q^{-2}-5q^{-4}+3q^{-6}-q^{-8})+
\ ^*{S_{[1,1,1]}}q^{2}\Big) \ee \be \frac{H_{[1,1]}}{\
^*{S_{[1,1]}}}=A^4q^{-8}\Big((q^{22}-2q^{20}-2q^{18}+5q^{16}-q^{14}-6q^{12}+5q^{10}+2q^{8}-5q^{6}+3q^{4}+q^{2}-2+q^{-2})A^{-4}+
\\
+(-q^{24}+q^{22}+4q^{20}-5q^{18}-4q^{16}+13q^{14}-q^{12}-14q^{10}+
\\
+10q^{8}+6q^{6}-12q^{4}+4q^{2}+4-6q^{-2}+q^{-4}+q^{-6}-q^{-8})A^{-2}+
\\
+(q^{24}-2q^{22}+6q^{18}-8q^{16}-7q^{14}+19q^{12}-4q^{10}-21q^{8}+
\\
+18q^{6}+8q^{4}-18q^{2}+9+6q^{-2}-8q^{-4}+4q^{-6}+2q^{-8}-2q^{-10}+q^{-12})+
\\
+(-q^{20}+q^{18}+3q^{16}-6q^{14}-2q^{12}+14q^{10}-6q^{8}-16q^{6}+
\\
+16q^{4}+4q^{2}-16+6q^{-2}+4q^{-4}-6q^{-6}+q^{-8}+q^{-10}-q^{-12})A^{2}+
\\
+(q^{14}-2q^{12}-q^{10}+6q^{8}-3q^{6}-7q^{4}+9q^{2}+1-7q^{-2}+4q^{-4}+q^{-6}-2q^{-8}+q^{-10})A^{4}
\Big)=
\\
=\frac{A^4q^{-8}}{\ ^*{S_{[1,1]}}}\Big(\ ^*{S_{[3,3]}}+ \
^*{S_{[3,2,1]}}(-q^{10}+3q^{8}-5q^{6}+6q^{4}-7q^{2}+6-5q^{-2}+3q^{-4}-q^{-6})+
\ ^*{S_{[3,1,1,1]}}q^{4}+ \ ^*{S_{[2,2,2]}}q^{4}+
\\
+\
^*{S_{[2,2,1,1]}}(q^{30}-3q^{28}+10q^{24}-11q^{22}-9q^{20}+29q^{18}-13q^{16}-30q^{14}+43q^{12}-q^{10}-47q^{8}+
\\
+40q^{6}+12q^{4}-46q^{2}+28+13q^{-2}-31q^{-4}+16q^{-6}+7q^{-8}-14q^{-10}+6q^{-12}+2q^{-14}-3q^{-16}+q^{-18})+
\\
+\
^*{S_{[2,1,1,1,1]}}(-q^{24}+3q^{20}-5q^{16}+6q^{12}-7q^{8}+6q^{4}-5+3q^{-4}-q^{-8})+
\ ^*{S_{[1,1,1,1,1,1]}}q^{12}\Big) \ee \be \frac{H_{[2]}}{\
^*{S_{[2]}}}=A^4q^{8}\Big((q^{2}-2+q^{-2}+3q^{-4}-5q^{-6}+2q^{-8}+5q^{-10}-6q^{-12}-q^{-14}+5q^{-16}-2q^{-18}-2q^{-20}+q^{-22})A^{-4}+
\\
+(-q^{8}+q^{6}+q^{4}-6q^{2}+4+4q^{-2}-12q^{-4}+6q^{-6}+10q^{-8}-14q^{-10}-
\\
-q^{-12}+13q^{-14}-4q^{-16}-5q^{-18}+4q^{-20}+q^{-22}-q^{-24})A^{-2}+
\\
+(q^{12}-2q^{10}+2q^{8}+4q^{6}-8q^{4}+6q^{2}+9-18q^{-2}+8q^{-4}+18q^{-6}-
\\
-21q^{-8}-4q^{-10}+19q^{-12}-7q^{-14}-8q^{-16}+6q^{-18}-2q^{-22}+q^{-24})+
\\
+(-q^{12}+q^{10}+q^{8}-6q^{6}+4q^{4}+6q^{2}-16+4q^{-2}+16q^{-4}-16q^{-6}-
\\
-6q^{-8}+14q^{-10}-2q^{-12}-6q^{-14}+3q^{-16}+q^{-18}-q^{-20})A^{2}+
\\
+(q^{10}-2q^{8}+q^{6}+4q^{4}-7q^{2}+1+9q^{-2}-7q^{-4}-3q^{-6}+6q^{-8}-q^{-10}-2q^{-12}+q^{-14})A^{4}
\Big)=
\\
=\frac{A^4q^{8}}{\ ^*{S_{[2]}}}\Big(\ ^*{S_{[6]}}q^{-12}+ \
^*{S_{[5,1]}}(-q^{8}+3q^{4}-5+6q^{-4}-7q^{-8}+6q^{-12}-5q^{-16}+3q^{-20}-q^{-24})+
\\
+\
^*{S_{[4,2]}}(q^{18}-3q^{16}+2q^{14}+6q^{12}-14q^{10}+7q^{8}+16q^{6}-31q^{4}+13q^{2}+28-46q^{-2}+12q^{-4}+
\\
+40q^{-6}-47q^{-8}-q^{-10}+43q^{-12}-30q^{-14}-13q^{-16}+29q^{-18}-9q^{-20}-11q^{-22}+10q^{-24}-3q^{-28}+q^{-30})+
\\
+\ ^*{S_{[4,1,1]}}q^{-4}+ \ ^*{S_{[3,3]}}q^{-4}+ \
^*{S_{[3,2,1]}}(-q^{6}+3q^{4}-5q^{2}+6-7q^{-2}+6q^{-4}-5q^{-6}+3q^{-8}-q^{-10})+
\ ^*{S_{[2,2,2]}}\Big) \ee

\paragraph{Alexander polynomials}

\be \mathfrak{A}_{[1]}=q^{6}-4q^{4}+8q^{2}-9+8q^{-2}-4q^{-4}+q^{-6}
\ee \be
\mathfrak{A}_{[1,1]}=\mathfrak{A}_{[2]}=\mathfrak{A}_{[1]}(q^2)=q^{12}-4q^{8}+8q^{4}-9+8q^{-4}-4q^{-8}+q^{-12}
\ee

\paragraph{Jones polynomials}

\be
J_{[1]}=-q^{12}+3q^{10}-5q^{8}+6q^{6}-6q^{4}+6q^{2}-4+3q^{-2}-q^{-4}
\ee \be J_{[1, 1]}=1 \ee \be
J_{[2]}=q^{34}-3q^{32}+2q^{30}+6q^{28}-15q^{26}+7q^{24}+19q^{22}-32q^{20}+8q^{18}+32q^{16}-41q^{14}+4q^{12}`+
\\
+38q^{10}-37q^{8}-3q^{6}+35q^{4}-25q^{2}-8+24q^{-2}-10q^{-4}-8q^{-6}+10q^{-8}-q^{-10}-3q^{-12}+q^{-14}
\ee

\paragraph{Special polynomials}

\be \mathfrak{H}_{[1]}=2A^{2}-A^{4} \ee \be \mathfrak{H}_{[1,
1]}=\mathfrak{H}_{[2]}=(\mathfrak{H}_{[1]})^2=4A^{4}-4A^{6}+A^{8}
\ee

\paragraph{Ooguri-Vafa polynomials}

\be f_{[1, 1]}=\frac{\{A\}^2\{A/q\}\{Aq\}(q^2-1+q^{-2})}{\{q\}}
\Big((q^{13}-q^{11}-3q^{9}+q^{7}+3q^{5}-4q^{3})A^{2}+
\\
+(-q^{9}-2q^{7}+6q^{5}-q^{3}-q^{1}-2q^{-1}+5q^{-3}-q^{-5}+q^{-9})A^4+
\\
+(-q^{7}+q^{5}+q^{1}-5q^{-1}+2q^{-3}+2q^{-5}-3q^{-7}+q^{-11}-q^{-13})A^{6}\Big)
\ee \be f_{[2]}=\frac{\{A\}^2\{A/q\}\{Aq\}(q^2-1+q^{-2})}{\{q\}}
\Big((4q^{1}-3q^{-1}-q^{-3}+3q^{-5}+q^{-7}-q^{-9})A^{2}+
\\
+(-q^{13}+q^{9}-5q^{7}+2q^{5}+q^{3}+q^{1}-6q^{-1}+2q^{-3}+q^{-5})A^4+(q^{17}-q^{15}+3q^{11}-2q^{9}-2q^{7}+5q^{5}-q^{3}-q^{-1}+q^{-3})A^{6}\Big)
\ee

\paragraph{Special Ooguri-Vafa polynomials}

\be` {\mathfrak f}_{[2]}=-{\mathfrak
f}_{[1,1]}=A^4(3A^2-4+3A^{-2})(A-A^{-1})^3 \ee

\subsubsection*{Numbers $N_{R,n,k}$}

\noindent
\begin{tabular}{cccc}
$N_{[1]}:$ &
\begin{tabular}{|c|cccc|}
\hline
&&&& \\
$k \backslash n=$& -1 & 1 & 3 & 5 \\
&&&& \\
\hline
&&&& \\
0 & 0 & -2 & 3 & -1 \\
&&&&\\
1 & 2 & -7 & 7 & -2 \\
&&&&\\
2 & 1 & -5 & 5 & -1 \\
&&&&\\
3 & 0 & -1 & 1 & 0 \\
&&&&\\
\hline
\end{tabular}
& $N_{[1,1]}:$ &
\begin{tabular}{|c|ccccccc|}
\hline
&&&&&&& \\
$k \backslash n=$& -2 & 0 & 2 & 4 & 6 & 8 & 10 \\
&&&&&&& \\
\hline
&&&&&&& \\
0 & -3 & -1 & 44 & -106 & 109 & -53 & 10 \\
&&&&&&&\\
1 & -4 & -43 & 305 & -685 & 721 & -368 & 74 \\
&&&&&&&\\
2 & 11 & -145 & 779 & -1824 & 2075 & -1135 & 239 \\
&&&&&&&\\
3 & 19 & -178 & 1012 & -2710 & 3408 & -1958 & 407 \\
&&&&&&&\\
4 & 8 & -96 & 757 & -2510 & 3493 & -2052 & 400 \\
&&&&&&&\\
5 & 1 & -23 & 345 & -1504 & 2294 & -1347 & 234 \\
&&&&&&&\\
6 & 0 & -2 & 96 & -579 & 957 & -551 & 79 \\
&&&&&&&\\
7 & 0 & 0 & 15 & -137 & 243 & -135 & 14 \\
&&&&&&&\\
8 & 0 & 0 & 1 & -18 & 34 & -18 & 1 \\
&&&&&&&\\
9 & 0 & 0 & 0 & -1 & 2 & -1 & 0 \\
&&&&&&&\\
\hline
\end{tabular}
\end{tabular}

\bigskip

\noindent
\begin{tabular}{cc}
$N_{[2]}:$ &
\begin{tabular}{|c|ccccccc|}
\hline
&&&&&&& \\
$k \backslash n=$& -2 & 0 & 2 & 4 & 6 & 8 & 10 \\
&&&&&&& \\
\hline
&&&&&&& \\
0 & -6 & 15 & 7 & -58 & 72 & -37 & 7 \\
&&&&&&&\\
1 & -14 & 20 & 126 & -395 & 447 & -229 & 45 \\
&&&&&&&\\
2 & 10 & -87 & 441 & -1013 & 1124 & -598 & 123 \\
&&&&&&&\\
3 & 43 & -219 & 659 & -1337 & 1530 & -848 & 172 \\
&&&&&&&\\
4 & 34 & -189 & 499 & -1013 & 1247 & -710 & 132 \\
&&&&&&&\\
5 & 10 & -75 & 198 & -456 & 624 & -357 & 56 \\
&&&&&&&\\
6 & 1 & -14 & 39 & -120 & 186 & -104 & 12 \\
&&&&&&&\\
7 & 0 & -1 & 3 & -17 & 30 & -16 & 1 \\
&&&&&&&\\
8 & 0 & 0 & 0 & -1 & 2 & -1 & 0 \\
&&&&&&&\\
\hline
\end{tabular}
\end{tabular}

\bigskip

\section*{\fbox{Knot $8_{17}$}}

{\large $(2,-1|1,-1|1,-2)$}

\bigskip

\paragraph{HOMFLY polynomials}
\be \frac{H_{[1]}}{\
^*{S_{[1]}}}=(q^{4}-2q^{2}+3-2q^{-2}+q^{-4})A^{-2}+
\\
+(-q^{6}+2q^{4}-4q^{2}+5-4q^{-2}+2q^{-4}-q^{-6})+(q^{4}-2q^{2}+3-2q^{-2}+q^{-4})A^{2}=
\\
=\frac{1}{\ ^*{S_{[1]}}}\Big(\ ^*{S_{[3]}}+ \
^*{S_{[2,1]}}(q^{8}-3q^{6}+5q^{4}-7q^{2}+7-7q^{-2}+5q^{-4}-3q^{-6}+q^{-8})+
\ ^*{S_{[1,1,1]}}\Big) \ee \be \frac{H_{[1,1]}}{\
^*{S_{[1,1]}}}=(q^{16}-2q^{14}+6q^{10}-5q^{8}-5q^{6}+10q^{4}-2q^{2}-6+5q^{-2}-2q^{-6}+q^{-8})A^{-4}+
\\
+(-q^{18}+q^{16}+2q^{14}-7q^{12}+q^{10}+13q^{8}-13q^{6}-10q^{4}+22q^{2}-
\\
-4-15q^{-2}+11q^{-4}+2q^{-6}-6q^{-8}+2q^{-10}+q^{-12}-q^{-14})A^{-2}+
\\
+(q^{18}-2q^{16}+q^{14}+5q^{12}-9q^{10}+q^{8}+18q^{6}-19q^{4}-10q^{2}+31-
\\
-10q^{-2}-19q^{-4}+18q^{-6}+q^{-8}-9q^{-10}+5q^{-12}+q^{-14}-2q^{-16}+q^{-18})+
\\
+(-q^{14}+q^{12}+2q^{10}-6q^{8}+2q^{6}+11q^{4}-15q^{2}-4+22q^{-2}-
\\
-10q^{-4}-13q^{-6}+13q^{-8}+q^{-10}-7q^{-12}+2q^{-14}+q^{-16}-q^{-18})A^{2}+
\\
+(q^{8}-2q^{6}+5q^{2}-6-2q^{-2}+10q^{-4}-5q^{-6}-5q^{-8}+6q^{-10}-2q^{-14}+q^{-16})A^{4}=
\\
=\frac{1}{\ ^*{S_{[1,1]}}}\Big(\ ^*{S_{[3,3]}}+ \
^*{S_{[3,2,1]}}(q^{8}-3q^{6}+5q^{4}-7q^{2}+7-7q^{-2}+5q^{-4}-3q^{-6}+q^{-8})+
\ ^*{S_{[3,1,1,1]}}+ \ ^*{S_{[2,2,2]}}+
\\
+\
^*{S_{[2,2,1,1]}}(q^{24}-3q^{22}+q^{20}+8q^{18}-14q^{16}+27q^{12}-30q^{10}-10q^{8}+53q^{6}-38q^{4}-28q^{2}+
\\
+66-28q^{-2}-38q^{-4}+53q^{-6}-10q^{-8}-30q^{-10}+27q^{-12}-14q^{-16}+8q^{-18}+q^{-20}-3q^{-22}+q^{-24})+
\\
+\
^*{S_{[2,1,1,1,1]}}(q^{16}-3q^{12}+5q^{8}-7q^{4}+7-7q^{-4}+5q^{-8}-3q^{-12}+q^{-16})+
\ ^*{S_{[1,1,1,1,1,1]}}\Big) \ee \be \frac{H_{[2]}}{\
^*{S_{[2]}}}=(q^{8}-2q^{6}+5q^{2}-6-2q^{-2}+10q^{-4}-5q^{-6}-5q^{-8}+6q^{-10}-2q^{-14}+q^{-16})A^{-4}+
\\
+(-q^{14}+q^{12}+2q^{10}-6q^{8}+2q^{6}+11q^{4}-15q^{2}-4+22q^{-2}-
\\
-10q^{-4}-13q^{-6}+13q^{-8}+q^{-10}-7q^{-12}+2q^{-14}+q^{-16}-q^{-18})A^{-2}+
\\
+(q^{18}-2q^{16}+q^{14}+5q^{12}-9q^{10}+q^{8}+18q^{6}-19q^{4}-10q^{2}+31-
\\
-10q^{-2}-19q^{-4}+18q^{-6}+q^{-8}-9q^{-10}+5q^{-12}+q^{-14}-2q^{-16}+q^{-18})+
\\
+(-q^{18}+q^{16}+2q^{14}-7q^{12}+q^{10}+13q^{8}-13q^{6}-10q^{4}+22q^{2}-
\\
-4-15q^{-2}+11q^{-4}+2q^{-6}-6q^{-8}+2q^{-10}+q^{-12}-q^{-14})A^{2}+
\\
+(q^{16}-2q^{14}+6q^{10}-5q^{8}-5q^{6}+10q^{4}-2q^{2}-6+5q^{-2}-2q^{-6}+q^{-8})A^{4}=
\\
=\frac{1}{\ ^*{S_{[2]}}}\Big(\ ^*{S_{[6]}}+ \
^*{S_{[5,1]}}(q^{16}-3q^{12}+5q^{8}-7q^{4}+7-7q^{-4}+5q^{-8}-3q^{-12}+q^{-16})+
\\
+\
^*{S_{[4,2]}}(q^{24}-3q^{22}+q^{20}+8q^{18}-14q^{16}+27q^{12}-30q^{10}-10q^{8}+53q^{6}-38q^{4}-28q^{2}+
\\
+66-28q^{-2}-38q^{-4}+53q^{-6}-10q^{-8}-30q^{-10}+27q^{-12}-14q^{-16}+8q^{-18}+q^{-20}-3q^{-22}+q^{-24})+
\\
+\ ^*{S_{[4,1,1]}}+ \ ^*{S_{[3,3]}}+ \
^*{S_{[3,2,1]}}(q^{8}-3q^{6}+5q^{4}-7q^{2}+7-7q^{-2}+5q^{-4}-3q^{-6}+q^{-8})+
\ ^*{S_{[2,2,2]}}\Big) \ee

\paragraph{Alexander polynomials}

\be
\mathfrak{A}_{[1]}=-q^{6}+4q^{4}-8q^{2}+11-8q^{-2}+4q^{-4}-q^{-6}
\ee \be
\mathfrak{A}_{[1,1]}=\mathfrak{A}_{[2]}=\mathfrak{A}_{[1]}(q^2)=-q^{12}+4q^{8}-8q^{4}+11-8q^{-4}+4q^{-8}-q^{-12}
\ee

\paragraph{Jones polynomials}

\be
J_{[1]}=q^{8}-3q^{6}+5q^{4}-6q^{2}+7-6q^{-2}+5q^{-4}-3q^{-6}+q^{-8}
\ee \be J_{[1, 1]}=1 \ee \be
J_{[2]}=q^{24}-3q^{22}+q^{20}+9q^{18}-14q^{16}-3q^{14}+28q^{12}-25q^{10}-14q^{8}+47q^{6}-29q^{4}-25q^{2}+
\\
+55-25q^{-2}-29q^{-4}+47q^{-6}-14q^{-8}-25q^{-10}+28q^{-12}-3q^{-14}-14q^{-16}+9q^{-18}+q^{-20}-3q^{-22}+q^{-24}
\ee

\paragraph{Special polynomials}

\be \mathfrak{H}_{[1]}=A^{-2}-1+A^{2} \ee \be
\mathfrak{H}_{[1,1]}=\mathfrak{H}_{[2]}=(\mathfrak{H}_{[1]})^2=A^{-4}-2A^{-2}+3-2A^{2}+A^{4}
\ee

\paragraph{Ooguri-Vafa polynomials}

\be f_{[1, 1]}=\frac{\{A\}^2\{A/q\}\{Aq\}(q^2-1+q^{-2})}{\{q\}}
\Big((q^{15}-q^{13}-q^{11}+4q^{9}-3q^{5}+4q^{1}-2q^{-1})A^{-2}+
\\
+(-q^{11}+q^{9}+2q^{7}-8q^{3}+8q^{1}-2q^{-3}-q^{-5}+q^{-7})+(2q^{5}-4q^{3}+3q^{-1}-4q^{-5}+q^{-7}+q^{-9}-q^{-11})A^{2}\Big)
\ee \be f_{[2]}=\frac{\{A\}^2\{A/q\}\{Aq\}(q^2-1+q^{-2})}{\{q\}}
\Big((2q^{5}-4q^{3}+3q^{-1}-4q^{-5}+q^{-7}+q^{-9}-q^{-11})A^{-2}+
\\
+(-q^{11}+q^{9}+2q^{7}-8q^{3}+8q^{1}-2q^{-3}-q^{-5}+q^{-7})+(q^{15}-q^{13}-q^{11}+4q^{9}-3q^{5}+4q^{1}-2q^{-1})A^{2}\Big)
\ee

\paragraph{Special Ooguri-Vafa polynomials}

\be {\mathfrak f}_{[2]}=-{\mathfrak
f}_{[1,1]}=2(A^2-A^{-2})(A-A^{-1})^3 \ee

\subsubsection*{Numbers $N_{R,n,k}$}

\noindent
\begin{tabular}{cc}
$N_{[1]}:$ &
\begin{tabular}{|c|cccc|}
\hline
&&&& \\
$k \backslash n=$&-3 & -1 & 1 & 3 \\
&&&& \\
\hline
&&&& \\
0 & -1 & 2 & -2 & 1 \\
&&&&\\
1 & -2 & 7 & -7 & 2 \\
&&&&\\
2 & -1 & 5 & -5 & 1 \\
&&&&\\
3 & 0 & 1 & -1 & 0 \\
&&&&\\
\hline
\end{tabular}
\end{tabular}

\bigskip

\noindent \hspace{-1cm}\begin{tabular}{cccc} $N_{[1,1]}:$ &
\begin{tabular}{|c|ccccccc|}
\hline
&&&&&&& \\
$k \backslash n=$& -6 & -4 & -2 & 0 & 2 & 4 & 6 \\
&&&&&&& \\
\hline
&&&&&&& \\
0 & 6 & -24 & 44 & -56 & 54 & -32 & 8 \\
&&&&&&&\\
1 & 29 & -128 & 256 & -350 & 345 & -198 & 46 \\
&&&&&&&\\
2 & 56 & -277 & 607 & -899 & 916 & -514 & 111 \\
&&&&&&&\\
3 & 60 & -322 & 753 & -1223 & 1328 & -749 & 153 \\
&&&&&&&\\
4 & 36 & -213 & 522 & -959 & 1148 & -658 & 124 \\
&&&&&&&\\
5 & 10 & -77 & 200 & -444 & 601 & -345 & 55 \\
&&&&&&&\\
6 & 1 & -14 & 39 & -119 & 184 & -103 & 12 \\
&&&&&&&\\
7 & 0 & -1 & 3 & -17 & 30 & -16 & 1 \\
&&&&&&&\\
8 & 0 & 0 & 0 & -1 & 2 & -1 & 0 \\
&&&&&&&\\
\hline
\end{tabular}
& $N_{[2]}:$ &
\begin{tabular}{|c|ccccccc|}
\hline
&&&&&&& \\
$k \backslash n=$& -6 & -4 & -2 & 0 & 2 & 4 & 6 \\
&&&&&&& \\
\hline
&&&&&&& \\
0 & 8 & -32 & 54 & -56 & 44 & -24 & 6 \\
&&&&&&&\\
1 & 46 & -198 & 345 & -350 & 256 & -128 & 29 \\
&&&&&&&\\
2 & 111 & -514 & 916 & -899 & 607 & -277 & 56 \\
&&&&&&&\\
3 & 153 & -749 & 1328 & -1223 & 753 & -322 & 60 \\
&&&&&&&\\
4 & 124 & -658 & 1148 & -959 & 522 & -213 & 36 \\
&&&&&&&\\
5 & 55 & -345 & 601 & -444 & 200 & -77 & 10 \\
&&&&&&&\\
6 & 12 & -103 & 184 & -119 & 39 & -14 & 1 \\
&&&&&&&\\
7 & 1 & -16 & 30 & -17 & 3 & -1 & 0 \\
&&&&&&&\\
8 & 0 & -1 & 2 & -1 & 0 & 0 & 0 \\
&&&&&&&\\
\hline
\end{tabular}
\end{tabular}

\bigskip

\section*{\fbox{Knot $8_{18}$}}

{\large $(1,-1|1,-1|1,-1|1,-1)$}

\bigskip

\paragraph{HOMFLY polynomials}
\be \frac{H_{[1]}}{\
^*{S_{[1]}}}=(q^{4}-3q^{2}+3-3q^{-2}+q^{-4})A^{-2}+
\\
+(-q^{6}+3q^{4}-4q^{2}+7-4q^{-2}+3q^{-4}-q^{-6})+(q^{4}-3q^{2}+3-3q^{-2}+q^{-4})A^{2}=
\\
=\frac{1}{\ ^*{S_{[1]}}}\Big(\ ^*{S_{[3]}}+ \
^*{S_{[2,1]}}(q^{8}-4q^{6}+6q^{4}-8q^{2}+9-8q^{-2}+6q^{-4}-4q^{-6}+q^{-8})+
\ ^*{S_{[1,1,1]}}\Big) \ee \be \frac{H_{[1,1]}}{\
^*{S_{[1,1]}}}=(q^{16}-3q^{14}+9q^{10}-8q^{8}-7q^{6}+15q^{4}-3q^{2}-9+8q^{-2}-3q^{-6}+q^{-8})A^{-4}+
\\
+(-q^{18}+2q^{16}+3q^{14}-10q^{12}+2q^{10}+19q^{8}-20q^{6}-15q^{4}+31q^{2}-
\\
-8-23q^{-2}+16q^{-4}+3q^{-6}-9q^{-8}+3q^{-10}+2q^{-12}-q^{-14})A^{-2}+
\\
+(q^{18}-3q^{16}+q^{14}+7q^{12}-14q^{10}+2q^{8}+27q^{6}-26q^{4}-13q^{2}+
\\
+47-13q^{-2}-26q^{-4}+27q^{-6}+2q^{-8}-14q^{-10}+7q^{-12}+q^{-14}-3q^{-16}+q^{-18})+
\\
+(-q^{14}+2q^{12}+3q^{10}-9q^{8}+3q^{6}+16q^{4}-23q^{2}-8+31q^{-2}-15q^{-4}-
\\
-20q^{-6}+19q^{-8}+2q^{-10}-10q^{-12}+3q^{-14}+2q^{-16}-q^{-18})A^{2}+
\\
+(q^{8}-3q^{6}+8q^{2}-9-3q^{-2}+15q^{-4}-7q^{-6}-8q^{-8}+9q^{-10}-3q^{-14}+q^{-16})A^{4}=
\\
=\frac{1}{\ ^*{S_{[1,1]}}}\Big(\ ^*{S_{[3,3]}}+ \
^*{S_{[3,2,1]}}(q^{8}-4q^{6}+6q^{4}-8q^{2}+9-8q^{-2}+6q^{-4}-4q^{-6}+q^{-8})+
\ ^*{S_{[3,1,1,1]}}+
\\
+\ ^*{S_{[2,2,2]}}+\
^*{S_{[2,2,1,1]}}(q^{24}-4q^{22}+2q^{20}+12q^{18}-21q^{16}+40q^{12}-44q^{10}-15q^{8}+76q^{6}-54q^{4}-40q^{2}+
\\
+94-40q^{-2}-54q^{-4}+76q^{-6}-15q^{-8}-44q^{-10}+40q^{-12}-21q^{-16}+12q^{-18}+2q^{-20}-4q^{-22}+q^{-24})+
\\
+\
^*{S_{[2,1,1,1,1]}}(q^{16}-4q^{12}+6q^{8}-8q^{4}+9-8q^{-4}+6q^{-8}-4q^{-12}+q^{-16})+
\ ^*{S_{[1,1,1,1,1,1]}}\Big) \ee \be \frac{H_{[2]}}{\
^*{S_{[2]}}}=(q^{8}-3q^{6}+8q^{2}-9-3q^{-2}+15q^{-4}-7q^{-6}-8q^{-8}+9q^{-10}-3q^{-14}+q^{-16})A^{-4}+
\\
+(-q^{14}+2q^{12}+3q^{10}-9q^{8}+3q^{6}+16q^{4}-23q^{2}-8+31q^{-2}-15q^{-4}-
\\
-20q^{-6}+19q^{-8}+2q^{-10}-10q^{-12}+3q^{-14}+2q^{-16}-q^{-18})A^{-2}+
\\
+(q^{18}-3q^{16}+q^{14}+7q^{12}-14q^{10}+2q^{8}+27q^{6}-26q^{4}-13q^{2}+
\\
+47-13q^{-2}-26q^{-4}+27q^{-6}+2q^{-8}-14q^{-10}+7q^{-12}+q^{-14}-3q^{-16}+q^{-18})+
\\
+(-q^{18}+2q^{16}+3q^{14}-10q^{12}+2q^{10}+19q^{8}-20q^{6}-15q^{4}+31q^{2}-
\\
-8-23q^{-2}+16q^{-4}+3q^{-6}-9q^{-8}+3q^{-10}+2q^{-12}-q^{-14})A^{2}+
\\
+(q^{16}-3q^{14}+9q^{10}-8q^{8}-7q^{6}+15q^{4}-3q^{2}-9+8q^{-2}-3q^{-6}+q^{-8})A^{4}=
\\
=\frac{1}{\ ^*{S_{[2]}}}\Big(\ ^*{S_{[6]}}+ \
^*{S_{[5,1]}}(q^{16}-4q^{12}+6q^{8}-8q^{4}+9-8q^{-4}+6q^{-8}-4q^{-12}+q^{-16})+
\\
+\
^*{S_{[4,2]}}(q^{24}-4q^{22}+2q^{20}+12q^{18}-21q^{16}+40q^{12}-44q^{10}-15q^{8}+76q^{6}-54q^{4}-40q^{2}+
\\
+94-40q^{-2}-54q^{-4}+76q^{-6}-15q^{-8}-44q^{-10}+40q^{-12}-21q^{-16}+12q^{-18}+2q^{-20}-4q^{-22}+q^{-24})+
\\
+\ ^*{S_{[4,1,1]}}+ \ ^*{S_{[3,3]}}+ \
^*{S_{[3,2,1]}}(q^{8}-4q^{6}+6q^{4}-8q^{2}+9-8q^{-2}+6q^{-4}-4q^{-6}+q^{-8})+
\ ^*{S_{[2,2,2]}}\Big) \ee

\paragraph{Alexander polynomials}

\be
\mathfrak{A}_{[1]}=-q^{6}+5q^{4}-10q^{2}+13-10q^{-2}+5q^{-4}-q^{-6}
\\
\mathfrak{A}_{[1,1]}=\mathfrak{A}_{[2]}=\mathfrak{A}_{[1]}(q^2)=-q^{12}+5q^{8}-10q^{4}+13-10q^{-4}+5q^{-8}-q^{-12}
\ee

\paragraph{Jones polynomials}

\be
J_{[1]}=q^{8}-4q^{6}+6q^{4}-7q^{2}+9-7q^{-2}+6q^{-4}-4q^{-6}+q^{-8}
\ee \be J_{[1, 1]}=1 \ee \be
J_{[2]}=q^{24}-4q^{22}+2q^{20}+13q^{18}-21q^{16}-4q^{14}+41q^{12}-38q^{10}-20q^{8}+69q^{6}-43q^{4}-36q^{2}+
\\
+81-36q^{-2}-43q^{-4}+69q^{-6}-20q^{-8}-38q^{-10}+41q^{-12}-4q^{-14}-21q^{-16}+13q^{-18}+2q^{-20}-4q^{-22}+q^{-24}
\ee

\paragraph{Special polynomials}

\be \mathfrak{H}_{[1]}=-A^{-2}+3-A^{2} \ee \be
\mathfrak{H}_{[1,1]}=\mathfrak{H}_{[2]}=(\mathfrak{H}_{[1]})^2=A^{-4}-6A^{-2}+11-6A^{2}+A^{4}
\ee

\paragraph{Ooguri-Vafa polynomials}

\be f_{[1, 1]}=\frac{\{A\}^2\{A/q\}\{Aq\}}{\{q\}}
\Big((q^{15}-3q^{13}+q^{11}+6q^{9}-8q^{7}+2q^{5}+4q^{3}+2q^{1}-9q^{-1}+9q^{-3}-3q^{-5})A^{-2}+
\\
+(-q^{11}+3q^{9}-2q^{5}-9q^{3}+23q^{1}-23q^{-1}+9q^{-3}+2q^{-5}-3q^{-9}+q^{-11})+
\\
+(3q^{5}-9q^{3}+9q^{1}-2q^{-1}-4q^{-3}-2q^{-5}+8q^{-7}-6q^{-9}-q^{-11}+3q^{-13}-q^{-15})A^{2}\Big)
\ee \be f_{[2]}=\frac{\{A\}^2\{A/q\}\{Aq\}}{\{q\}}
\Big((3q^{5}-9q^{3}+9q^{1}-2q^{-1}-4q^{-3}-2q^{-5}+8q^{-7}-6q^{-9}-q^{-11}+3q^{-13}-q^{-15})A^{-2}+
\\
+(-q^{11}+3q^{9}-2q^{5}-9q^{3}+23q^{1}-23q^{-1}+9q^{-3}+2q^{-5}-3q^{-9}+q^{-11})+
\\
+(q^{15}-3q^{13}+q^{11}+6q^{9}-8q^{7}+2q^{5}+4q^{3}+2q^{1}-9q^{-1}+9q^{-3}-3q^{-5})A^{2}\Big)
\ee

\paragraph{Special Ooguri-Vafa polynomials}

\be {\mathfrak f}_{[2]}=-{\mathfrak
f}_{[1,1]}=2(A^2-A^{-2})(A-A^{-1})^3 \ee

\subsubsection*{Numbers $N_{R,n,k}$}

\noindent
\begin{tabular}{cc}
$N_{[1]}:$ &
\begin{tabular}{|c|cccc|}
\hline
&&&& \\
$k \backslash n=$&-3 & -1 & 1 & 3 \\
&&&& \\
\hline
&&&& \\
0 & 1 & -4 & 4 & -1 \\
&&&&\\
1 & -1 & 2 & -2 & 1 \\
&&&&\\
2 & -1 & 4 & -4 & 1 \\
&&&&\\
3 & 0 & 1 & -1 & 0 \\
&&&&\\
\hline
\end{tabular}
\end{tabular}

\bigskip

\noindent
\begin{tabular}{cccc}
$N_{[1,1]}:$ &
\begin{tabular}{|c|ccccccc|}
\hline
&&&&&&& \\
$k \backslash n=$& -6 & -4 & -2 & 0 & 2 & 4 & 6 \\
&&&&&&& \\
\hline
&&&&&&& \\
0 & 2 & -6 & 8 & -12 & 18 & -14 & 4 \\
&&&&&&&\\
1 & 4 & -2 & -32 & 54 & -18 & -12 & 6 \\
&&&&&&&\\
2 & 10 & -42 & 54 & -46 & 58 & -44 & 10 \\
&&&&&&&\\
3 & 26 & -131 & 288 & -406 & 388 & -211 & 46 \\
&&&&&&&\\
4 & 26 & -138 & 328 & -542 & 593 & -338 & 71 \\
&&&&&&&\\
5 & 9 & -63 & 161 & -327 & 421 & -244 & 43 \\
&&&&&&&\\
6 & 1 & -13 & 36 & -102 & 154 & -87 & 11 \\
&&&&&&&\\
7 & 0 & -1 & 3 & -16 & 28 & -15 & 1 \\
&&&&&&&\\
8 & 0 & 0 & 0 & -1 & 2 & -1 & 0 \\
&&&&&&&\\
\hline
\end{tabular}
& $N_{[2]}:$ &
\begin{tabular}{|c|ccccccc|}
\hline
&&&&&&& \\
$k \backslash n=$& -6 & -4 & -2 & 0 & 2 & 4 & 6 \\
&&&&&&& \\
\hline
&&&&&&& \\
0 & 4 & -14 & 18 & -12 & 8 & -6 & 2 \\
&&&&&&&\\
1 & 6 & -12 & -18 & 54 & -32 & -2 & 4 \\
&&&&&&&\\
2 & 10 & -44 & 58 & -46 & 54 & -42 & 10 \\
&&&&&&&\\
3 & 46 & -211 & 388 & -406 & 288 & -131 & 26 \\
&&&&&&&\\
4 & 71 & -338 & 593 & -542 & 328 & -138 & 26 \\
&&&&&&&\\
5 & 43 & -244 & 421 & -327 & 161 & -63 & 9 \\
&&&&&&&\\
6 & 11 & -87 & 154 & -102 & 36 & -13 & 1 \\
&&&&&&&\\
7 & 1 & -15 & 28 & -16 & 3 & -1 & 0 \\
&&&&&&&\\
8 & 0 & -1 & 2 & -1 & 0 & 0 & 0 \\
&&&&&&&\\
\hline
\end{tabular}
\end{tabular}

\bigskip

\section*{\fbox{Knot $8_{19}$}}

{\large $(1,3|1,3)=(1,1|1,1|1,1|1,1)$}

\bigskip

\paragraph{HOMFLY polynomials}
\be \frac{H_{[1]}}{\
^*{S_{[1]}}}=A^{-8}\Big(A^{-2}+(-q^{4}-q^{2}-1-q^{-2}-q^{-4})+(q^{6}+q^{2}+1+q^{-2}+q^{-6})A^{2}
\Big)=
\\
=\frac{A^{-8}}{\ ^*{S_{[1]}}}\Big(\ ^*{S_{[3]}}q^{8}- \
^*{S_{[2,1]}}+ \ ^*{S_{[1,1,1]}}q^{-8}\Big) \ee \be
\frac{H_{[1,1]}}{\
^*{S_{[1,1]}}}=A^{-16}q^{32}\Big(q^{-16}A^{-4}+(-q^{-12}-2q^{-14}-q^{-16}-q^{-18}-2q^{-20}-q^{-22}-q^{-24}-q^{-26})A^{-2}+
\\
+(2q^{-10}+2q^{-12}+3q^{-14}+3q^{-16}+5q^{-18}+4q^{-20}+4q^{-22}+3q^{-24}+3q^{-26}+2q^{-28}+2q^{-30}+q^{-32}+q^{-34})+
\\
+(-q^{-8}-2q^{-10}-2q^{-12}-3q^{-14}-5q^{-16}-5q^{-18}-4q^{-20}-5q^{-22}-5q^{-24}-
\\
-4q^{-26}-4q^{-28}-3q^{-30}-2q^{-32}-2q^{-34}-q^{-36}-q^{-38}-q^{-40})A^{2}+
\\
+(q^{-8}+q^{-12}+2q^{-14}+2q^{-16}+q^{-18}+3q^{-20}+2q^{-22}+2q^{-24}+2q^{-26}+
\\
+2q^{-28}+q^{-30}+2q^{-32}+q^{-34}+q^{-36}+q^{-38}+q^{-44})A^{4}
\Big)=
\\
=\frac{A^{-16}q^{32}}{\ ^*{S_{[1,1]}}}\Big(\ ^*{S_{[3,3]}}- \
^*{S_{[3,2,1]}}q^{-8}+ \ ^*{S_{[3,1,1,1]}}q^{-16}+ \
^*{S_{[2,2,2]}}q^{-16}- \ ^*{S_{[2,1,1,1,1]}}q^{-32}+ \
^*{S_{[1,1,1,1,1,1]}}q^{-48}\Big) \ee \be \frac{H_{[2]}}{\
^*{S_{[2]}}}=A^{-16}q^{-32}\Big(q^{16}A^{-4}+(-q^{26}-q^{24}-q^{22}-2q^{20}-q^{18}-q^{16}-2q^{14}-q^{12})A^{-2}+
\\
+(q^{34}+q^{32}+2q^{30}+2q^{28}+3q^{26}+3q^{24}+4q^{22}+4q^{20}+5q^{18}+3q^{16}+3q^{14}+2q^{12}+2q^{10})+
\\
+(-q^{40}-q^{38}-q^{36}-2q^{34}-2q^{32}-3q^{30}-4q^{28}-4q^{26}-5q^{24}-5q^{22}-4q^{20}-5q^{18}-5q^{16}-3q^{14}-2q^{12}-2q^{10}-q^{8})A^{2}+
\\
+(q^{44}+q^{38}+q^{36}+q^{34}+2q^{32}+q^{30}+2q^{28}+2q^{26}+2q^{24}+2q^{22}+3q^{20}+q^{18}+2q^{16}+2q^{14}+q^{12}+q^{8})A^{4}
\Big)=
\\
=\frac{A^{-16}q^{-32}}{\ ^*{S_{[2]}}}\Big(\ ^*{S_{[6]}}q^{48}- \
^*{S_{[5,1]}}q^{32}+ \ ^*{S_{[4,1,1]}}q^{16}+ \ ^*{S_{[3,3]}}q^{16}-
\ ^*{S_{[3,2,1]}}q^{8}+ \ ^*{S_{[2,2,2]}}\Big) \ee

\paragraph{Alexander polynomials}

\be \mathfrak{A}_{[1]}=q^{6}-q^{4}+1-q^{-4}+q^{-6} \ee \be
\mathfrak{A}_{[1,
1]}=\mathfrak{A}_{[2]}=\mathfrak{A}_{[1]}(q^2)=q^{12}-q^{8}+1-q^{-8}+q^{-12}
\ee

\paragraph{Jones polynomials}

\be J_{[1]}=q^{-6}+q^{-10}-q^{-16} \ee \be J_{[1, 1]}=1 \ee \be
J_{[2]}=q^{-12}+q^{-18}+q^{-24}-q^{-26}-q^{-32}-q^{-38}+q^{-40}-q^{-44}+q^{-46}
\ee

\paragraph{Special polynomials}

\be \mathfrak{H}_{[1]}=A^{-10}-5A^{-8}+5A^{-6} \ee \be
\mathfrak{H}_{[1,1]}=\mathfrak{H}_{[2]}=(\mathfrak{H}_{[1]})^2=A^{-20}-10A^{-18}+35A^{-16}-50A^{-14}+25A^{-12}
\ee

\paragraph{Ooguri-Vafa polynomials}

\be f_{[1, 1]}=\frac{\{A\}^2\{A/q\}\{Aq\}(q^2+q^{-2})}{\{q\}}
\Big((q^{13}+q^{5})A^{-18}+(-q^{17}-q^{15}-q^{11}-3q^{9}-2q^{7}-q^{5}-2q^{3}-2q^{1}-q^{-1})A^{-16}+
\\
+(q^{19}+q^{15}+2q^{13}+3q^{11}+q^{9}+3q^{7}+3q^{5}+3q^{3}+q^{1}+2q^{-1}+q^{-3}+q^{-5})A^{-14}\Big)
\ee \be f_{[2]}=\frac{\{A\}^2\{A/q\}\{Aq\}(q^2+q^{-2})}{\{q\}}
\Big((-q^{-5}-q^{-13})A^{-18}+
\\
+(q^{1}+2q^{-1}+2q^{-3}+q^{-5}+2q^{-7}+3q^{-9}+q^{-11}+q^{-15}+q^{-17})A^{-16}+
\\
+(-q^{5}-q^{3}-2q^{1}-q^{-1}-3q^{-3}-3q^{-5}-3q^{-7}-q^{-9}-3q^{-11}-2q^{-13}-q^{-15}-q^{-19})A^{-14}\Big)
\ee

\paragraph{Special Ooguri-Vafa polynomials}

\be {\mathfrak f}_{[2]}=-{\mathfrak
f}_{[1,1]}=-\frac{4(11A^4-7A^2+1)(A-A^{-1})^3}{A^{18}} \ee

\subsubsection*{Numbers $N_{R,n,k}$}

\noindent
\begin{tabular}{cccc}
$N_{[1]}:$ &
\begin{tabular}{|c|cccc|}
\hline
&&&& \\
$k \backslash n=$&-11 & -9 & -7 & -5 \\
&&&& \\
\hline
&&&& \\
0 & -1 & 6 & -10 & 5 \\
&&&&\\
1 & 0 & 5 & -15 & 10 \\
&&&&\\
2 & 0 & 1 & -7 & 6 \\
&&&&\\
3 & 0 & 0 & -1 & 1 \\
&&&&\\
\hline
\end{tabular}
& $N_{[1,1]}:$ &
\begin{tabular}{|c|ccccccc|}
\hline
&&&&&&& \\
$k \backslash n=$& -22 & -20 & -18 & -16 & -14 & -12 & -10 \\
&&&&&&& \\
\hline
&&&&&&& \\
0 & 16 & -146 & 540 & -1020 & 1040 & -546 & 116 \\
&&&&&&&\\
1 & 80 & -775 & 3060 & -6090 & 6410 & -3395 & 710 \\
&&&&&&&\\
2 & 148 & -1709 & 7503 & -15954 & 17363 & -9219 & 1868 \\
&&&&&&&\\
3 & 128 & -2001 & 10254 & -23804 & 26999 & -14365 & 2789 \\
&&&&&&&\\
4 & 56 & -1365 & 8580 & -22281 & 26597 & -14182 & 2595 \\
&&&&&&&\\
5 & 12 & -560 & 4570 & -13657 & 17333 & -9247 & 1549 \\
&&&&&&&\\
6 & 1 & -136 & 1556 & -5557 & 7575 & -4029 & 590 \\
&&&&&&&\\
7 & 0 & -18 & 328 & -1485 & 2196 & -1159 & 138 \\
&&&&&&&\\
8 & 0 & -1 & 39 & -250 & 405 & -211 & 18 \\
&&&&&&&\\
9 & 0 & 0 & 2 & -24 & 43 & -22 & 1 \\
&&&&&&&\\
10 & 0 & 0 & 0 & -1 & 2 & -1 & 0 \\
&&&&&&&\\
\hline
\end{tabular}
\end{tabular}

\bigskip

\noindent
\begin{tabular}{cc}
$N_{[2]}:$ &
\begin{tabular}{|c|ccccccc|}
\hline
&&&&&&& \\
$k \backslash n=$& -22 & -20 & -18 & -16 & -14 & -12 & -10 \\
&&&&&&& \\
\hline
&&&&&&& \\
0 & 20 & -190 & 720 & -1380 & 1420 & -750 & 160 \\
&&&&&&&\\
1 & 130 & -1245 & 4890 & -9710 & 10220 & -5425 & 1140 \\
&&&&&&&\\
2 & 314 & -3403 & 14459 & -30206 & 32619 & -17325 & 3542 \\
&&&&&&&\\
3 & 367 & -4996 & 24019 & -53906 & 60220 & -32020 & 6316 \\
&&&&&&&\\
4 & 230 & -4367 & 24779 & -61035 & 71182 & -37920 & 7131 \\
&&&&&&&\\
5 & 79 & -2380 & 16653 & -46053 & 56575 & -30177 & 5303 \\
&&&&&&&\\
6 & 14 & -816 & 7422 & -23680 & 30910 & -16472 & 2622 \\
&&&&&&&\\
7 & 1 & -171 & 2177 & -8320 & 11638 & -6175 & 850 \\
&&&&&&&\\
8 & 0 & -20 & 404 & -1963 & 2967 & -1561 & 173 \\
&&&&&&&\\
9 & 0 & -1 & 43 & -297 & 489 & -254 & 20 \\
&&&&&&&\\
10 & 0 & 0 & 2 & -26 & 47 & -24 & 1 \\
&&&&&&&\\
11 & 0 & 0 & 0 & -1 & 2 & -1 & 0 \\
&&&&&&&\\
\hline
\end{tabular}
\end{tabular}

\bigskip

\section*{\fbox{Knot $8_{20}$}}

{\large $(-1,-3|-1,3)$}

\bigskip

\paragraph{HOMFLY polynomials}
\be \frac{H_{[1]}}{\
^*{S_{[1]}}}=A^2\Big((-q^{2}+1-q^{-2})A^{-2}+(q^{4}+2+q^{-4})+(-q^{2}-q^{-2})A^{2}
\Big)=
\\
=\frac{A^2}{\ ^*{S_{[1]}}}\Big(\ ^*{S_{[3]}}q^{-2}+ \
^*{S_{[2,1]}}(-q^{6}+q^{4}-q^{2}+1-q^{-2}+q^{-4}-q^{-6})+ \
^*{S_{[1,1,1]}}q^{2}\Big) \ee \be \frac{H_{[1,1]}}{\
^*{S_{[1,1]}}}=A^4q^{-8}\Big((-q^{12}+q^{10}+q^{8}-q^{6}+q^{2})A^{-4}+
\\
+(-q^{12}-q^{10}+q^{8}-q^{6}-3q^{4}-q^{2}-q^{-2}-q^{-4})A^{-2}+(q^{16}+q^{12}+q^{10}+2q^{8}+3q^{6}+3q^{4}+q^{2}+3+3q^{-2}+q^{-4}+q^{-8})+
\\
+(-q^{12}-q^{10}-q^{8}-2q^{6}-2q^{4}-2q^{2}-3-2q^{-2}-q^{-6}-q^{-8})A^{2}+(q^{6}+q^{2}+1+q^{-6})A^{4}
\Big)=
\\
=\frac{A^4q^{-8}}{\ ^*{S_{[1,1]}}}\Big(\ ^*{S_{[3,3]}}+ \
^*{S_{[3,2,1]}}(-q^{8}+q^{6}-q^{4}+q^{2}-1+q^{-2}-q^{-4})+ \
^*{S_{[3,1,1,1]}}q^{4}+ \ ^*{S_{[2,2,2]}}q^{4}+
\\
+\
^*{S_{[2,2,1,1]}}(q^{22}-q^{20}+q^{16}-q^{14}-q^{2}+1+q^{-2}-2q^{-4}+2q^{-8}-q^{-10}-q^{-12}+q^{-14})+
\\
+\ ^*{S_{[2,1,1,1,1]}}(-q^{20}+q^{16}-q^{12}+q^{8}-q^{4}+1-q^{-4})+
\ ^*{S_{[1,1,1,1,1,1]}}q^{12}\Big) \ee \be \frac{H_{[2]}}{\
^*{S_{[2]}}}=A^4q^{8}\Big((q^{-2}-q^{-6}+q^{-8}+q^{-10}-q^{-12})A^{-4}+(-q^{4}-q^{2}-q^{-2}-3q^{-4}-q^{-6}+q^{-8}-q^{-10}-q^{-12})A^{-2}+
\\
+(q^{8}+q^{4}+3q^{2}+3+q^{-2}+3q^{-4}+3q^{-6}+2q^{-8}+q^{-10}+q^{-12}+q^{-16})+
\\
+(-q^{8}-q^{6}-2q^{2}-3-2q^{-2}-2q^{-4}-2q^{-6}-q^{-8}-q^{-10}-q^{-12})A^{2}+(q^{6}+1+q^{-2}+q^{-6})A^{4}
\Big)=
\\
=\frac{A^4q^{8}}{\ ^*{S_{[2]}}}\Big(\ ^*{S_{[6]}}q^{-12}+ \
^*{S_{[5,1]}}(-q^{4}+1-q^{-4}+q^{-8}-q^{-12}+q^{-16}-q^{-20})+
\\
+\
^*{S_{[4,2]}}(q^{14}-q^{12}-q^{10}+2q^{8}-2q^{4}+q^{2}+1-q^{-2}-q^{-14}+q^{-16}-q^{-20}+q^{-22})+
\ ^*{S_{[4,1,1]}}q^{-4}+ \ ^*{S_{[3,3]}}q^{-4}+
\\
+\ ^*{S_{[3,2,1]}}(-q^{4}+q^{2}-1+q^{-2}-q^{-4}+q^{-6}-q^{-8})+ \
^*{S_{[2,2,2]}}\Big) \ee

\paragraph{Alexander polynomials}

\be \mathfrak{A}_{[1]}=q^{8}-2q^{4}+3-2q^{-4}+q^{-8} \ee \be
\mathfrak{A}_{[1,
1]}=\mathfrak{A}_{[2]}=\mathfrak{A}_{[1]}(q^2)=q^{8}-2q^{4}+3-2q^{-4}+q^{-8}
\ee

\paragraph{Jones polynomials}

\be J_{[1]}=-q^{10}+q^{8}-q^{6}+2q^{4}-q^{2}+2-q^{-2} \ee \be J_{[1,
1]}=1 \ee \be
J_{[2]}=q^{30}-q^{28}-q^{26}+2q^{24}-q^{22}-2q^{20}+2q^{18}-2q^{14}+2q^{12}+q^{10}-2q^{8}+q^{6}+2q^{4}-2q^{2}+1+q^{-2}-q^{-4}
\ee

\paragraph{Special polynomials}

\be \mathfrak{H}_{[1]}=-1+4A^{2}-2A^{4} \ee \be \mathfrak{H}_{[1,
1]}=\mathfrak{H}_{[2]}=(\mathfrak{H}_{[1]})^2=1-8A^{2}+20A^{4}-16A^{6}+4A^{8}
\ee

\paragraph{Ooguri-Vafa polynomials}

\be f_{[1, 1]}=\frac{\{A\}^2\{A/q\}\{Aq\}}{\{q\}}
\Big((-q^{3}+q^{1}-q^{-1})A^{2}+
\\
+(q^{1}+q^{-1}+q^{-3}+q^{-5}+q^{-7}+q^{-9})A^4+(-q^{3}-2q^{-1}-q^{-3}-2q^{-5}-q^{-7}-q^{-9}-q^{-13})A^{6}\Big)
\ee \be f_{[2]}=\frac{\{A\}^2\{A/q\}\{Aq\}}{\{q\}}
\Big((q^{1}-q^{-1}+q^{-3})A^{2}+
\\
+(-q^{9}-q^{7}-q^{5}-q^{3}-q^{1}-q^{-1})A^4+(q^{13}+q^{9}+q^{7}+2q^{5}+q^{3}+2q^{1}+q^{-3})A^{6}\Big)
\ee

\paragraph{Special Ooguri-Vafa polynomials}

\be {\mathfrak f}_{[2]}=-{\mathfrak
f}_{[1,1]}=A^4(3A^2-A^{-2})^2(A-A^{-1})^3 \ee

\subsubsection*{Numbers $N_{R,n,k}$}

\noindent
\begin{tabular}{cc}
$N_{[1]}:$ &
\begin{tabular}{|c|cccc|}
\hline
&&&& \\
$k \backslash n=$&-1 & 1 & 3 & 5 \\
&&&& \\
\hline
&&&& \\
0 & 1 & -5 & 6 & -2 \\
&&&&\\
1 & 1 & -5 & 5 & -1 \\
&&&&\\
2 & 0 & -1 & 1 & 0 \\
&&&&\\
\hline
\end{tabular}
\end{tabular}

\bigskip

\noindent
\begin{tabular}{cccc}
$N_{[1,1]}:$ &
\begin{tabular}{|c|ccccccc|}
\hline
&&&&&&& \\
$k \backslash n=$& -2 & 0 & 2 & 4 & 6 & 8 & 10 \\
&&&&&&& \\
\hline
&&&&&&& \\
0 & 0 & -15 & 85 & -190 & 210 & -115 & 25 \\
&&&&&&&\\
1 & 0 & -35 & 250 & -645 & 775 & -440 & 95 \\
&&&&&&&\\
2 & 0 & -28 & 302 & -953 & 1267 & -743 & 155 \\
&&&&&&&\\
3 & 0 & -9 & 193 & -781 & 1148 & -680 & 129 \\
&&&&&&&\\
4 & 0 & -1 & 69 & -377 & 607 & -354 & 56 \\
&&&&&&&\\
5 & 0 & 0 & 13 & -106 & 185 & -104 & 12 \\
&&&&&&&\\
6 & 0 & 0 & 1 & -16 & 30 & -16 & 1 \\
&&&&&&&\\
7 & 0 & 0 & 0 & -1 & 2 & -1 & 0 \\
&&&&&&&\\
\hline
\end{tabular}
& $N_{[2]}:$ &
\begin{tabular}{|c|ccccccc|}
\hline
&&&&&&& \\
$k \backslash n=$& -2 & 0 & 2 & 4 & 6 & 8 & 10 \\
&&&&&&& \\
\hline
&&&&&&& \\
0 & 0 & -5 & 46 & -114 & 131 & -73 & 16 \\
&&&&&&&\\
1 & 0 & -10 & 111 & -319 & 400 & -231 & 50 \\
&&&&&&&\\
2 & 0 & -6 & 104 & -373 & 521 & -309 & 63 \\
&&&&&&&\\
3 & 0 & -1 & 48 & -231 & 359 & -212 & 37 \\
&&&&&&&\\
4 & 0 & 0 & 11 & -79 & 135 & -77 & 10 \\
&&&&&&&\\
5 & 0 & 0 & 1 & -14 & 26 & -14 & 1 \\
&&&&&&&\\
6 & 0 & 0 & 0 & -1 & 2 & -1 & 0 \\
&&&&&&&\\
\hline
\end{tabular}
\end{tabular}

\bigskip

\section*{\fbox{Knot $8_{21}$}}

{\large $(-2,2|-1,-3)$}

\bigskip

\paragraph{HOMFLY polynomials}
\be \frac{H_{[1]}}{\
^*{S_{[1]}}}=A^4\Big((2q^{2}-1+2q^{-2})A^{-2}+(-q^{4}+q^{2}-3+q^{-2}-q^{-4})+(q^{2}-1+q^{-2})A^{2}
\Big)=
\\
=\frac{A^4}{\ ^*{S_{[1]}}}\Big(\ ^*{S_{[3]}}q^{-4}+ \
^*{S_{[2,1]}}(q^{6}-2q^{4}+2q^{2}-3+2q^{-2}-2q^{-4}+q^{-6})+ \
^*{S_{[1,1,1]}}q^{4}\Big) \ee
\be \frac{H_{[1,1]}}{\
^*{S_{[1,1]}}}=A^8q^{-16}\Big((q^{22}+3q^{20}-2q^{18}+5q^{14}-q^{10}+3q^{8})A^{-4}+
\\
+(-q^{24}-2q^{22}+q^{20}-8q^{16}-2q^{14}+4q^{12}-6q^{10}-4q^{8}+2q^{6}-q^{4}-q^{2})A^{-2}+
\\
+(q^{22}+3q^{18}+3q^{16}-3q^{14}+3q^{12}+8q^{10}-3q^{8}-q^{6}+4q^{4}-1+q^{-2})+
\\
+(-q^{18}-4q^{12}+3q^{8}-3q^{6}-2q^{4}+2q^{2}-q^{-2})A^{2}+(q^{12}-q^{10}+2q^{6}-q^{4}-q^{2}+1)A^{4}
\Big)=
\\
=\frac{A^8q^{-16}}{\ ^*{S_{[1,1]}}}\Big(\ ^*{S_{[3,3]}}+ \
^*{S_{[3,2,1]}}(q^{10}-2q^{8}+2q^{6}-3q^{4}+2q^{2}-2+q^{-2})+ \
^*{S_{[3,1,1,1]}}q^{8}+ \ ^*{S_{[2,2,2]}}q^{8}+
\\
+\
^*{S_{[2,2,1,1]}}(q^{28}-2q^{26}+3q^{22}-5q^{20}+2q^{18}+5q^{16}-8q^{14}+3q^{12}+
\\
+7q^{10}-9q^{8}+8q^{4}-5q^{2}-3+5q^{-2}-q^{-4}-2q^{-6}+q^{-8})+
\\
+\
^*{S_{[2,1,1,1,1]}}(q^{28}-2q^{24}+2q^{20}-3q^{16}+2q^{12}-2q^{8}+q^{4})+
\ ^*{S_{[1,1,1,1,1,1]}}q^{24}\Big) \ee
\be \frac{H_{[2]}}{\
^*{S_{[2]}}}=A^8q^{16}\Big((3q^{-8}-q^{-10}+5q^{-14}-2q^{-18}+3q^{-20}+q^{-22})A^{-4}+
\\
+(-q^{-2}-q^{-4}+2q^{-6}-4q^{-8}-6q^{-10}+4q^{-12}-2q^{-14}-8q^{-16}+q^{-20}-2q^{-22}-q^{-24})A^{-2}+
\\
+(q^{2}-1+4q^{-4}-q^{-6}-3q^{-8}+8q^{-10}+3q^{-12}-3q^{-14}+3q^{-16}+3q^{-18}+q^{-22})+
\\
+(-q^{2}+2q^{-2}-2q^{-4}-3q^{-6}+3q^{-8}-4q^{-12}-q^{-18})A^{2}+
\\
+(1-q^{-2}-q^{-4}+2q^{-6}-q^{-10}+q^{-12})A^{4} \Big)= \ee
\be
=\frac{A^8q^{16}}{\ ^*{S_{[2]}}}\Big(\ ^*{S_{[6]}}q^{-24}+ \
^*{S_{[5,1]}}(q^{-4}-2q^{-8}+2q^{-12}-3q^{-16}+2q^{-20}-2q^{-24}+q^{-28})+
\\
+\
^*{S_{[4,2]}}(q^{8}-2q^{6}-q^{4}+5q^{2}-3-5q^{-2}+8q^{-4}-9q^{-8}+7q^{-10}+
\\
+3q^{-12}-8q^{-14}+5q^{-16}+2q^{-18}-5q^{-20}+3q^{-22}-2q^{-26}+q^{-28})+
\\
+\ ^*{S_{[4,1,1]}}q^{-8}+ \ ^*{S_{[3,3]}}q^{-8}+ \
^*{S_{[3,2,1]}}(q^{2}-2+2q^{-2}-3q^{-4}+2q^{-6}-2q^{-8}+q^{-10})+ \
^*{S_{[2,2,2]}}\Big)\nn \ee

\paragraph{Alexander polynomials}

\be \mathfrak{A}_{[1]}=-q^{4}+4q^{2}-5+4q^{-2}-q^{-4} \ee \be
\mathfrak{A}_{[1,
1]}=\mathfrak{A}_{[2]}=\mathfrak{A}_{[1]}(q^2)=-q^{8}+4q^{4}-5+4q^{-4}-q^{-8}
\ee

\paragraph{Jones polynomials}

\be J_{[1]}=q^{14}-2q^{12}+2q^{10}-3q^{8}+3q^{6}-2q^{4}+2q^{2} \ee
\be J_{[1, 1]}=1 \ee \be
J_{[2]}=q^{40}-2q^{38}-q^{36}+5q^{34}-3q^{32}-4q^{30}+8q^{28}-2q^{26}-8q^{24}+
\\
+10q^{22}-q^{20}-10q^{18}+10q^{16}-8q^{12}+6q^{10}+q^{8}-4q^{6}+2q^{4}+q^{2}
\ee

\paragraph{Special polynomials}

\be \mathfrak{H}_{[1]}=3A^{2}-3A^{4}+A^{6} \ee \be
\mathfrak{H}_{[1,1]}=\mathfrak{H}_{[2]}=(\mathfrak{H}_{[1]})^2=9A^{4}-18A^{6}+15A^{8}-6A^{10}+A^{12}
\ee

\paragraph{Ooguri-Vafa polynomials}

\be f_{[1, 1]}=\frac{\{A\}^2\{A/q\}\{Aq\}}{\{q\}}
\Big((q^{5}-2q^{-1}+2q^{-3}-3q^{-5})A^{6}+
\\
+(q^{5}+2q^{1}-2q^{-1}+3q^{-3}+q^{-5}-q^{-7}+q^{-9}+q^{-11})A^8+(-q^{-1}-q^{-9}+q^{-13}-q^{-15})A^{10}\Big)
\ee \be f_{[2]}=\frac{\{A\}^2\{A/q\}\{Aq\}}{\{q\}}
\Big((3q^{5}-2q^{3}+2q^{1}-q^{-5})A^{6}+
\\
+(-q^{11}-q^{9}+q^{7}-q^{5}-3q^{3}+2q^{1}-2q^{-1}-q^{-5})A^8+(q^{15}-q^{13}+q^{9}+q^{1})A^{10}\Big)
\ee

\paragraph{Special Ooguri-Vafa polynomials}

\be {\mathfrak f}_{[2]}=-{\mathfrak
f}_{[1,1]}=2A^8(A^2-3+A^{-2})(A-A^{-1})^3 \ee

\subsubsection*{Numbers $N_{R,n,k}$}

\noindent
\begin{tabular}{cc}
$N_{[1]}:$ &
\begin{tabular}{|c|cccc|}
\hline
&&&& \\
$k \backslash n=$& 1 & 3 & 5 & 7 \\
&&&& \\
\hline
&&&& \\
0 & -3 & 6 & -4 & 1 \\
&&&&\\
1 & -2 & 5 & -4 & 1 \\
&&&&\\
2 & 0 & 1 & -1 & 0 \\
&&&&\\
\hline
\end{tabular}
\end{tabular}

\bigskip

\noindent
\begin{tabular}{cccc}
$N_{[1,1]}:$ &
\begin{tabular}{|c|ccccccc|}
\hline
&&&&&&& \\
$k \backslash n=$& 2 & 4 & 6 & 8 & 10 & 12 & 14 \\
&&&&&&& \\
\hline
&&&&&&& \\
0 & 9 & -48 & 109 & -136 & 99 & -40 & 7 \\
&&&&&&&\\
1 & 11 & -104 & 348 & -582 & 529 & -250 & 48 \\
&&&&&&&\\
2 & 3 & -95 & 526 & -1193 & 1320 & -708 & 147 \\
&&&&&&&\\
3 & 0 & -46 & 468 & -1424 & 1846 & -1062 & 218 \\
&&&&&&&\\
4 & 0 & -11 & 253 & -1034 & 1519 & -893 & 166 \\
&&&&&&&\\
5 & 0 & -1 & 81 & -458 & 743 & -431 & 66 \\
&&&&&&&\\
6 & 0 & 0 & 14 & -120 & 211 & -118 & 13 \\
&&&&&&&\\
7 & 0 & 0 & 1 & -17 & 32 & -17 & 1 \\
&&&&&&&\\
8 & 0 & 0 & 0 & -1 & 2 & -1 & 0 \\
&&&&&&&\\
\hline
\end{tabular}
& $N_{[2]}:$ &
\begin{tabular}{|c|ccccccc|}
\hline
&&&&&&& \\
$k \backslash n=$& 2 & 4 & 6 & 8 & 10 & 12 & 14 \\
&&&&&&& \\
\hline
&&&&&&& \\
0 & 7 & -34 & 71 & -84 & 61 & -26 & 5 \\
&&&&&&&\\
1 & 7 & -58 & 185 & -314 & 301 & -152 & 31 \\
&&&&&&&\\
2 & 1 & -36 & 219 & -542 & 642 & -360 & 76 \\
&&&&&&&\\
3 & 0 & -10 & 148 & -521 & 723 & -425 & 85 \\
&&&&&&&\\
4 & 0 & -1 & 58 & -289 & 453 & -266 & 45 \\
&&&&&&&\\
5 & 0 & 0 & 12 & -91 & 157 & -89 & 11 \\
&&&&&&&\\
6 & 0 & 0 & 1 & -15 & 28 & -15 & 1 \\
&&&&&&&\\
7 & 0 & 0 & 0 & -1 & 2 & -1 & 0 \\
&&&&&&&\\
\hline
\end{tabular}
\end{tabular}

\bigskip

\section*{\fbox{Knot $10_{139}$}}

{\large $(2,3|1,4)$}

\bigskip

\paragraph{HOMFLY polynomials}
\be \frac{H_{[1]}}{\
^*{S_{[1]}}}=A^{-10}\Big((q^{2}-1+q^{-2})A^{-2}+(-q^{6}-q^{4}-2-q^{-4}-q^{-6})+(q^{8}+q^{4}+q^{2}+q^{-2}+q^{-4}+q^{-8})A^{2}
\Big)=
\\
=\frac{A^{-10}}{\ ^*{S_{[1]}}}\Big(\ ^*{S_{[3]}}q^{10}+ \
^*{S_{[2,1]}}(-q^{4}+q^{2}-1+q^{-2}-q^{-4})+ \
^*{S_{[1,1,1]}}q^{-10}\Big) \ee \be \frac{H_{[1,1]}}{\
^*{S_{[1,1]}}}=A^{-20}q^{40}\Big((q^{-16}-q^{-18}+2q^{-22}-q^{-24}-q^{-26}+q^{-28})A^{-4}+
\\
+(-q^{-12}-q^{-14}-q^{-18}-3q^{-20}-q^{-22}+q^{-24}-2q^{-26}-2q^{-28}-q^{-36}-q^{-38})A^{-2}+
\\
+(2q^{-10}+q^{-12}+2q^{-14}+3q^{-16}+4q^{-18}+4q^{-20}+4q^{-22}+3q^{-24}+6q^{-26}+
\\
+3q^{-28}+q^{-30}+3q^{-32}+4q^{-34}+2q^{-36}+q^{-38}+q^{-40}+2q^{-42}+q^{-44}+q^{-46})+
\\
+(-q^{-8}-2q^{-10}-q^{-12}-3q^{-14}-5q^{-16}-3q^{-18}-5q^{-20}-6q^{-22}-5q^{-24}-6q^{-26}-4q^{-28}-4q^{-30}-6q^{-32}-
\\
-4q^{-34}-q^{-36}-3q^{-38}-5q^{-40}-2q^{-42}-q^{-44}-2q^{-46}-q^{-48}-q^{-50}-q^{-52})A^{2}+
\\
+(q^{-8}+q^{-12}+2q^{-14}+q^{-16}+q^{-18}+4q^{-20}+q^{-22}+3q^{-24}+3q^{-26}+q^{-28}+3q^{-30}+
\\
+3q^{-32}+2q^{-36}+3q^{-38}+q^{-40}+2q^{-44}+q^{-46}+q^{-48}+q^{-50}+q^{-56})A^{4}
\Big)=
\\
=\frac{A^{-20}q^{40}}{\ ^*{S_{[1,1]}}}\Big(\ ^*{S_{[3,3]}}+ \
^*{S_{[3,2,1]}}(-q^{-6}+q^{-8}-q^{-10}+q^{-12}-q^{-14})+ \
^*{S_{[3,1,1,1]}}q^{-20}+ \ ^*{S_{[2,2,2]}}q^{-20}+
\\
+\
^*{S_{[2,2,1,1]}}(q^{-14}-q^{-16}-q^{-18}+2q^{-20}-2q^{-24}+q^{-26}+q^{-28}-2q^{-30}+q^{-32}-q^{-36}+q^{-38})+
\\
+\ ^*{S_{[2,1,1,1,1]}}(-q^{-32}+q^{-36}-q^{-40}+q^{-44}-q^{-48})+ \
^*{S_{[1,1,1,1,1,1]}}q^{-60}\Big) \ee \be \frac{H_{[2]}}{\
^*{S_{[2]}}}=A^{-20}q^{-40}\Big((q^{28}-q^{26}-q^{24}+2q^{22}-q^{18}+q^{16})A^{-4}+
\\
+(-q^{38}-q^{36}-2q^{28}-2q^{26}+q^{24}-q^{22}-3q^{20}-q^{18}-q^{14}-q^{12})A^{-2}+
\\
+(q^{46}+q^{44}+2q^{42}+q^{40}+q^{38}+2q^{36}+4q^{34}+3q^{32}+q^{30}+3
\\
+q^{28}+6q^{26}+3q^{24}+4q^{22}+4q^{20}+4q^{18}+3q^{16}+2q^{14}+q^{12}+2q^{10})+
\\
+(-q^{52}-q^{50}-q^{48}-2q^{46}-q^{44}-2q^{42}-5q^{40}-3q^{38}-q^{36}-4q^{34}-6q^{32}-4q^{30}-
\\
L-4q^{28}-6q^{26}-5q^{24}-6q^{22}-5q^{20}-3q^{18}-5q^{16}-3q^{14}-q^{12}-2q^{10}-q^{8})A^{2}+
\\
+(q^{56}+q^{50}+q^{48}+q^{46}+2q^{44}+q^{40}+3q^{38}+2q^{36}+3q^{32}+3q^{30}+
\\
+q^{28}+3q^{26}+3q^{24}+q^{22}+4q^{20}+q^{18}+q^{16}+2q^{14}+q^{12}+q^{8})A^{4}
\Big)=
\\
=\frac{A^{-20}q^{-40}}{\ ^*{S_{[2]}}}\Big(\ ^*{S_{[6]}}q^{60}+ \
^*{S_{[5,1]}}(-q^{48}+q^{44}-q^{40}+q^{36}-q^{32})+
\\
+\
^*{S_{[4,2]}}(q^{38}-q^{36}+q^{32}-2q^{30}+q^{28}+q^{26}-2q^{24}+2q^{20}-q^{18}-q^{16}+q^{14})+
\\
+\ ^*{S_{[4,1,1]}}q^{20}+ \ ^*{S_{[3,3]}}q^{20}+ \
^*{S_{[3,2,1]}}(-q^{14}+q^{12}-q^{10}+q^{8}-q^{6})+ \
^*{S_{[2,2,2]}}\Big) \ee

\paragraph{Alexander polynomials}

\be \mathfrak{A}_{[1]}=q^{8}-q^{6}+2q^{2}-3+2q^{-2}-q^{-6}+q^{-8}
\ee \be \mathfrak{A}_{[1,
1]}=\mathfrak{A}_{[2]}=\mathfrak{A}_{[1]}(q^2)=q^{16}-q^{12}+2q^{4}-3+2q^{-4}-q^{-12}+q^{-16}
\ee

\paragraph{Jones polynomials}

\be J_{[1]}=q^{-8}+q^{-12}-q^{-16}+q^{-18}-q^{-20}+q^{-22}-q^{-24}
\ee \be J_{[1, 1]}=1 \ee \be
J_{[2]}=q^{-16}+q^{-22}+q^{-28}-2q^{-30}+2q^{-34}-2q^{-36}-q^{-38}+3q^{-40}-3q^{-44}+
\\
+2q^{-46}+q^{-48}-4q^{-50}+2q^{-52}+q^{-54}-3q^{-56}+q^{-58}+2q^{-60}-q^{-62}-q^{-64}+q^{-66}
\ee

\paragraph{Special polynomials}

\be \mathfrak{H}_{[1]}=A^{-12}-6A^{-10}+6A^{-8} \ee \be
\mathfrak{H}_{[1,1]}=\mathfrak{H}_{[2]}=(\mathfrak{H}_{[1]})^2=A^{-24}-12A^{-22}+48A^{-20}-72A^{-18}+36A^{-16}
\ee

\paragraph{Ooguri-Vafa polynomials}

\be f_{[1, 1]}=\frac{\{A\}^2\{A/q\}\{Aq\}}{\{q\}}
\Big((q^{23}-q^{21}+q^{19}+q^{17}+q^{11}+q^{7}+q^{1})A^{-22}+
\\
+(-q^{27}-q^{23}-2q^{21}-2q^{19}-2q^{17}-3q^{15}-4q^{13}-4q^{11}-
\\
-4q^{9}-3q^{7}-4q^{5}-4q^{3}-2q^{1}-2q^{-1}-2q^{-3}-q^{-5}-q^{-7})A^{-20}+
\\
+(q^{29}+2q^{25}+2q^{23}+3q^{21}+3q^{19}+7q^{17}+4q^{15}+9q^{13}+6q^{11}+8q^{9}+
\\
+8q^{7}+8q^{5}+5q^{3}+7q^{1}+4q^{-1}+4q^{-3}+2q^{-5}+2q^{-7}+q^{-9}+q^{-11})A^{-18}\Big)
\ee \be f_{[2]}=\frac{\{A\}^2\{A/q\}\{Aq\}}{\{q\}}
\Big((-q^{-1}-q^{-7}-q^{-11}-q^{-17}-q^{-19}+q^{-21}-q^{-23})A^{-22}+
\\
+(q^{7}+q^{5}+2q^{3}+2q^{1}+2q^{-1}+4q^{-3}+4q^{-5}+3q^{-7}+4q^{-9}+4q^{-11}+
\\
+4q^{-13}+3q^{-15}+2q^{-17}+2q^{-19}+2q^{-21}+q^{-23}+q^{-27})A^{-20}+
\\
+(-q^{11}-q^{9}-2q^{7}-2q^{5}-4q^{3}-4q^{1}-7q^{-1}-5q^{-3}-8q^{-5}-8q^{-7}-8q^{-9}-6q^{-11}-
\\
-9q^{-13}-4q^{-15}-7q^{-17}-3q^{-19}-3q^{-21}-2q^{-23}-2q^{-25}-q^{-29})A^{-18}\Big)
\ee

\paragraph{Special Ooguri-Vafa polynomials}

\be {\mathfrak f}_{[2]}=-{\mathfrak
f}_{[1,1]}=-\frac{(159A^8-360A^6+272A^4-76A^2+7)(A-A^{-1})^3}{A^{26}}
\ee

\subsubsection*{Numbers $N_{R,n,k}$}

\noindent
\begin{tabular}{cccc}
$N_{[1]}:$ &
\begin{tabular}{|c|cccc|}
\hline
&&&& \\
$k \backslash n=$& -13 & -11 & -9 & -7 \\
&&&& \\
\hline
&&&& \\
0 & -1 & 7 & -12 & 6 \\
&&&&\\
1 & -1 & 14 & -34 & 21 \\
&&&&\\
2 & 0 & 7 & -28 & 21 \\
&&&&\\
3 & 0 & 1 & -9 & 8 \\
&&&&\\
4 & 0 & 0 & -1 & 1 \\
&&&&\\
\hline
\end{tabular}
& $N_{[1,1]}:$ &
\begin{tabular}{|c|ccccccc|}
\hline
&&&&&&& \\
$k \backslash n=$& -26 & -24 & -22 & -20 & -18 & -16 & -14 \\
&&&&&&& \\
\hline
&&&&&&& \\
0 & 26 & -271 & 1149 & -2406 & 2644 & -1467 & 325 \\
&&&&&&&\\
1 & 283 & -2908 & 12519 & -26725 & 29712 & -16483 & 3602 \\
&&&&&&&\\
2 & 1231 & -13215 & 59305 & -130641 & 147643 & -81894 & 17571 \\
&&&&&&&\\
3 & 2846 & -33702 & 161724 & -372044 & 429324 & -237978 & 49830 \\
&&&&&&&\\
4 & 3939 & -54000 & 283511 & -689018 & 815314 & -451252 & 91506 \\
&&&&&&&\\
5 & 3445 & -57565 & 338189 & -878533 & 1070393 & -590788 & 114859 \\
&&&&&&&\\
6 & 1939 & -42074 & 283375 & -797042 & 1004073 & -551784 & 101513 \\
&&&&&&&\\
7 & 697 & -21353 & 169548 & -524186 & 685766 & -374583 & 64111 \\
&&&&&&&\\
8 & 154 & -7507 & 72733 & -251892 & 343856 & -186357 & 29013 \\
&&&&&&&\\
9 & 19 & -1792 & 22182 & -88275 & 126395 & -67847 & 9318 \\
&&&&&&&\\
10 & 1 & -277 & 4692 & -22269 & 33635 & -17852 & 2070 \\
&&&&&&&\\
11 & 0 & -25 & 654 & -3931 & 6302 & -3302 & 302 \\
&&&&&&&\\
12 & 0 & -1 & 54 & -460 & 788 & -407 & 26 \\
&&&&&&&\\
13 & 0 & 0 & 2 & -32 & 59 & -30 & 1 \\
&&&&&&&\\
14 & 0 & 0 & 0 & -1 & 2 & -1 & 0 \\
&&&&&&&\\
\hline
\end{tabular}
\end{tabular}

\bigskip

\noindent
\begin{tabular}{cc}
$N_{[2]}:$ &
\begin{tabular}{|c|ccccccc|}
\hline
&&&&&&& \\
$k \backslash n=$& -26 & -24 & -22 & -20 & -18 & -16 & -14 \\
&&&&&&& \\
\hline
&&&&&&& \\
0 & 31 & -333 & 1434 & -3026 & 3339 & -1857 & 412 \\
&&&&&&&\\
1 & 396 & -4107 & 17757 & -37965 & 42249 & -23472 & 5142 \\
&&&&&&&\\
2 & 2031 & -21561 & 95976 & -210260 & 237023 & -131551 & 28342 \\
&&&&&&&\\
3 & 5548 & -63635 & 299280 & -680083 & 780367 & -432718 & 91241 \\
&&&&&&&\\
4 & 9162 & -118624 & 602538 & -1436213 & 1684844 & -933053 & 191346 \\
&&&&&&&\\
5 & 9738 & -148552 & 831465 & -2101076 & 2529808 & -1397916 & 276533 \\
&&&&&&&\\
6 & 6853 & -129374 & 814438 & -2205943 & 2736313 & -1506797 & 284510 \\
&&&&&&&\\
7 & 3212 & -79780 & 577794 & -1698237 & 2178428 & -1193693 & 212276 \\
&&&&&&&\\
8 & 988 & -35018 & 299618 & -969806 & 1291661 & -703214 & 115771 \\
&&&&&&&\\
9 & 191 & -10859 & 113471 & -412141 & 572417 & -309146 & 46067 \\
&&&&&&&\\
10 & 21 & -2323 & 31010 & -129656 & 188672 & -100927 & 13203 \\
&&&&&&&\\
11 & 1 & -326 & 5950 & -29730 & 45559 & -24104 & 2650 \\
&&&&&&&\\
12 & 0 & -27 & 760 & -4821 & 7823 & -4088 & 353 \\
&&&&&&&\\
13 & 0 & -1 & 58 & -523 & 904 & -466 & 28 \\
&&&&&&&\\
14 & 0 & 0 & 2 & -34 & 63 & -32 & 1 \\
&&&&&&&\\
15 & 0 & 0 & 0 & -1 & 2 & -1 & 0 \\
&&&&&&&\\
\hline
\end{tabular}
\end{tabular}
\end{footnotesize}

\end{document}